\documentclass[english]{elsarticle}
\usepackage[LGR,T1]{fontenc}
\usepackage[latin9]{inputenc}
\usepackage{geometry}
\usepackage{color}
\usepackage{float}
\usepackage{amsmath}
\usepackage{amssymb}
\usepackage{stackrel}
\usepackage{graphicx}

\makeatletter

\DeclareRobustCommand{\greektext}{%
  \fontencoding{LGR}\selectfont\def\encodingdefault{LGR}}
\DeclareRobustCommand{\textgreek}[1]{\leavevmode{\greektext #1}}
\ProvideTextCommand{\~}{LGR}[1]{\char126#1}

\providecommand{\tabularnewline}{\\}

\@ifundefined{showcaptionsetup}{}{%
 \PassOptionsToPackage{caption=false}{subfig}}
\usepackage{subfig}
\makeatother

\usepackage{babel}
\begin{document}

\begin{frontmatter}{}

\title{Incoherent transport on the $\nu=2/3$ quantum Hall edge}

\author{Casey Nosiglia$^{1}$, Jinhong Park$^{1}$, Bernd Rosenow$^{2}$,
Yuval Gefen\corref{@}$^{1}$}

\address{$^{1}$Dept. of Condensed Matter Physics, Weizmann Institute of Science,
Rehovot 76100, Israel\\
$^{2}$Institut f\"{u}r Theoretische Physik, Universit\"{a}t Leipzig,
D-04103 Leipzig, Germany}
\begin{abstract}
The nature of edge state transport in quantum Hall systems has been
studied intensely ever since Halperin \cite{=000023Halperin1982}
noted its importance for the quantization of the Hall conductance.
Since then, there have been many developments in the study of edge
states in the quantum Hall effect, including the prediction of multiple
counter-propagating modes in the fractional quantum Hall regime, the
prediction of edge mode renormalization due to disorder, and studies
of how the sample confining potential affects the edge state structure
(edge reconstruction), among others. In this paper, we study edge
transport for the \textcolor{black}{$\nu_{\text{bulk}}=2/3$ }edge
in the disordered, fully incoherent transport regime. To do so, we
use a hydrodynamic approximation for the calculation of voltage and
temperature profiles along the edge of the sample. Within this formalism,
we study two different bare mode structures with tunneling: the original
edge structure predicted by Wen \cite{=000023Wen1991} and MacDonald
\cite{=000023MacDonald1990}, and the more complicated edge structure
proposed by Meir \cite{=000023Meir1994}, whose renormalization and
transport characteristics were discussed by Wang, Meir and Gefen (WMG)
\cite{=000023WMG2013}. \textcolor{black}{We find that in the fully
incoherent regime, the topological characteristics of transport (quantized
electrical and heat conductance) are intact, with finite size corrections
which are determined by the extent of equilibration. In particular,
our calculation of conductance for the WMG model in a double QPC geometry
reproduce conductance results of a recent experiment by Sabo et al.
\cite{=000023Heiblum2017}, which are inconsistent with the model
of MacDonald. Our results can be explained in the charge/neutral mode
picture, with incoherent analogues of the renormalization fixed points
of Ref. \cite{=000023WMG2013}. Additionally, we find diffusive $(\sim1/L)$
conductivity corrections to the heat conductance in the fully incoherent
regime for both models of the edge. }
\end{abstract}

\cortext[@]{yuval.gefen@weizman.ac.il}

\end{frontmatter}{}

\section{Introduction}

The edge states in a quantum Hall system play an essential role in
defining the low energy dynamics of the system, specifically for low-frequency
and d.c. transport \cite{=000023Halperin1982}. In particular, the
formation and remarkable quantization of Hall conductance plateaus
in the integer quantum Hall effect (IQHE) and fractional quantum Hall
effect (FQHE) provides evidence for transport through such gapless
edge states, with localized states in the bulk. Furthermore, it was
shown by Wen \cite{=000023Wen1990} that the gapless edge states consist
of one-dimensional chiral bosonic modes, which are described in the
framework of the Luttinger Liquid theory. These concepts have been
successful in explaining transport properties in various quantum Hall
geometries, including constrictions formed at quantum point contacts
(QPCs) \cite{=000023Reznikov1999,=000023Kane1992,=000023Milliken1996}.

While one can easily characterize the edge states of the Laughlin
series $1/(2m+1)$ in terms of single branches of a chiral Luttinger
Liquid, other fractional filling factors different from the Laughlin
series are more complicated. A representative example of such a system
is the hole-conjugate state $\nu_{\text{bulk}}=2/3$. MacDonald \cite{=000023MacDonald1990},
following an extension of the Haldane-Halperin hierarchy \cite{=000023Haldane1983,=000023Girvin1984},
predicted the edge of the $\nu_{\text{bulk}}=2/3$ state to consist
of two counter-propagating modes, a ``downstream'' $\delta\nu=+1$
mode and an ``upstream'' $\delta\nu=-1/3$ mode, a picture that
was corroborated by Wen \cite{=000023Wen1991} {[}see Fig. $1(a)${]}.
However, this picture predicts a 2-terminal conductance of $G=(4/3)e^{2}/h$
universally, which is clearly in contradiction with the experimentally
observed $G=(2/3)e^{2}/h$; additionally, the upstream $1/3$ mode
was never detected \cite{=000023Ashoori1992}. The resolution of this
problem was described in a pioneering paper of Kane, Fischer and Polchinski
\cite{=000023KFP1994} (KFP), which considered random disorder-induced
tunneling between the modes. \textcolor{black}{Interplay between tunneling
and the interaction drives the system into a disorder-dominated phase,
described by a downstream charge mode $\delta\nu=+2/3$ and an upstream
neutral mode, predicting a conductance of $G=(2/3)e^{2}/h$.} These
ideas have recently received experimental support with the discovery
of neutral modes at several filling factors \cite{=000023Heiblum2010}.
Additionally, neutral modes have attracted a lot of interest because
they carry energy and a quantum number without carrying charge. Moreover,
they may drastically suppress anyonic interference \cite{Jinhong 2015,Goldstein 2016},
and are stable in the presence of a noisy environment, with the environment
acting to renormalize the edge state tunneling exponents \cite{Carrega 2012,Braggio 2012}.

\textcolor{black}{Yet despite the experimental success of Ref. \cite{=000023Heiblum2010},
a major observation contradicts the KFP analysis: the conductance
through a quantum point contact exhibits a plateau at $G=(1/3)e^{2}/h$
as a function of the gate voltage of the QPC \cite{=000023Chang1992,=000023Heiblum2009,=000023Heiblum2017}.
This, along with the evidence for neutral modes \cite{=000023Heiblum2010}
and a crossover of the effective charge as a function of temperature
\cite{=000023Heiblum2009} motivated Wang, Meir and Gefen (WMG) \cite{=000023WMG2013}
to propose a different edge structure, based on earlier work of edge
reconstruction in a shallow confining potential \cite{=000023Meir1994}.
The proposed bare edge structure consists of four edge modes corresponding
to filling factor discontinuities (from the bulk to the edge) of $\delta\nu=-1/3,+1,-1/3$,
and $+1/3$ {[}see Fig. $1(b)${]}. Including interactions and disorder
induced tunneling between channels, WMG apply renormalization group
techniques \cite{=000023Moore1998} in a manner analogous to KFP,
finding stable phases consisting of charge/neutral modes, which they
argue account for the experimental observations cited above. In particular,
in the case that the outermost edge is clearly decoupled from the
inner three modes, WMG find an intermediate fixed point phase consisting
of two downstream modes with $\delta\nu=+1/3,+1/3$ and two upstream
neutral modes. At the intermediate fixed point, the emergence of the
conductance plateau of $G=(1/3)e^{2}/h$ is clear since the outer-most
mode is fully transmitted, while the inner three modes are fully reflected.}

\textcolor{black}{From the above experimental considerations, it is
clear that the MacDonald edge structure does not provide the full
picture of electrical transport, leaving the question of the correct
edge structure for $\nu_{\text{bulk}}=2/3$ unresolved. Here, we produce
calculations for both the WMG and MacDonald models with disorder in
the incoherent regime, where tunneling events are uncorrelated and
phase interference effects are washed out. The approach of fully incoherent
transport in FQH edges was first investigated by Kane and Fisher in
the context of equilibration of the edge states with contacts \cite{=000023Kane1995}.
In later works, it was shown that the incoherent regime is important
for explaining conductance quantization in disordered quantum Hall
edges near the KFP fixed point, whereas the comparable setting for
the coherent regime exhibits mesoscopic conductance fluctuations which
are dependent upon the disorder realization \cite{=000023Rosenow2010,=000023Gefen2017}.
Thus, the incoherent regime is essential for conductance quantization
with charge and neutral modes. Previous works on the incoherent regime
have only been done in the context of the MacDonald edge \cite{=000023Rosenow2010,=000023Kane1995,=000023Kane1997,=000023Sen2008};
our calculations in the incoherent regime provide evidence for the
WMG model, qualitatively matching the results of a recent experiment
\cite{=000023Heiblum2017} measuring electrical transport in a two
QPC geometry.}

\textcolor{black}{Additionally, there has been increased experimental
activity in the study of heat transport in the FQH regime \cite{Venkatachalam 2012,Altimiras 2012,Banerjee2017,Rosenblatt and Lafont 2017},
which is of interest for the study of neutral modes \cite{Takei 2011,Takei 2015,Shtanko 2014},
and may prove a useful tool for characterizing the topological order
in edge structures of various filling factors: for the $\nu_{\text{bulk}}=5/2$
state, see Ref. \cite{Levin2007,Lee 2007}. In particular, recent
papers by Banerjee et al. \cite{Banerjee2017,Banerjee 2018} report
measurements of the thermal conductance for several different filling
factors in the FQH regime. Motivated by this recent experimental progress,
we study heat transport for the MacDonald and WMG models within the
incoherent transport framework, applying it to the single QPC geometry.}

\textcolor{black}{There is also a recent theoretical paper \cite{=000023Gefen2017}
by one of the authors, which is a comprehensive study of the electrical
and thermal conductances for the $\nu_{\text{bulk}}=2/3$ state with
counter-propagating $\delta\nu=+1$ and $\delta\nu=-1/3$ edges. This
work provides a microscopic calculation for the thermal and electrical
conductances, and examines their dependence on length and temperature
near the KFP fixed point, and in the weakly interacting regime. The
authors find that, beyond some length and temperature scales, incoherent
transport becomes universal in both cases, and provide calculations
for the electrical and thermal conductances in this regime. The present
work extends the results of Ref. \cite{=000023Gefen2017} in the incoherent
regime in several respects. Beyond the two terminal setups examined
in Ref. \cite{=000023Gefen2017}, we calculate the electrical and
thermal conductances for the MacDonald model in several experimentally
relevant geometries, including the line junction, low-density constriction,
and a single QPC. We also calculate local temperature and voltage
profiles $V_{1\left(\nu\right)}(x),T_{1\left(\nu\right)}(x)$ (where
the index $1(\nu)$ indicates the filling factor discontinuity $\delta\nu=+1$
$\left(\delta\nu=-\nu\right)$), which provide some insight into how
inter-channel equilibration occurs. In addition to the MacDonald model,
we provide calculations for the WMG model which are directly relevant
to experiment \cite{=000023Heiblum2017}, as mentioned above. Furthermore,
the downstream channel $\delta\nu=-1/3$ can be generalized to any
Laughlin edge state, with $\delta\nu=-1/\left(2p+1\right)$, for any
positive integer $p$.}

\textcolor{black}{In addition to extending the calculations on incoherent
transport in \cite{=000023Gefen2017}, the present work was partly
inspired by previous papers on incoherent transport by several authors
\cite{=000023Rosenow2010,=000023Kane1995,=000023Kane1997,=000023Sen2008},
and our results overlap with some of these works. Overlap includes
the form for the conductance of a single edge \cite{=000023Rosenow2010,=000023Sen2008},
vanishing heat conductance with diffusive corrections \cite{=000023Kane1997},
and a finite neutral mode decay length \cite{=000023Kane1995}. However,
our work goes beyond the previously mentioned works in several respects.
The new results of this work include:}
\begin{itemize}
\item \textcolor{black}{Electrical and thermal conductances for the MacDonald
model in several geometries (line junction, low-density constriction,
single QPC).}
\item Calculation of the local temperature and voltage profiles $V_{1\left(\nu\right)}(x),T_{1\left(\nu\right)}(x)$
in the line junction {[}see Fig. $2(b)${]}, including the voltage
induced temperature profiles. In particular, the voltage profiles
have an exponential dependence on $x$, with exponential decay near
$x=L$, rather than the usual Ohmic behavior in diffusive conductors
\cite{=000023Nagaev 1995}; this exponential decay near the drain
reflects the presence of the neutral mode. Additionally, the voltage
induced temperature profiles show rapid heating near the drain due
to the presence of the neutral mode. 
\item Electrical conductances for the WMG model in the single and double
QPC configurations with the outer-most channel fully transmitted.
In particular, in the two QPC configuration we see the transition
from $\left(1/3\right)e^{2}/h$ conductance to $\left(1/6\right)e^{2}/h$
conductance as a function of the separation of the QPCs, matching
the experimental results of \cite{=000023Heiblum2017}.
\item Thermal conductance for the WMG model in the single QPC configuration
with the outer-most channel fully transmitted. We also find vanishing
thermal conductance with diffusive conductivity corrections $\left(\sim1/L\right)$,
as previously noted in Ref. \cite{=000023Kane1997}.
\end{itemize}
\textcolor{black}{The organization of this article is as follows.
In Section 2, we explain the basic assumptions of our incoherent model
for the edge, and derive the steady state heat and electrical transport
equations for the MacDonald edge, with applications to various geometries.
In Section 3, we apply the same formalism to the WMG edge in single
and double QPC geometries. In Section 4, we summarize the results
of the previous sections, and discuss the implications of our work
and possible future directions.}

\begin{figure}[h]
\begin{centering}
\subfloat[\label{fig: 1a edge structure KFP}]{\includegraphics[width=6cm]{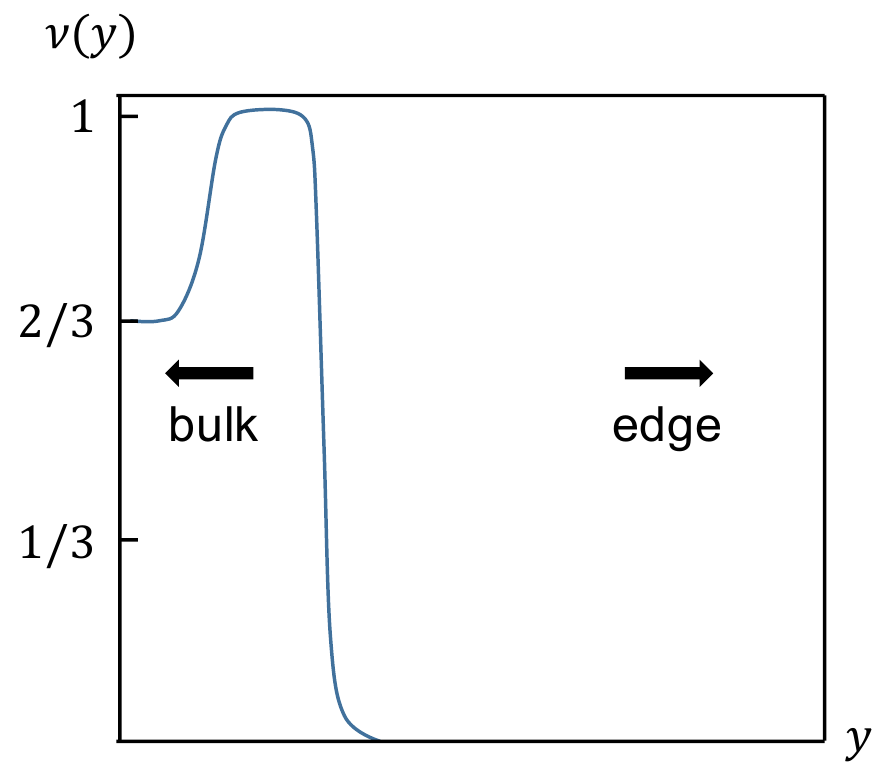}

}\qquad{}\subfloat[\label{fig: 1b edge structure WMG}]{\includegraphics[width=6cm]{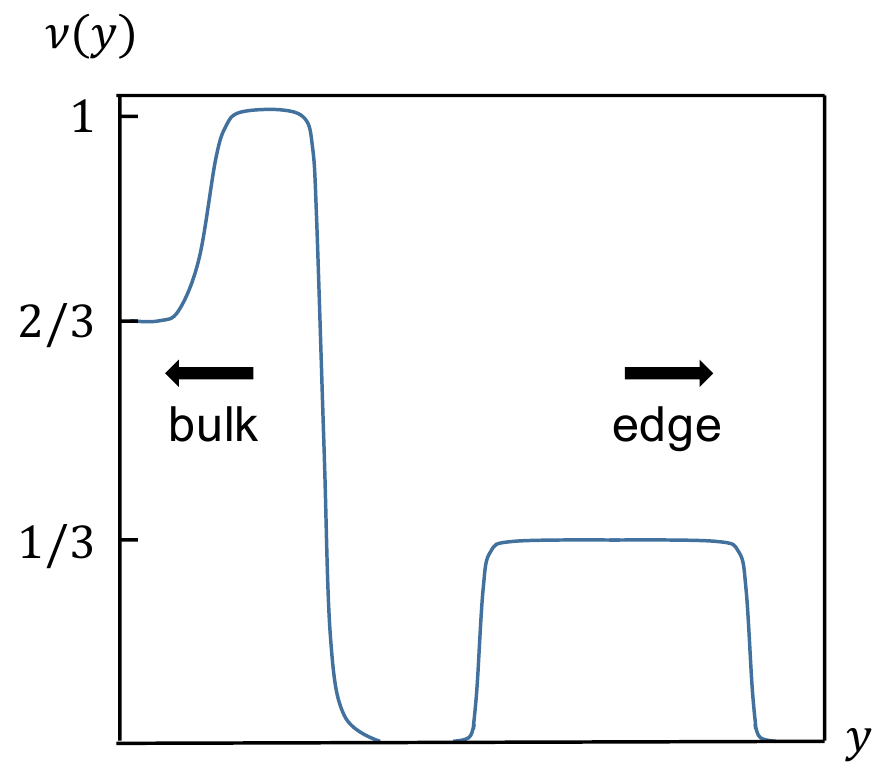}

}
\par\end{centering}
\caption{\textcolor{black}{$(a)$ Schematic plot of the filling factor $\nu(y)$
as a function of the coordinate $y$ perpendicular to the edge for
the KFP edge structure as proposed by MacDonald \cite{=000023MacDonald1990}.
The boundary between regions of different filling factor are where
the edge modes reside. $(b)$ Filling factor $\nu(y)$ as a function
of the coordinate $y$ perpendicular to the edge for the WMG edge
structure. Due to the shallow confining potential, a wide $\nu=1/3$
incompressible strip is created near the edge of the sample \cite{=000023Meir1994,=000023WMG2013}.}}
\end{figure}

\section{Description of the model (MacDonald edge with tunneling)}

\begin{table}

\begin{centering}
\begin{tabular}{|l|l|l|}
\hline 
symbol & short description  & section\tabularnewline
\hline 
$g$ & tunneling probability between counter- & \tabularnewline
 & propagating $\delta\nu=+1,-\nu$ modes & $2.1$\tabularnewline
$a$ & separation between impurities & $2.1$\tabularnewline
$\ell$ & scattering length & $2.1$\tabularnewline
$L$ & length of tunneling region & $2.1$\tabularnewline
$\nu_{0}\;(\nu)$ & filling factor of the downstream & \tabularnewline
 & (upstream) mode & $2.1$\tabularnewline
$T_{0}$ & ambient temperature & $2.1$\tabularnewline
$\gamma$ & deviation from the Wiedemann-Franz & \tabularnewline
 & law at a single tunneling bridge & $2.1$\tabularnewline
$\bar{\ell}$ & modified scattering length & $2.1$\tabularnewline
$\Delta V_{1(\nu)},\Delta T_{1(\nu)}$ & temperature and voltage biases applied & \tabularnewline
 & to the corresponding sources & $2.1$\tabularnewline
$\Delta I_{1(\nu)},\Delta J_{1(\nu)}$ & electrical and heat currents induced by & \tabularnewline
 & biases at corresponding drains & $2.1$\tabularnewline
$G_{S_{1(\nu)}D_{1(\nu)}}$ & electrical conductances for voltage  & \tabularnewline
 & biasing at $S_{1},S_{\nu}$ & $2.1$\tabularnewline
$K_{S_{1(\nu)},D_{1(\nu)}}$ & thermal conductances for temperature  & \tabularnewline
 & biasing at $S_{1},S_{\nu}$ & $2.1$\tabularnewline
\hline 
$g_{1}\;(g_{2})$ & tunneling probability for current between & \tabularnewline
 & inner-three (outer-two) modes  & $3.1$\tabularnewline
$\ell_{1}\;(\ell_{2})$ & scattering lengths for  & \tabularnewline
 & inner-three (outer-two) modes  & $3.1$\tabularnewline
$\alpha$ & ratio of scattering lengths $\ell_{1}/\ell_{2}$ & $3.1$\tabularnewline
$\tilde{\gamma}$ & deviation from the Wiedemann-Franz & \tabularnewline
 & law at a single tunneling bridge for  & \tabularnewline
 & tunneling between $\delta\nu=+\nu,-\nu$ modes & $3.1$\tabularnewline
$\bar{\ell}_{1}\;(\bar{\ell}_{2})$ & scattering lengths for thermal transport & \tabularnewline
 & for inner-three (outer-two) modes & $3.1$\tabularnewline
$\bar{\alpha}$ & ratio of modified scattering lengths $\bar{\ell}_{1}/\bar{\ell}_{2}$ & $3.1$\tabularnewline
$L_{V}$ & width of QPC constriction & $3.2$\tabularnewline
$G_{S_{1},D_{2,3,4}}$ & conductance from $S_{1}$ to drains & $3.2$\tabularnewline
$K_{S_{1}D_{2,3,4}}$ & thermal conductance from $S_{1}$ to drains & $3.2$\tabularnewline
$K_{{\rm back}}$ & thermal conductance of backscattered & \tabularnewline
 & heat current & $3.2$\tabularnewline
\hline 
$v_{1}$ & velocity of the upper $\left(\delta\nu=+1\right)$ mode & app. A\tabularnewline
$v_{\nu}$ & velocity of the lower $\left(\delta\nu=-\nu\right)$ mode & app. A\tabularnewline
$\Gamma_{0}$ & tunneling strength at a single impurity & app. A\tabularnewline
$b$ & ultraviolet spatial cutoff & app. A\tabularnewline
\hline 
\end{tabular}\caption{List of symbols, their brief description, and the section where they
are defined}
\par\end{centering}
\end{table}

\textcolor{black}{In this section, we study the steady state transport
characteristics of counter-propagating $\delta\nu=+1$ and $\delta\nu=-\nu$
modes with impurity mediated tunneling between them. We treat the
problem in the incoherent regime, where phase coherence and quantum
interference effects between successive tunneling events are ignored.
In addition, we assume that electrons equilibrate between tunneling
events, resulting in local electro-chemical potential $\mu_{0}+eV_{1(\nu)}(x)$
and local temperature $T_{1(\nu)}(x)$ on the edge mode $1(\nu)$}\textcolor{blue}{{}
}\textcolor{black}{{[}see Fig. $2(a)${]}.}\textcolor{blue}{{} }\textcolor{black}{Here
$\mu_{0}$ is chemical potential in the absence of bias voltage. Effectively
introducing local voltage and temperature probes between consecutive
tunneling bridges on each edge mode, the local chemical potential
and temperature are determined such that no net electric current and
energy current flow between the system and the voltage contacts \cite{Buttiker 1988}.}\textcolor{red}{{}
}\textcolor{black}{While we implicitly assume inelastic scattering
within each mode to facilitate such equilibration, we take this as
an assumption and will not specify any detailed mechanisms for equilibration
in this paper.}\textcolor{blue}{{} }We note that while our general emphasis
in this paper is on the $\nu_{\text{\text{bulk}}}=2/3$ state, the
calculations in this section can be applied to filling fractions of
the form $\nu_{\text{bulk}}=2p/(2p+1)$, with positive integer $p$\textcolor{black}{.
We note that the results of this section may be relevant to recent
experiments by Ronen et al. \cite{Y. Ronen 2018}, who were able to
engineer counter-propagating modes with tunable filling factors in
both the integer and fractional quantum Hall regimes, using a double-quantum-well
structure in a GaAs-based system. }\\
\\
\begin{figure}
\begin{centering}
\subfloat[\label{fig: 2a tunneling bridge}]{\includegraphics[width=6cm]{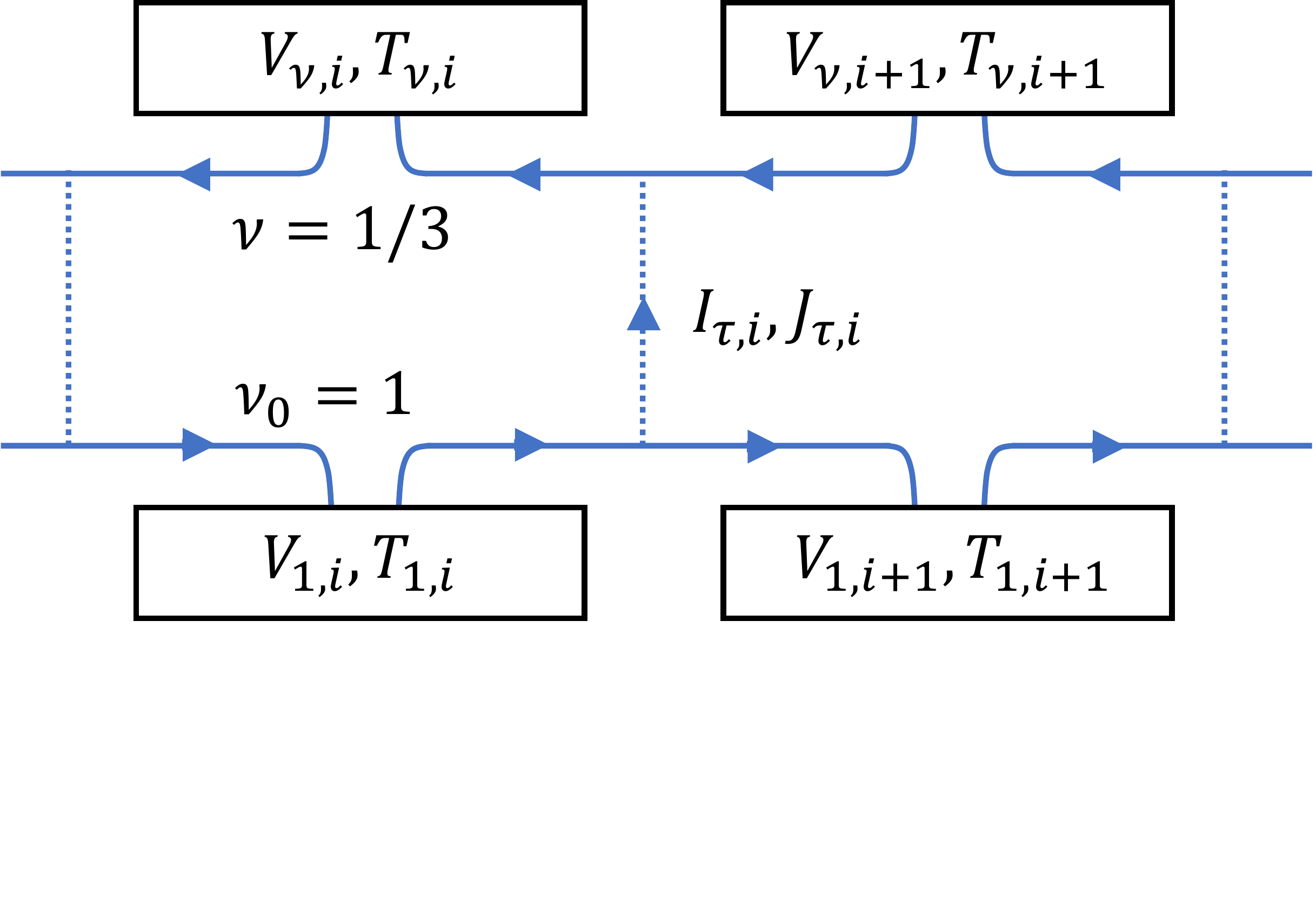}

}\qquad{}\subfloat[\label{fig: 2b line junction}]{\includegraphics[width=6cm]{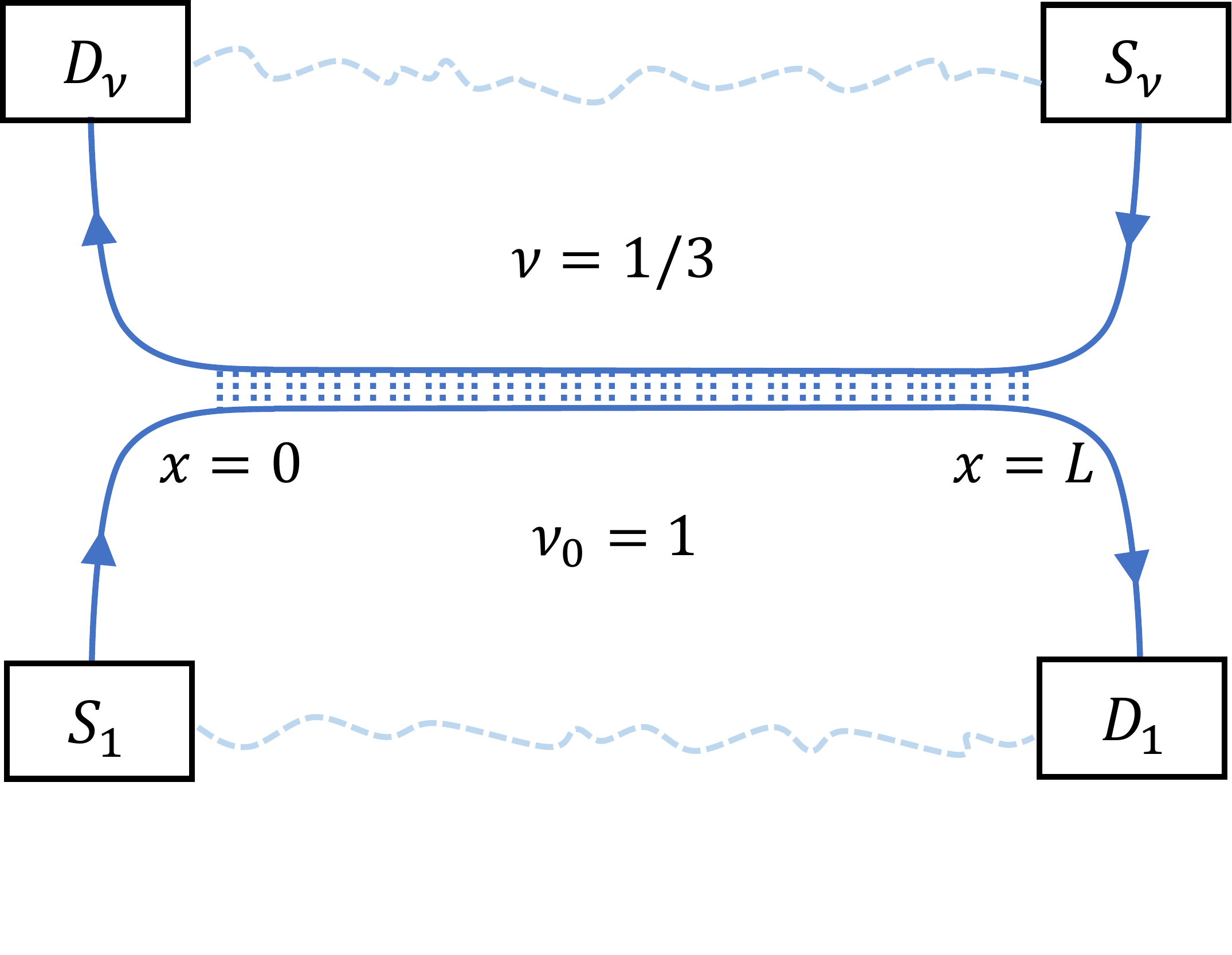}

}\caption{\textcolor{black}{$(a)$ Discrete tunneling bridge for calculation
of the steady-state current/voltage distribution. The dashed line
indicates the impurity-mediated tunneling, with associated currents
$I_{\tau,i},J_{\tau,i}$. The length of each segment $i$ is $a$.
The voltage and temperature of the probes are adjusted in such a way
that no net electrical or thermal current flows between the probes
and the edge. In this way the system equilibrates in between tunneling
events, such that it can be characterized by a local chemical potential
and temperature. $(b)$ A line junction consisting of counter-propagating
$\delta\nu=+1$ and $\delta\nu=-\nu$ edge modes, formed at the interface
between $\nu_{0}=1$ and $\nu=1/3$ FQH fluids. The heat/energy current
can be injected at $S_{1}$ or $S_{\nu}$ and is measured at the drains.
The dashed lines are schematic, indicating the continuation of the
QH fluid not shown in the figure. The short dotted lines represent
tunneling between the counter-propagating modes.}}
\par\end{centering}
\end{figure}

\subsection{Derivation of transport equations and application to the line junction}

We first take the simplest case of tunneling between counter-propagating
$1$ and $\nu$ modes in a line junction as our starting point {[}see
Fig. $2(b)${]}. The model consists of tunneling bridges evenly spaced
with separation $a$. For each segment, we impose the local constitutive
relations for coupling voltage to current for each mode on a FQH edge:
\begin{equation}
\frac{e^{2}}{h}V_{1,i}=I_{1,i},\ \ \ -\nu\frac{e^{2}}{h}V_{\nu,i}=I_{\nu,i}\label{eq:1}
\end{equation}
Additionally, we have conservation of current at each tunneling bridge,
with our sign convention such that current flowing \textit{downstream}
on the $1$-channel (we use the words ``mode'' and ``channel''
interchangeably) is positive, and current flowing \textit{downstream}
on the $\nu$-channel is negat\textcolor{black}{ive {[}see Fig. }$2(a)${]}:
\begin{equation}
I_{1,i+1}=I_{1,i}-I_{\tau,i},\ \ \ I_{\nu,i}=I_{\nu,i+1}-I_{\tau,i},\label{eq:2}
\end{equation}
where $I_{\tau,i}$ is the tunneling current between segment $i$
and $i+1$. The tunneling current is of the form:
\begin{equation}
I_{\tau,i}=g\frac{e^{2}}{h}(V_{1,i}-V_{\nu,i+1}),\label{eq:3}
\end{equation}
 where the coefficient $g$ is the tunneling probability (\textcolor{black}{see
Appendix B}). Applying Eqs. $(1)$ and $\left(3\right)$ to the conservation
condition Eq. $(2)$, we obtain the following iterative equations
at a given tunneling bridge:

\begin{equation}
\frac{1}{\left(1-g/\nu\right)}\begin{pmatrix}\left[1-g\left(1+1/\nu\right)\right] & g\\
-g/\nu & 1
\end{pmatrix}\begin{pmatrix}V_{1,i}\\
V_{\nu,i}
\end{pmatrix}=\begin{pmatrix}V_{1,i+1}\\
V_{\nu,i+1}
\end{pmatrix}\label{eq:4}
\end{equation}

Making use of the assumption that $g\ll1$ by keeping terms only up
to $O(g)$, taking the distance between impurities $a$ as the smallest
scale of variation (so that $ia\rightarrow x$) and taking the number
of tunneling bridges $n\rightarrow\infty$, we obtain the following
differential equations:
\begin{equation}
\frac{g}{a}\begin{pmatrix}-1 & 1\\
-\nu^{-1} & \nu^{-1}
\end{pmatrix}\begin{pmatrix}V_{1}(x)\\
V_{\nu}(x)
\end{pmatrix}=\frac{d}{dx}\begin{pmatrix}V_{1}(x)\\
V_{\nu}(x)
\end{pmatrix}\label{eq:5}
\end{equation}
The eigenvectors and corresponding eigenvalues are $\left[\left(\nu^{-1}-1\right)g/a,0\right]$
and $\begin{pmatrix}1\\
\nu^{-1}
\end{pmatrix},\begin{pmatrix}1\\
1
\end{pmatrix}$, respectively; thus the solution can be written in the form:

\begin{equation}
\vec{V}(x)=A\begin{pmatrix}1\\
\nu^{-1}
\end{pmatrix}e^{\frac{(\nu^{-1}-1)gx}{a}}+B\begin{pmatrix}1\\
1
\end{pmatrix}\label{eq:6}
\end{equation}
Biasing the $1$-channel by $V_{0}$ and grounding the $\nu$-channel,
we obtain the following solutions:
\begin{equation}
V_{1}(x)=\frac{V_{0}\left(1-\nu e^{\frac{(\nu^{-1}-1)(x-L)}{\ell}}\right)}{\left(1-\nu e^{-\frac{(\nu^{-1}-1)L}{\ell}}\right)},\ \ \ V_{\nu}(x)=\frac{V_{0}\left(1-e^{\frac{(\nu^{-1}-1)(x-L)}{\ell}}\right)}{\left(1-\nu e^{-\frac{(\nu^{-1}-1)L}{\ell}}\right)},\label{eq:7}
\end{equation}
\textcolor{black}{where $\ell\equiv a/g$ is the scattering length,
which also characterizes inter-channel equilibration of voltages.}\textcolor{red}{{}
}From this, we can easily calculate the conductance of the current
transmitted from the source of the $1$-channel to its drain \cite{footnote-1}:
\begin{equation}
G_{S_{1}D_{1}}=\frac{I_{1}(L)}{V_{0}}=\frac{e^{2}}{h}\frac{\left(1-\nu\right)}{\left(1-\nu e^{-\frac{(\nu^{-1}-1)L}{\ell}}\right)}\label{eq:8}
\end{equation}
Additionally, we have the conductance of the current backscattered
to the drain of the $\nu$-channel:
\begin{equation}
G_{S_{1}D_{\nu}}=\frac{\left|I_{\nu}(0)\right|}{V_{0}}=\nu\frac{e^{2}}{h}\frac{\left(1-e^{-\frac{(\nu^{-1}-1)L}{\ell}}\right)}{\left(1-\nu e^{-\frac{(\nu^{-1}-1)L}{\ell}}\right)},\label{eq:9}
\end{equation}
so that $G_{S_{1}D_{1}}+G_{S_{1}D_{\nu}}=e^{2}/h$, as required by
current conservation. Here, we see that in the limit that $\ell\ll L$,
we obtain near quantization of the transmitted and backscattered conductances,
with exponentially small corrections: $G_{S_{1}D_{1}}\approx(1-\nu)\left(e^{2}/h\right)\left(1+\nu e^{-\frac{(\nu^{-1}-1)L}{\ell}}\right),G_{S_{1}D_{\nu}}\approx\nu\left(e^{2}/h\right)\left(1-(1-\nu)e^{-\frac{(\nu^{-1}-1)L}{\ell}}\right)$.
\\
We can perform a similar analysis for the temperature profiles induced
by the applied voltage bias $V_{0}$, through enforcing local conservation
of energy. The energy tunneling current calculated from the Keldysh
formalism (see Appendix B) is of the form:
\begin{equation}
J_{\tau,i}=g\frac{e^{2}}{2h}\left(V_{1,i}^{2}-V_{\nu,i+1}^{2}\right)+\gamma g\frac{\pi^{2}k^{2}}{6h}\left(T_{1,i}^{2}-T_{\nu,i+1}^{2}\right),\label{eq:10}
\end{equation}
where $\gamma\equiv3/(2\nu+1)\xrightarrow[\nu\rightarrow1/3]{}9/5$
in the absence of interactions between the modes; $\gamma$ is a constant
which measures the deviation from the Wiedemann-Franz for the tunneling
currents at a single impurity (see Appendix B), and $k$ is Boltzmann's
constant. Additionally, we have the local constitutive relations for
the energy current for each mode on the edge:
\begin{equation}
J_{1,i}=\frac{e^{2}}{h}\frac{V_{1,i}^{2}}{2}+\frac{\pi^{2}k^{2}}{6h}T_{1,i}^{2},\ J_{\nu,i}=-\nu\frac{e^{2}}{h}\frac{V_{\nu,i}^{2}}{2}-\frac{\pi^{2}k^{2}}{6h}T_{\nu,i}^{2}.\label{eq:11}
\end{equation}
Applying conservation of energy current at each tunneling bridge,
we obtain the following equations for the temperatures $T_{1},T_{\nu}$:

\begin{align}
\frac{e^{2}}{h}\frac{V_{1,i+1}^{2}}{2}+\frac{\pi^{2}k^{2}}{6h}T_{1,i+1}^{2} & =(1-g)\frac{e^{2}}{h}\frac{V_{1,i}^{2}}{2}+g\frac{e^{2}}{h}\frac{V_{\nu,i+1}^{2}}{2}+(1-\gamma g)\frac{\pi^{2}k^{2}}{6h}T_{1,i}^{2}+\gamma g\frac{\pi^{2}k^{2}}{6h}T_{\nu,i+1}^{2}\label{eq:12}\\
\nu\frac{e^{2}}{h}\frac{V_{\nu,i}^{2}}{2}+\frac{\pi^{2}k^{2}}{6h}T_{\nu,i}^{2} & =(\nu-g)\frac{e^{2}}{h}\frac{V_{\nu,i+1}^{2}}{h}+g\frac{e^{2}}{h}\frac{V_{1,i}^{2}}{2}+(1-\gamma g)\frac{\pi^{2}k^{2}}{6h}T_{\nu,i+1}^{2}+\gamma g\frac{\pi^{2}k^{2}}{6h}T_{1,i}^{2}\label{eq:13}
\end{align}

Once again keeping terms only up to $O(g)$ and going to the continuum
limit, we obtain the following differential equations:

\begin{equation}
\Rightarrow\partial_{x}\begin{pmatrix}T_{1}^{2}(x)\\
T_{\nu}^{2}(x)
\end{pmatrix}=\frac{\gamma}{\ell}\begin{pmatrix}-1 & 1\\
-1 & 1
\end{pmatrix}\begin{pmatrix}T_{1}^{2}(x)\\
T_{\nu}^{2}(x)
\end{pmatrix}+\frac{1}{\ell}\frac{3e^{2}(V_{\nu}(x)-V_{1}(x))^{2}}{\pi^{2}k^{2}}\begin{pmatrix}-1\\
1
\end{pmatrix}\label{eq:14}
\end{equation}

\textcolor{black}{These coupled differential equations are inhomogeneous
linear differential equations (i.e. of the form $A\bar{x}(t)+\bar{f}(t)=\partial_{t}\bar{x}(t)$),
and can be solved generally with a matrix exponential. The boundary
conditions are specified by applying a voltage bias $V_{0}$ on either
source $S_{1}$ or source $S_{\nu}$ in the presence of an ambient
temperature $T_{0}$. One finds that the difference $(V_{\nu}(x)-V_{1}(x))^{2}$
is the same in either case; the solution of Eq. $(13)$} \textcolor{black}{is
thus the same for both cases, and is of the form:}
\begin{align}
T_{1}^{2}(x) & =T_{0}^{2}+\frac{3\nu\left(eV_{0}/k\right)^{2}}{2\pi^{2}\left(1-\nu e^{-\frac{(\nu^{-1}-1)L}{\ell}}\right)^{2}}\left\{ \frac{\gamma x}{\ell+\gamma L}\left[\left(1-\nu+\gamma\nu\right)+\left(1-\nu-\gamma\nu\right)e^{-\frac{2(\nu^{-1}-1)L}{\ell}}\right]\right.\nonumber \\
 & \left.+\left[1-\nu-\gamma\nu\right]\left[e^{\frac{2(\nu^{-1}-1)(x-L)}{\ell}}-e^{-\frac{2(\nu^{-1}-1)L}{\ell}}\right]\right\} \nonumber \\
T_{\nu}^{2}(x) & =T_{0}^{2}+\frac{3\nu\left(eV_{0}/k\right)^{2}}{2\pi^{2}\left(1-\nu e^{-\frac{(\nu^{-1}-1)L}{\ell}}\right)^{2}}\left\{ \frac{\gamma\left(x-L\right)}{\ell+\gamma L}\left[\left(1-\nu+\gamma\nu\right)+\left(1-\nu-\gamma\nu\right)e^{-\frac{2(\nu^{-1}-1)L}{\ell}}\right]\right.\nonumber \\
 & \left.-\left[1-\nu+\gamma\nu\right]\left[e^{\frac{2(\nu^{-1}-1)(x-L)}{\ell}}-1\right]\right\} \label{eq:15}
\end{align}
\\
With Eq. $(14)$, we can also calculate heat transport along this
edge in the presence of a temperature bias $\Delta T$ applied to
the drain of the $1$-channel, above an ambient temperature $T_{0}$.
Noting the form of the energy tunneling current (in the absence of
a voltage bias $V_{0}$) at a tunneling bridge $i$, $J_{\tau,i}=\gamma g\left(\pi^{2}k^{2}/6h\right)(T_{1,i}^{2}-T_{\nu,i+1}^{2})$,
and the local constitutive relations for the heat current on a FQH
edge as $J_{1,i}=\left(\pi^{2}k^{2}/6h\right)T_{1,i}^{2},\;J_{\nu,i}=-\left(\pi^{2}k^{2}/6h\right)T_{\nu,i}^{2}$,
we locally apply conservation of energy current at each tunneling
bridge and obtain the following temperature profiles:
\begin{align}
T_{1}^{2}(x) & =\left(T_{0}+\Delta T\right)^{2}-\Delta T(2T_{0}+\Delta T)\frac{(x/\bar{\ell})}{(1+L/\bar{\ell})},\nonumber \\
T_{\nu}^{2}(x) & =\frac{(T_{0}^{2}+(L/\bar{\ell})\left(T_{0}+\Delta T\right)^{2})}{(1+L/\bar{\ell})}-\Delta T(2T_{0}+\Delta T)\frac{(x/\bar{\ell})}{(1+L/\bar{\ell})},
\end{align}
with $\bar{\ell}\equiv\ell/\gamma$. Additionally, one can calculate
the thermal conductances associated with the transmitted and backscattered
heat currents:
\begin{align}
K_{S_{1}D_{1}} & =\lim_{\Delta T\rightarrow0}\frac{J_{1}(L,\Delta T+T_{0})-J_{1}(L,T_{0})}{\Delta T}=\frac{\pi^{2}k^{2}T_{0}}{3h}\frac{1}{(1+L/\bar{\ell})},\label{eq:17}\\
K_{S_{1}D_{\nu}} & =\lim_{\Delta T\rightarrow0}\frac{J_{\nu}(0,\Delta T+T_{0})-J_{\nu}(0,T_{0})}{\Delta T}=\frac{\pi^{2}k^{2}T_{0}}{3h}\frac{L/\text{\ensuremath{\bar{\ell}}}}{(1+L/\bar{\ell})},\label{eq:18}
\end{align}
with $K_{S_{1}D_{1}}+K_{S_{1}D_{\nu}}=\pi^{2}k^{2}/3h,$ as required
by conservation of energy. Finally, one can apply temperature and
voltage biases to the $1$ and $\nu$-channel sources individually,
described by a $4\times4$ linear response matrix:
\begin{equation}
\begin{pmatrix}\Delta I_{1}\\
\Delta I_{\nu}\\
\Delta J_{1}\\
\Delta J_{\nu}
\end{pmatrix}=\begin{pmatrix}L_{11}^{11} & L_{11}^{12} & L_{12}^{11} & L_{12}^{12}\\
L_{11}^{21} & L_{11}^{22} & L_{12}^{21} & L_{12}^{22}\\
L_{21}^{11} & L_{21}^{12} & L_{22}^{11} & L_{22}^{12}\\
L_{21}^{21} & L_{21}^{22} & L_{22}^{21} & L_{22}^{22}
\end{pmatrix}\begin{pmatrix}\Delta V_{1}\\
\Delta V_{\nu}\\
\Delta T_{1}\\
\Delta T_{\nu}
\end{pmatrix},\label{eq:19}
\end{equation}
where $\Delta V_{1},\Delta V_{\nu},\Delta T_{1},\Delta T_{\nu}$ are
the individual biases on the drains and $\Delta I_{1},\Delta I_{\nu},\Delta J_{1},\Delta J_{\nu}$
are the induced currents. The general form of the matrix turns out
to be:
\begin{equation}
L=\begin{pmatrix}L_{11} & 0\\
0 & L_{22}
\end{pmatrix},\label{eq:20}
\end{equation}
where the matrices $L_{11}$ and $L_{22}$ are of the form 
\begin{equation}
L_{11}=\frac{e^{2}/h}{\left(1-\nu e^{-\frac{(\nu^{-1}-1)L}{\ell}}\right)}\begin{bmatrix}\left(1-\nu\right) & \nu\left(1-e^{-\frac{(\nu^{-1}-1)L}{\ell}}\right)\\
\nu\left(1-e^{-\frac{(\nu^{-1}-1)L}{\ell}}\right) & \nu\left(1-\nu\right)e^{-\frac{(\nu^{-1}-1)L}{\ell}}
\end{bmatrix},\;\;\;L_{22}=\frac{\pi^{2}k^{2}}{3h}\frac{T_{0}}{\left(1+L/\bar{\ell}\right)}\begin{bmatrix}1 & L/\bar{\ell}\\
L/\bar{\ell} & 1
\end{bmatrix}.\label{eq:21}
\end{equation}
We see in the limit that $\bar{\ell}/L,\ell/L\rightarrow0$, we obtain
the following quantized values:
\begin{equation}
L_{11}=\frac{e^{2}}{h}\begin{bmatrix}\left(1-\nu\right) & \nu\\
\nu & 0
\end{bmatrix},\;\;\;L_{22}=\frac{\pi^{2}k^{2}}{3h}T_{0}\begin{bmatrix}0 & 1\\
1 & 0
\end{bmatrix}.\label{eq:22}
\end{equation}
\\
\begin{figure}[H]
\begin{centering}
\subfloat[\label{fig: 3a voltage profiles}]{\includegraphics[width=8cm]{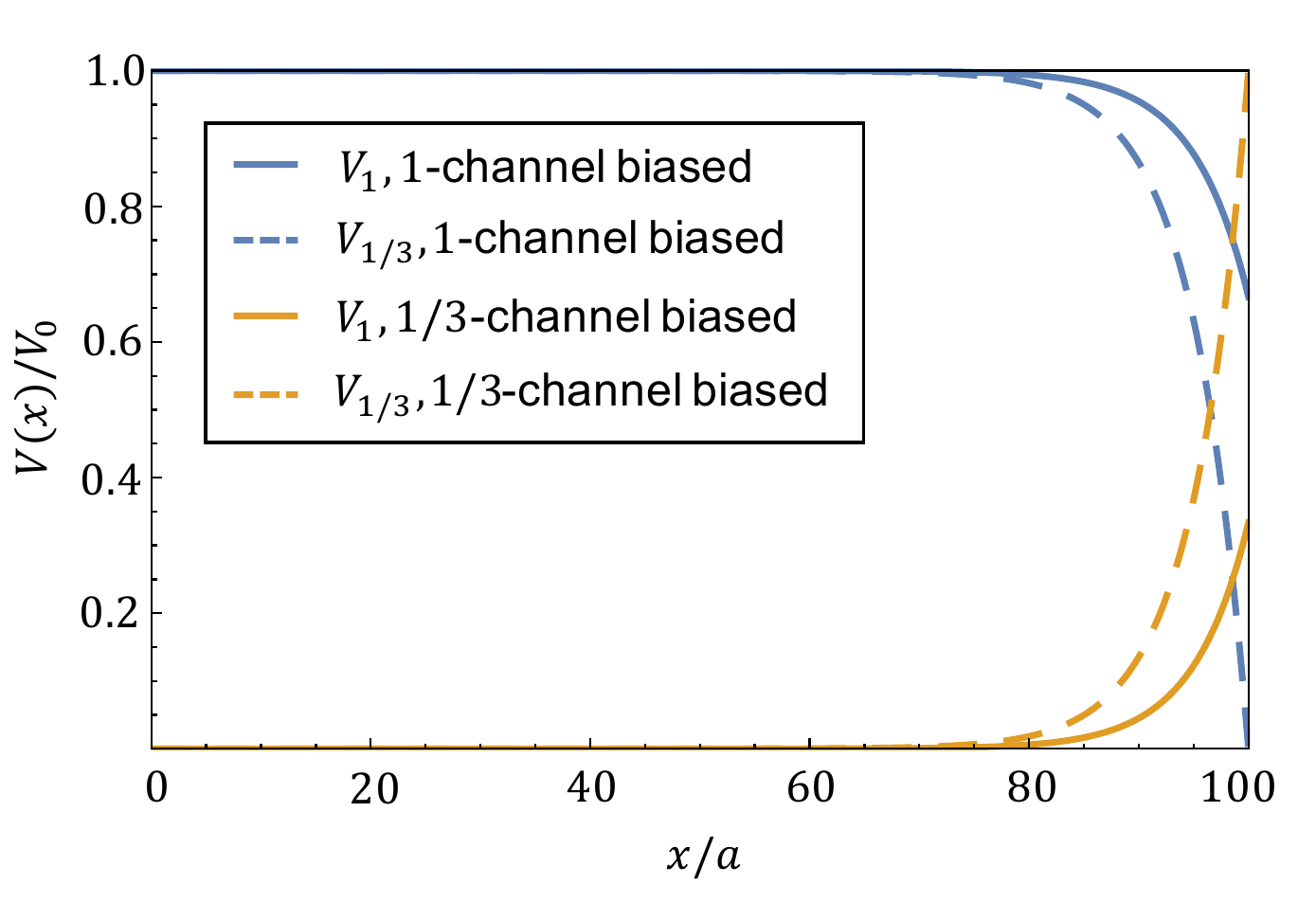}

}\hfill{}\subfloat[\label{fig: 3b temperature profiles}]{\includegraphics[width=8cm]{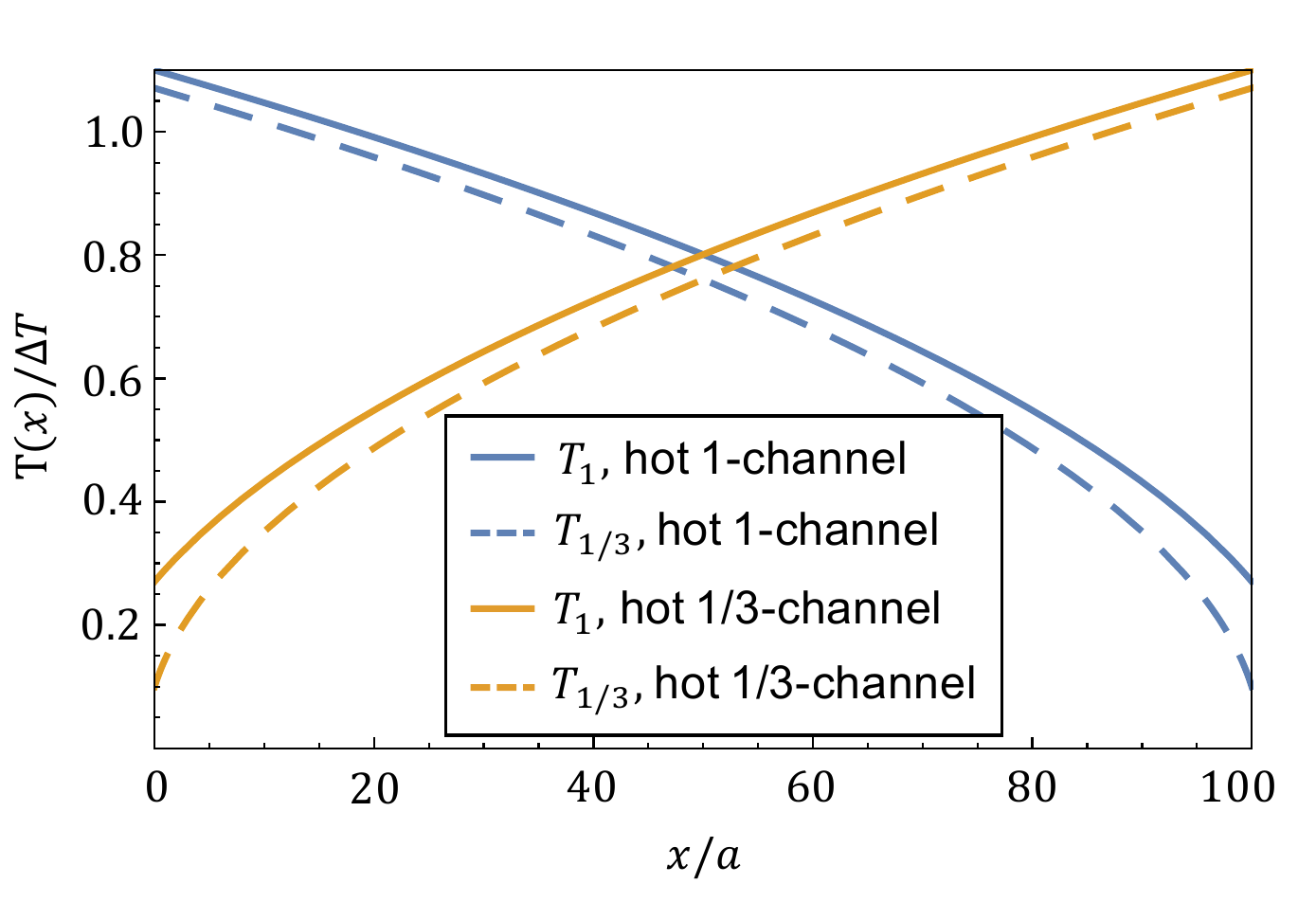}

}\\
\subfloat[\label{fig: 3c volt induced temp profiles}]{\includegraphics[width=8cm]{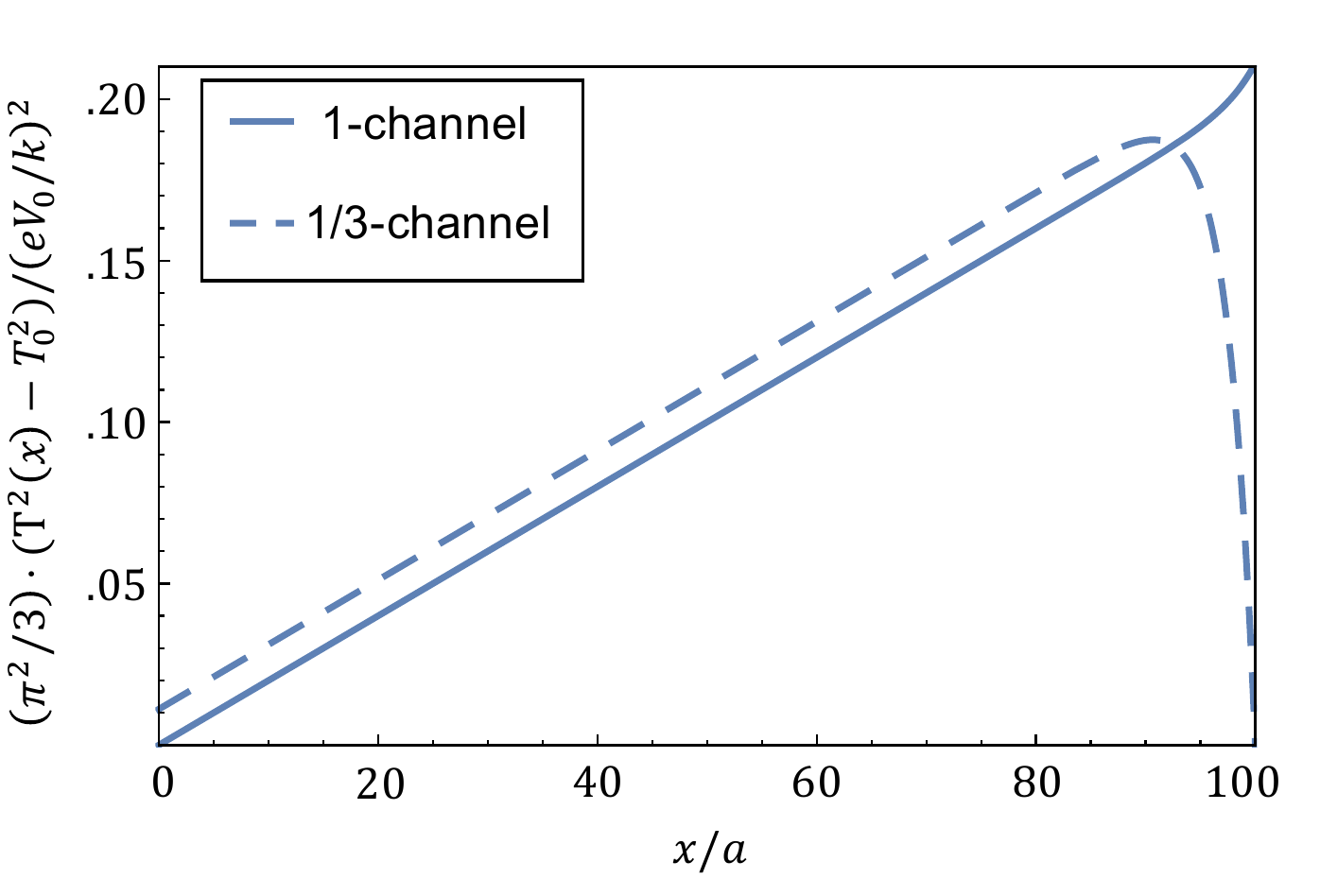}}\caption{\textcolor{black}{$(a)$ Voltage profiles for the $1-1/3$ line junction,
with $\ell=10a,L=100a$. We see explicitly that the equilibration
of the two channels takes place within the length $\ell$ $(b)$ Plot
of the temperature profiles, with $\Delta T=1,T_{0}=.1,\bar{\ell}=5.55a,L=100a$
$(c)$ Plot of the ratio of the net heat current to the total injected
energy current $\left(\pi^{2}/3\right)\cdot\left(T^{2}(x)-T_{0}^{2}\right)/\left(eV_{0}/k\right)^{2}$
for $\ell=10a,L=100a$. This ratio is independent of the experimental
parameters $T_{0},V_{0}$. }}
\par\end{centering}
\end{figure}

\subsection{Discussion of line junction results}

The results of the line junction calculations are shown in Figure
$3$. Fig. $3(a)$ shows the voltage profiles of Eq. $(7)$ induced
by an applied voltage $V_{0}$. These profiles exhibit exponential
decay near the drain, decaying with the scattering length $\ell$,
which characterizes inter-channel equilibration. Fig. $3(b)$ shows
the temperature profiles of Eq. $(16)$ induced by an applied temperature
bias $\Delta T$. These profiles exhibit inter-channel equilibration
on a modified scale $\bar{\ell}$, but with power-law behavior which
is characteristic of diffusive transport. Fig. $3(c)$ shows the ratio
of the net heat current $J_{1,\nu}=\left(\pi^{2}k^{2}/6h\right)\left(T_{1,\nu}^{2}(x)-T_{0}^{2}\right)$
induced by an applied voltage $V_{0}$ to the total injected energy
current $\left(e^{2}/2h\right)V_{0}^{2}$ (the voltage-induced temperature
profiles are given in Eq. $(15)$). The inter-channel equilibration
is controlled by the scattering length $\ell$, and we see that the
$\nu$ channel is heated up near the drain; these points are a reflection
of the inter-channel equilibration of the voltages, which is the underlying
mechanism for generating heat in this case.\\
It is interesting to contrast electrical transport in the case of
counter-propagating $\delta\nu=+1$ and $\delta\nu=-\nu$ modes with
that of counter-propagating $\delta\nu=+1$ and $\delta\nu=-1$ modes,
which is the more familiar case of a quantum wire. In the $\delta\nu=+1$
and $\delta\nu=-1$ case, one finds that the voltage profiles are
linear in $x$, and the conductance of the transmitted channel has
a power-law decay as $G_{S_{1}\rightarrow D_{1}}\sim\ell/L$, which
is characteristic of diffusive Ohmic transport\cite{=000023Nagaev 1995}:
\begin{align}
V_{1,+}(x) & =\frac{V_{0}}{\left(1+L/\ell\right)}\left[1+\left(\frac{L-x}{\ell}\right)\right],\ \ \ V_{1,-}(x)=\frac{V_{0}}{\left(1+L/\ell\right)}\left(\frac{L-x}{\ell}\right),\label{eq:23}\\
L_{11} & =\frac{e^{2}/h}{\left(1+L/\ell\right)}\begin{bmatrix}1 & L/\ell\\
L/\ell & 1
\end{bmatrix},\label{eq:24}
\end{align}
where $V_{1,+}(x)$ is the voltage profile associated with the downstream
mode and $V_{1,-}(x)$ is the voltage profile associated with the
upstream mode. However, if we look at Eqs. $(8)$ and $(21)$, we
see an \textit{exponential} dependence on $x$ and exponentially small
corrections to the quantized $(2/3)e^{2}/h$ value for the conductance,
as mentioned before. Additionally, we can transform into the charge/neutral
mode basis. Noting that the transformation into the charge and neutral
mode basis is defined as $\phi_{\rho}=\phi_{1}+\phi_{\nu},\phi_{\sigma}=\phi_{1}+\phi_{\nu}/\nu$,
\cite{=000023WMG2013} the currents in the charge/neutral mode basis
become (see Appendix E):
\begin{equation}
I_{\rho}\equiv I_{1}+I_{\nu}=\frac{e^{2}}{h}V_{0}\frac{\left(1-\nu\right)}{\left(1-\nu e^{-\frac{(\nu^{-1}-1)L}{\ell}}\right)},\ \ \ I_{\sigma}\equiv I_{1}+\frac{I_{\nu}}{\nu}=\frac{e^{2}}{h}V_{0}\frac{\left(1-\nu\right)}{\left(1-\nu e^{-\frac{(\nu^{-1}-1)L}{\ell}}\right)}e^{\frac{(\nu^{-1}-1)(x-L)}{\ell}},\label{eq:25}
\end{equation}
In this basis, we see that the charge mode is constant, reproducing
the conductance of Eq. $(8)$, and that the neutral mode decays exponentially
from the drain to the source $\left(\text{see Fig. }3(a)\right)$;
additionally, the direction of decay is predicted by the hierarchy
scheme, as noted in Ref. \cite{=000023Kane1995}.\textcolor{black}{{}
The transformation into the charge/neutral basis can be given a more
physical interpretation. The current corresponding to the charge field
$\phi_{\rho}=\phi_{1}+\phi_{\nu}$ is the total current $I_{1}+I_{\nu}$,
which is conserved along the line junction; note that this is not
the total injected current $\left(e^{2}/h\right)V_{0}$ due to the
backscattering to the drain of the $\nu$-channel. The current corresponding
to the neutral field $\phi_{\sigma}=\phi_{1}+\phi_{\nu}/\nu$}\textcolor{red}{{}
}\textcolor{black}{is proportional to the tunneling current in the
basis of the bare modes, i.e. $I_{\sigma}(x)=I_{\tau}(x)/g$;}\textcolor{red}{{}
}\textcolor{black}{thus, in the incoherent regime, the decay of the
current $I_{\sigma}$ directly reflects the equilibration between
the bare modes \cite{=000023Kane1995}. }\\
\textcolor{black}{With regards to heat transport $\left(\text{see Fig. }3(b)\right)$,
we note the same behavior (up to constants) in the $1$ and $1$ case
as in the $1$ and $\nu$ case, which is power-law dependence in the
thermal conductance, with a slightly modified scattering length; thus
in the limit of $\bar{\ell}/L\rightarrow\infty$, we have a vanishing
thermal conductance but diffusive ($\sim1/L$) thermal }\textit{\textcolor{black}{conductivity}}\textcolor{black}{{}
corrections. }\\
We also note that the off-diagonal terms (thermoelectric contributions)
are identically $0$. At the individual tunneling bridge level, there
are no $L_{12}$ (electrical current induced by temperature difference)
components due to the linearization of the electron energy spectrum
(See Appendix B). We can see this because a temperature bias induces
electron-hole excitations, and if the spectra of the channels are
linear, there is no current generated, i.e. $\Delta I_{1}=\Delta I_{\nu}=0\rightarrow\Delta I_{\tau}=0$;
this is more generally a reflection of the fact that the spectra have
particle-hole symmetry. That the $L_{21}$ elements (heat current
induced by voltage difference) are identically zero can be seen as
a consequence of Onsager symmetry.

\textcolor{black}{We also note the voltage-induced temperature profiles
$\left(\text{see Fig. }3(c)\right)$ are different from the typical
profile in a diffusive conductor, where temperature scales as $T^{2}(x)\propto x\left(L-x\right)$
\cite{=000023Nagaev 1995}. The spatial dependence of our calculated
temperature profiles scales as $T^{2}(x)\sim x$, with exponential
corrections within a length $\sim\ell$ from the drain. Noting that
the inhomogeneous term in Eq. $(14)$ is proportional to $I_{\sigma}^{2}$,
we see that the exponential corrections reflect the decay of the neutral
mode from the drain. }\\
\begin{figure}[H]
\centering{}\includegraphics[width=6cm]{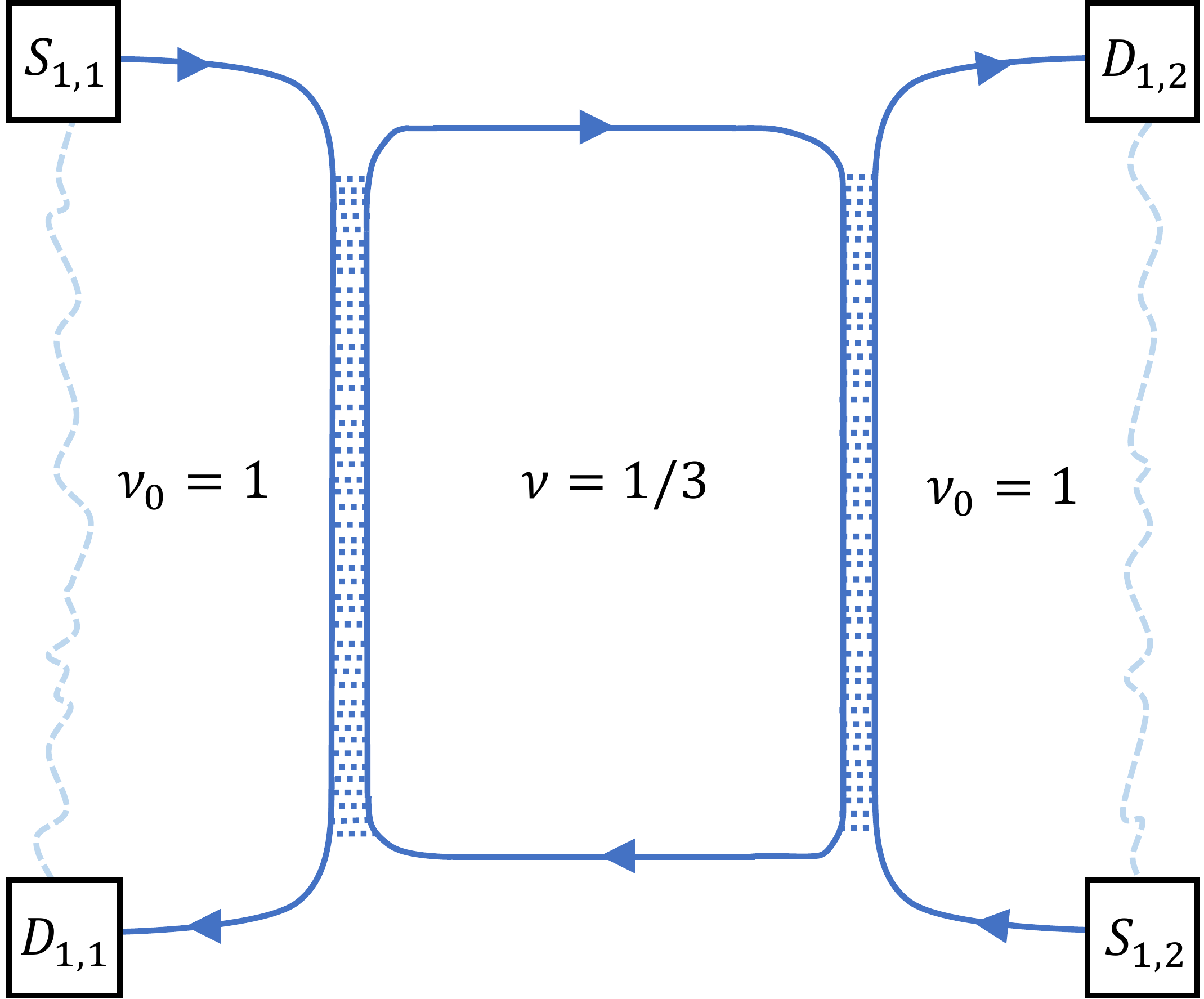}\caption{\textcolor{black}{Geometry of a low-density constriction. \label{fig:4 low density constriction}}}
\end{figure}

\subsection{Low-density constriction and single-QPC geometries}

To make a connection with a more common experimental setup and with
the work of Ref.\textcolor{blue}{{} }\textcolor{black}{\cite{=000023Rosenow2010},}
we studied the same problem in a constriction geometry {[}\textcolor{black}{see
Fig.} $4${]}. In this geometry, the filling factor $\nu_{C}$ in
the constriction is less than that of the bulk $\nu_{B}$, leading
to tunneling at the interface of the two QH droplets in a manner similar
to the problem of the counter-propagating $\nu_{0}=1$ and $\nu$
modes; this is a possible model for a QPC. Our setup matches with
Fig. $1(b)$ of Ref. \cite{=000023Rosenow2010}, with the exception
that we do not consider inter-edge scattering across the constricted
QH droplet. All of the machinery from the previous section is applicable
here; the problem is effectively that of two coupled line junctions,
where the $\nu$ modes are connected through the boundary conditions
on their ends. Applying the appropriate boundary conditions, we calculate
the response matrix $L$:
\begin{align}
L_{12}=L_{21}=0,\;\;\; & L_{11}=\frac{e^{2}/h}{\left(1+e^{\frac{(\nu^{-1}-1)L}{\ell}}-2\nu\right)}\begin{bmatrix}\left(1-\nu\right)\left(1+e^{\frac{(\nu^{-1}-1)L}{\ell}}\right) & \nu\left(-1+e^{\frac{(\nu^{-1}-1)L}{\ell}}\right)\\
\nu\left(-1+e^{\frac{(\nu^{-1}-1)L}{\ell}}\right) & \left(1-\nu\right)\left(1+e^{\frac{(\nu^{-1}-1)L}{\ell}}\right)
\end{bmatrix},\nonumber \\
 & L_{22}=\frac{\pi^{2}k^{2}}{3h\left(1+\frac{L/\bar{\ell}}{2}\right)}\begin{bmatrix}1 & \frac{L/\bar{\ell}}{2}\\
\frac{L/\bar{\ell}}{2} & 1
\end{bmatrix}\label{eq:26}
\end{align}
We see that the results are very similar to those of the line junction
in the $1$-biased case; they conform to the same quantized values
for heat and electrical conductance in the $\ell/L,\bar{\ell}/L\rightarrow0$
limit, and the finite-size corrections have the same scaling with
$\ell/L,\bar{\ell}/L$ as in the $1$-biased case. \\

\section{The WMG edge}

In this section, we extend our results to the WMG edge, analyzing
the single and double QPC geometries with only the outermost channel
transmitted. Since the KFP picture cannot account for the $G=(1/3)e^{2}/h$
conductance plateau \cite{=000023Chang1992,=000023Heiblum2009,=000023Heiblum2017}
in a QPC, analysis of the WMG edge here becomes more relevant and
gives rise to several different behaviors. Additionally, the recent
transport experiment by Sabo et al. \cite{=000023Heiblum2017} in
a two QPC geometry shows a conductance crossover from $\left(1/3\right)e^{2}/h$
to $\left(1/6\right)e^{2}/h$ as a function of the separation between
the QPCs, which cannot be explained in the more conventional MacDonald
picture and provides the impetus for studying the WMG model in this
geometry. We also examine heat transport in the single QPC and compare
the results to the MacDonald model.

\begin{figure}
\centering{}\subfloat[\label{fig: 5a tunneling bridges}]{\includegraphics[width=6cm]{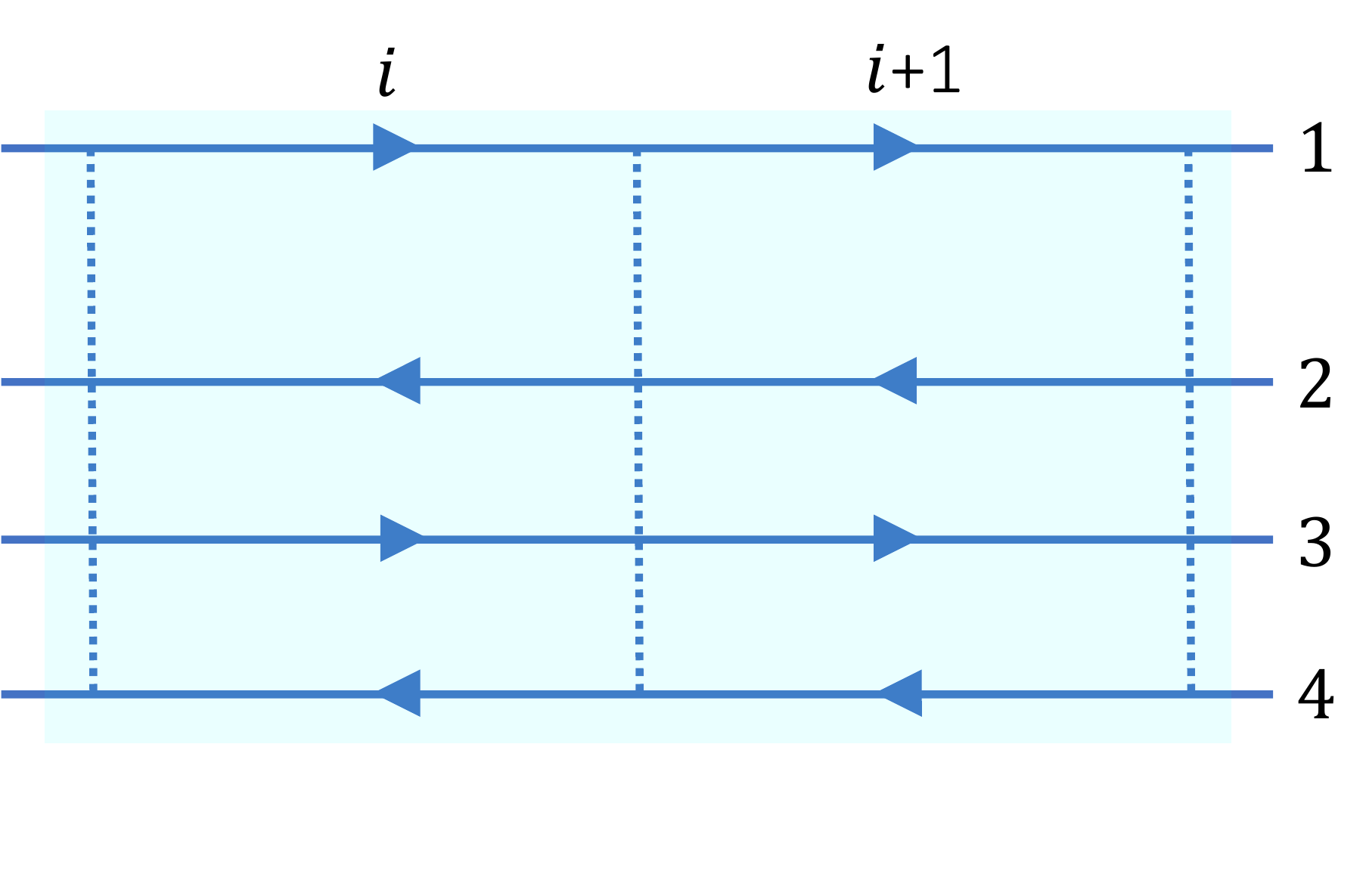}

}\qquad{}\subfloat[\label{fig: 5b WMG single QPC with size bars}]{\includegraphics[width=8cm]{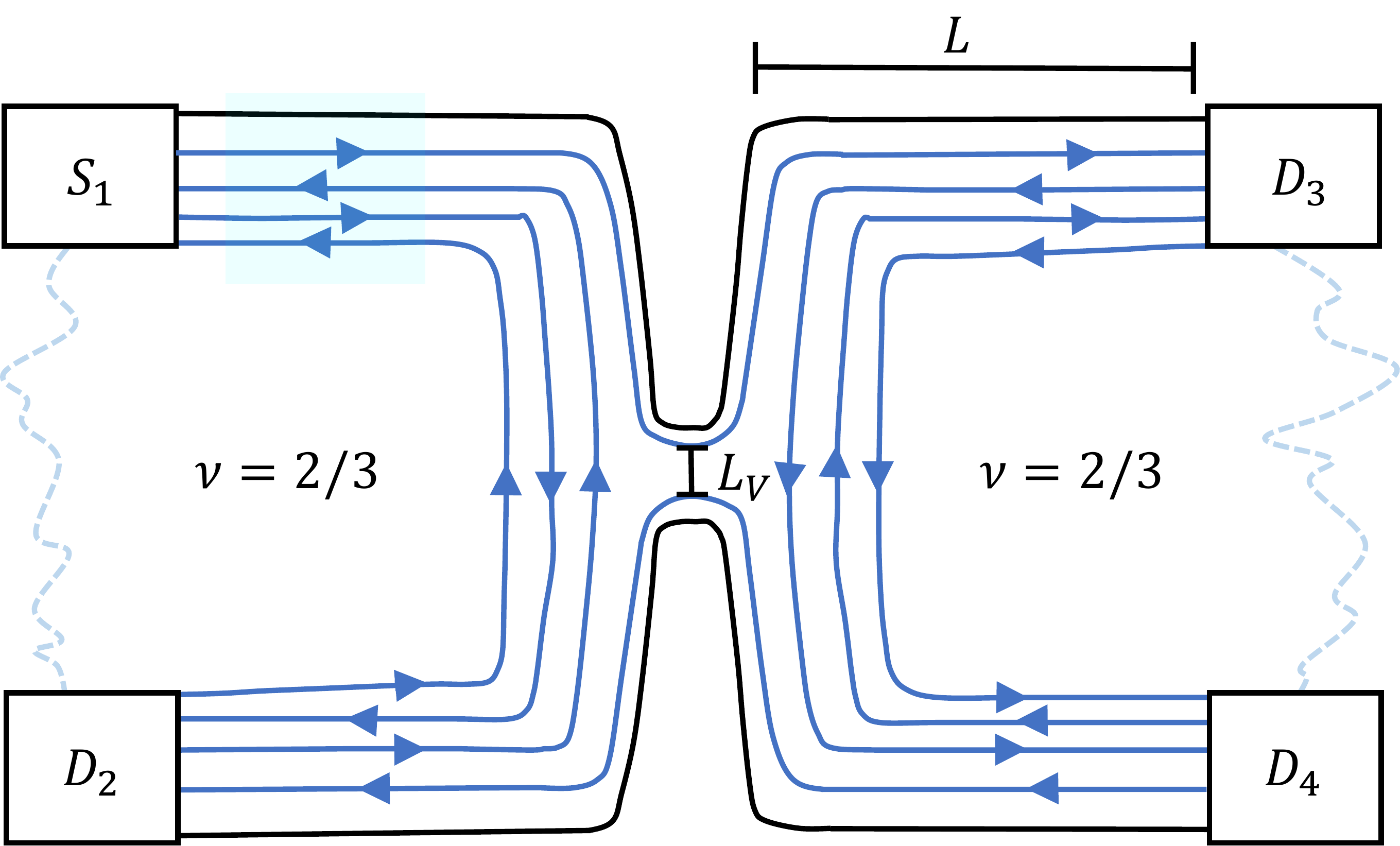}}\caption{\textcolor{black}{$(a)$ Discrete tunneling bridge for the WMG edge
structure. The inner three channels are characterized by the tunneling
probability $g_{1}$, while the outer-channel processes are characterized
by the tunneling parameter $g_{2}$ $(b)$ WMG edge structure in the
QPC geometry (tunneling not shown) }}
\end{figure}

\subsection{Derivation of the transport equations}

\subsubsection*{Electrical transport}

We now generalize our calculations to the WMG picture of the edge.
In this scheme, we take the bare modes of the WMG picture as a starting
point (labelling the outermost mode $1$, and the inner modes $2,3,4$
from the edge to the bulk \textcolor{black}{{[}see Fig. $5(a)$}{]}),
and add single-electron tunneling processes between inner counter-propagating
modes, in addition to a two-electron process which involves all three
inner modes. The operators for these processes are of the form $\hat{O}_{i}=e^{i\sum_{m}n_{i,m}\cdot\phi_{m}}$,
where $\phi_{m=2,3,4}$ are boson fields describing the inner-three
modes, and the vectors ${\bf n}_{i}$ are of the following form \cite{=000023WMG2013}:

\begin{equation}
\mathbf{n}_{1}=\begin{pmatrix}3\\
1\\
0
\end{pmatrix},\mathbf{n}_{2}=\begin{pmatrix}0\\
1\\
3
\end{pmatrix},\mathbf{n}_{3}=\begin{pmatrix}3\\
2\\
3
\end{pmatrix}.\label{eq:27}
\end{equation}
While in principle these different processes will all have different
tunneling probabilities $g_{n_{1}},g_{n_{2}},g_{n_{3}}$ associated
with them, for simplicity we take the associated tunneling coefficients
to all be the same magnitude, $g_{1}\ll1$, where $g_{1}$ is the
probability for tunneling for the inner-three mode processes. In addition,
we couple mode $1$ to mode $2$ through $1/3$-charged quasiparticle
tunneling, which has the form $I_{\tau1,2,i}=g_{2}\left(e^{2}/h\right)(V_{1,i}-V_{2,i+1})$,
$g_{2}\ll1$, where $g_{2}$ is the probability for $1/3$-charged
quasiparticle tunneling. From here, we can now derive the transport
equations. We first divide the current into normal current and tunneling
current as before:
\begin{align}
I_{1,i+1} & =I_{1,i}-I_{\tau1,i}\nonumber \\
I_{2,i+1} & =I_{2,i}-I_{\tau2,i}\nonumber \\
I_{3,i+1} & =I_{3,i}-I_{\tau3,i}\nonumber \\
I_{4,i+1} & =I_{4,i}-I_{\tau4,i}\label{eq:28}
\end{align}
From current conservation, we must have that 
\begin{equation}
I_{\tau1,i}+I_{\tau2,i}+I_{\tau3,i}+I_{\tau4,i}=0\label{eq:29}
\end{equation}
and in particular, for any individual tunneling process, we must also
have current conservation, e.g.
\begin{equation}
I_{n_{3}2,i}+I_{n_{3}3,i}+I_{n_{3}4,i}=0,\label{eq:30}
\end{equation}
where $I_{n_{3}2,i}$ is the tunneling current from channel $2$ associated
with the $\mathbf{n}_{3}$ process. The tunneling processes $\mathbf{n}_{1},\mathbf{n}_{2}$
are simply the same tunneling processes we accounted for in the discussion
of the KFP edge, and therefore have the same form for the tunneling
current. However, $\mathbf{n}_{3}$ is a two-electron process and
is more complicated. The tunneling current $I_{n_{3}3,i}$ is of the
following form:
\begin{equation}
I_{n_{3}3,i}=\left(2g_{1}\frac{e^{2}}{h}\right)\times\left[\left(V_{3,i}-V_{2,i+1}\right)+\left(V_{3,i}-V_{4,i+1}\right)\right],\label{eq:31}
\end{equation}
where the factor of $2$ comes from the fact that it is a two-electron
process \cite{=000023Kane1995}. With the use of Eq. $(28)$, we can
now write down the current conservation equations:

\begin{align}
I_{1,i+1} & =I_{1,i}-g_{2}\frac{e^{2}}{3h}\left(V_{1,i}-V_{2,i+1}\right)\nonumber \\
I_{2,i+1} & =I_{2,i}+g_{2}\frac{e^{2}}{3h}\left(V_{1,i}-V_{2,i+1}\right)-g_{1}\frac{e^{2}}{h}\left(V_{2,i+1}-V_{3,i}\right)-g_{1}\frac{e^{2}}{h}\left(V_{2,i+1}-2V_{3,i}+V_{4,i+1}\right)\nonumber \\
I_{3,i+1} & =I_{3,i}+g_{1}\frac{e^{2}}{h}\left(V_{2,i+1}-V_{3,i}\right)+2g_{1}\frac{e^{2}}{h}\left(V_{2,i+1}-2V_{3,i}+V_{4,i+1}\right)+g_{1}\frac{e^{2}}{h}\left(V_{4,i+1}-V_{3,i}\right)\nonumber \\
I_{4,i+1} & =I_{4,i}-g_{1}\frac{e^{2}}{h}\left(V_{4,i+1}-V_{3,i}\right)-g_{1}\frac{e^{2}}{h}\left(V_{2,i+1}-2V_{3,i}+V_{4,i+1}\right)\label{eq:32}
\end{align}

Converting voltages to currents and taking the continuum limit, we
obtain the following differential equations:
\begin{equation}
\frac{1}{\ell_{1}}\begin{pmatrix}-\alpha & -\alpha & 0 & 0\\
\alpha & \left(\alpha+6\right) & 3 & 3\\
0 & -9 & -6 & -9\\
0 & 3 & 3 & 6
\end{pmatrix}\begin{pmatrix}I_{1}\left(x\right)\\
I_{2}\left(x\right)\\
I_{3}\left(x\right)\\
I_{4}\left(x\right)
\end{pmatrix}=\partial_{x}\begin{pmatrix}I_{1}\left(x\right)\\
I_{2}\left(x\right)\\
I_{3}\left(x\right)\\
I_{4}\left(x\right)
\end{pmatrix},\label{eq:33}
\end{equation}
with $\ell_{1}\equiv a/g_{1},\ell_{2}\equiv a/g_{2},\alpha\equiv\ell_{1}/\ell_{2}$.
From $(33)$, one can directly see that current conservation is obeyed,
i.e. $\partial_{x}\left[I_{1}(x)+I_{2}(x)+I_{3}(x)+I_{4}(x)\right]=0$.
The solution to the differential equations $(33)$ is of the form
\begin{align}
\vec{I}\left(x\right) & =a_{1}\begin{pmatrix}0\\
0\\
-1\\
1
\end{pmatrix}e^{\frac{3x}{\ell_{1}}}+a_{2}\begin{pmatrix}-1\\
1\\
-3\\
1
\end{pmatrix}+a_{3}\begin{pmatrix}\frac{1}{6}\left(3+\sqrt{3\left(3+8\alpha\right)}\right)\\
\frac{1}{6}\left(9-\sqrt{3\left(3+8\alpha\right)}\right)\\
-3\\
1
\end{pmatrix}e^{\frac{1}{2}\left(3-\sqrt{3\left(3+8\alpha\right)}\right)\frac{x}{\ell_{1}}}\nonumber \\
 & +a_{4}\begin{pmatrix}\frac{1}{6}\left(3-\sqrt{3\left(3+8\alpha\right)}\right)\\
\frac{1}{6}\left(9+\sqrt{3\left(3+8\alpha\right)}\right)\\
-3\\
1
\end{pmatrix}e^{\frac{1}{2}\left(3+\sqrt{3\left(3+8\alpha\right)}\right)\frac{x}{\ell_{1}}},\label{eq:34}
\end{align}
where $a_{1},a_{2},a_{3},a_{4}$ are constants to be determined by
the boundary conditions. In addition, for the analysis of the QPC
with the outermost channel transmitted, we need the solution for the
inner-three channel mixing without any coupling to the outermost mode:
\begin{align}
\frac{1}{\ell_{1}}\begin{pmatrix}6 & 3 & 3\\
-9 & -6 & -9\\
3 & 3 & 6
\end{pmatrix}\begin{pmatrix}I_{2}\\
I_{3}\\
I_{4}
\end{pmatrix} & =\partial_{x}\begin{pmatrix}I_{2}\\
I_{3}\\
I_{4}
\end{pmatrix}\nonumber \\
\Rightarrow\begin{pmatrix}I_{2}(x)\\
I_{3}(x)\\
I_{4}(x)
\end{pmatrix} & =b_{1}\begin{pmatrix}-1\\
0\\
1
\end{pmatrix}e^{\frac{3x}{\ell_{1}}}+b_{2}\begin{pmatrix}-1\\
1\\
0
\end{pmatrix}e^{\frac{3x}{\ell_{1}}}+b_{3}\begin{pmatrix}1\\
-3\\
1
\end{pmatrix},\label{eq:35}
\end{align}
where $b_{1},b_{2},b_{3}$ are constants to be determined by boundary
conditions. 

\subsubsection*{Charge/neutral mode basis\textcolor{red}{{} }}

The differential equation Eq. $(33)$ can be transformed into the
corresponding charge/neutral mode basis for $\alpha=0$ (i.e. the
charge/neutral mode basis for the inner-three modes, as in Ref. \cite{=000023WMG2013});
this corresponds to obtaining the diagonal matrix $D=UAU^{-1}$:

\begin{equation}
A\vec{I}=\partial_{x}\vec{I}\rightarrow D\vec{\tilde{I}}=\partial_{x}\vec{\tilde{I}},\label{eq:36}
\end{equation}
with $\vec{\tilde{I}}=U^{-1}\vec{I}$. At the level of the $\phi$
fields, this corresponds to the transformation $\tilde{\phi}_{i}=\left(U^{-1}\right)_{ij}\phi_{j}$,
where $\tilde{\phi}_{i}$ are the new fields in the charge/neutral
mode basis. For $\alpha=0$, $D$ and $\left(U_{ij}\right)^{-1}$
are given as:
\begin{equation}
D=diag\left[\frac{3}{\ell_{1}},\frac{3}{\ell_{1}},0,0\right],\left(U^{-1}\right)_{ij}=\begin{pmatrix}0 & 1 & 1 & 2\\
0 & 3 & 2 & 3\\
0 & 1 & 1 & 1\\
1 & 0 & 0 & 0
\end{pmatrix}\label{eq:37}
\end{equation}
The first two rows correspond to the transformations for the neutral
modes modes (with a finite decay length $\ell_{1}/3$), while the
latter two correspond to the transformations for the charge modes
(with eigenvalue $0$); the neutrality of the neutral modes can be
established by the fact that the creation operators carry zero net
charge (See Appendix A). The diagonal basis thus corresponds to the
incoherent analogue of the intermediate fixed point. 

We can also find the incoherent analogue of the KFP fixed point by
looking at the $\alpha\rightarrow\infty$ limit. Diagonalizing the
matrix of Eq. $(30)$, we get the following solution:
\begin{equation}
\vec{\tilde{I}}(x)=a_{1}\begin{pmatrix}1\\
0\\
0\\
0
\end{pmatrix}e^{\frac{3x}{\ell_{1}}}+a_{2}\begin{pmatrix}0\\
1\\
0\\
0
\end{pmatrix}+a_{3}\begin{pmatrix}0\\
0\\
1\\
0
\end{pmatrix}e^{\frac{\left(3-\sqrt{3(3+8\alpha)}\right)x}{2\ell_{1}}}+a_{4}\begin{pmatrix}0\\
0\\
0\\
1
\end{pmatrix}e^{\frac{\left(3+\sqrt{3(3+8\alpha)}\right)x}{2\ell_{1}}}.\label{eq:38}
\end{equation}
In the case where $\alpha\rightarrow\infty$ (by fixing $\ell_{1}$
and taking $\ell_{2}\rightarrow0$), modes $3$ and $4$ in the new
basis become completely localized near the boundaries, while mode
$1$ is a neutral mode with decay length $\ell_{1}/3$ and mode $2$
is a charge mode. 

\subsubsection*{Heat transport}

In the case of heat transport, we find the following form of the tunneling
current between counter-propagating $\delta\nu=+1$ and $\delta\nu=-\nu$
modes, and between counter-propagating $\delta\nu=+\nu$ and $\delta\nu=-\nu$
modes:
\begin{equation}
J_{\tau\nu_{1}-\nu_{2},i}=\tilde{\gamma}g_{2}\frac{\pi^{2}k^{2}}{6h}\left(T_{\nu_{1},i}^{2}-T_{\nu_{2},i+1}^{2}\right),\ \ \ J_{\tau1-\nu,i}=\gamma g_{1}\frac{\pi^{2}k^{2}}{6h}\left(T_{1,i}^{2}-T_{\nu,i+1}^{2}\right),\label{eq:39}
\end{equation}

with $\tilde{\gamma}\equiv3\nu/(2\nu+1)\xrightarrow[\nu\rightarrow1/3]{}3/5$
in the absence of interaction between the modes and $\gamma$ as before;
$\tilde{\gamma}$ is a constant which measures the deviation from
the Wiedemann-Franz law for the tunneling current at a single impurity
in the case of counter-propagating $\delta\nu=+\nu$ and $\delta\nu=-\nu$
modes (see Appendix B). For the $\mathbf{n}_{3}$ process, we assume
the following form for the tunneling current:
\begin{equation}
J_{\tau3,i}^{n_{3}}=A(\overline{T}_{i})(T_{3,i}^{2}-T_{2,i+i}^{2})+B(\overline{T}_{i})(T_{3,i}^{2}-T_{4,i+1}^{2}),\label{eq:40}
\end{equation}
where the coefficients $A(\overline{T}_{i}),B(\overline{T}_{i})$
can in principle have temperature dependence, with $\overline{T}_{i}$
as the average temperature of the incoming modes at a tunneling bridge.
We come to this form through the following considerations. We expect
the heat tunneling current to depend on the difference of the squares
of the temperatures of the individual modes, since heat current should
depend on the square of the temperature, and should not have a contribution
if adjacent channels have the same temperature. Also, by the symmetry
of the problem, we set $A=B$, and as a simplifying assumption, $A=B=\gamma g_{1}$.
The coefficients can in principle be different, and without doing
a microscopic calculation we cannot know their form precisely, but
since we are interested broadly in seeing the behavior of equilibration,
we suspect that minute changes in the coefficients will not make much
difference in the overall behavior of the system.

Putting this all together, we obtain the following equations for heat
transport:
\begin{equation}
\frac{\gamma}{\ell_{1}}\begin{bmatrix}-\bar{\alpha} & \bar{\alpha} & 0 & 0\\
-\bar{\alpha} & (\frac{3}{2}+\bar{\alpha}) & -2 & \frac{1}{2}\\
0 & 2 & -4 & 2\\
0 & \frac{1}{2} & -2 & \frac{3}{2}
\end{bmatrix}\begin{bmatrix}T_{1}^{2}(x)\\
T_{2}^{2}(x)\\
T_{3}^{2}(x)\\
T_{4}^{2}(x)
\end{bmatrix}=\partial_{x}\begin{bmatrix}T_{1}^{2}(x)\\
T_{2}^{2}(x)\\
T_{3}^{2}(x)\\
T_{4}^{2}(x)
\end{bmatrix},\ \ \ \frac{\gamma}{\ell_{1}}\begin{bmatrix}\frac{3}{2} & -2 & \frac{1}{2}\\
2 & -4 & 2\\
\frac{1}{2} & -2 & \frac{3}{2}
\end{bmatrix}\begin{bmatrix}T_{2}^{2}(x)\\
T_{3}^{2}(x)\\
T_{4}^{2}(x)
\end{bmatrix}=\partial_{x}\begin{bmatrix}T_{2}^{2}(x)\\
T_{3}^{2}(x)\\
T_{4}^{2}(x)
\end{bmatrix},\label{eq:41}
\end{equation}
for the four-channel and three-channel mixing, respectively, where
$\bar{\alpha}\equiv(\tilde{\gamma}/\gamma)\alpha=(\tilde{\gamma}/\gamma)\ell_{1}/\ell_{2}$.
Taking into account the chirality of the channels, from Eqs. $(38)$
one can easily see that energy is conserved. The resulting solutions
are
\begin{align}
\overrightarrow{T^{2}}(x) & =c_{1}\begin{pmatrix}1\\
1\\
1\\
1
\end{pmatrix}+c_{2}\left[\frac{x}{\ell_{1}}\begin{pmatrix}1\\
1\\
1\\
1
\end{pmatrix}-\frac{1}{\bar{\alpha}\eta}\begin{pmatrix}(1+\bar{\alpha})\\
\bar{\alpha}\\
3\bar{\alpha}/4\\
0
\end{pmatrix}\right]+\frac{c_{3}}{\left(9+\eta\right)}\begin{pmatrix}-6\bar{\alpha}\\
3-6\bar{\alpha}+3\eta\\
4(3+\eta)\\
(9+\eta)
\end{pmatrix}e^{\frac{-(1+\eta)\gamma x}{2\ell_{1}}}\nonumber \\
 & +\frac{c_{4}}{\left(-9+\eta\right)}\begin{pmatrix}6\bar{\alpha}\\
-3+6\bar{\alpha}+3\eta\\
4(-3+\eta)\\
\left(-9+\eta\right)
\end{pmatrix}e^{\frac{(-1+\eta)\gamma x}{2\ell_{1}}},\nonumber \\
\overrightarrow{T^{2}}(x) & {\color{black}=d_{1}\begin{pmatrix}1\\
2\\
1
\end{pmatrix}e^{\frac{-2\gamma x}{\ell_{1}}}+d_{2}\begin{pmatrix}-1\\
0\\
1
\end{pmatrix}e^{\frac{\gamma x}{\ell_{1}}}+d_{3}\begin{pmatrix}1\\
1\\
1
\end{pmatrix},}
\end{align}
with $\eta\equiv\sqrt{9+6\bar{\alpha}}$ and $c_{1},c_{2},c_{3},c_{4},d_{1},d_{2},d_{3}$
to be determined by the boundary conditions.

\subsection{\textcolor{black}{Transport in a single QPC}}

\subsubsection*{\textcolor{black}{Electrical transport}}

The problem is now to match the boundary conditions to conform with
the geometry of the QPC \textcolor{black}{with the outermost channel
transmitted {[}see Fig.} $5(b)${]}. This requires matching 22 boundary
conditions, which in principle is straightforward but leaves us with
lengthy analytical\textcolor{black}{{} expressions (see Appendix D).
}We numerically solve this linear system of 22 equations in 22 variables
for different values of the parameters $L,L_{V},\ell_{1},\ell_{2}$
and examine the conductances $G_{S_{1}D_{2}},G_{S_{1}D_{3}},G_{S_{1}D_{4}}$
as functions of these parameters, where $L_{V}$ is vertical separation
of transmitted channels in the QPC, $L$ is the length of an arm of
the QPC, and $\ell_{1},\ell_{2}$ are the scattering lengths associated
with the inner three channels and the outer channel tunneling, respectively.
We note that the following discussion is always in the regime where
$\ell_{1}\ll L$. Within this regime, we always have the condition
that $G_{S_{1}D_{2}}+G_{S_{1}D_{3}}=(2/3)e^{2}/h$ (up to exponentially
small corrections in $L/\ell_{1}$); we see numerically that this
condition is met.\\
\begin{figure}[H]
\centering{}\subfloat[\label{fig: 6a conductance plot single QPC}]{\includegraphics[width=8cm]{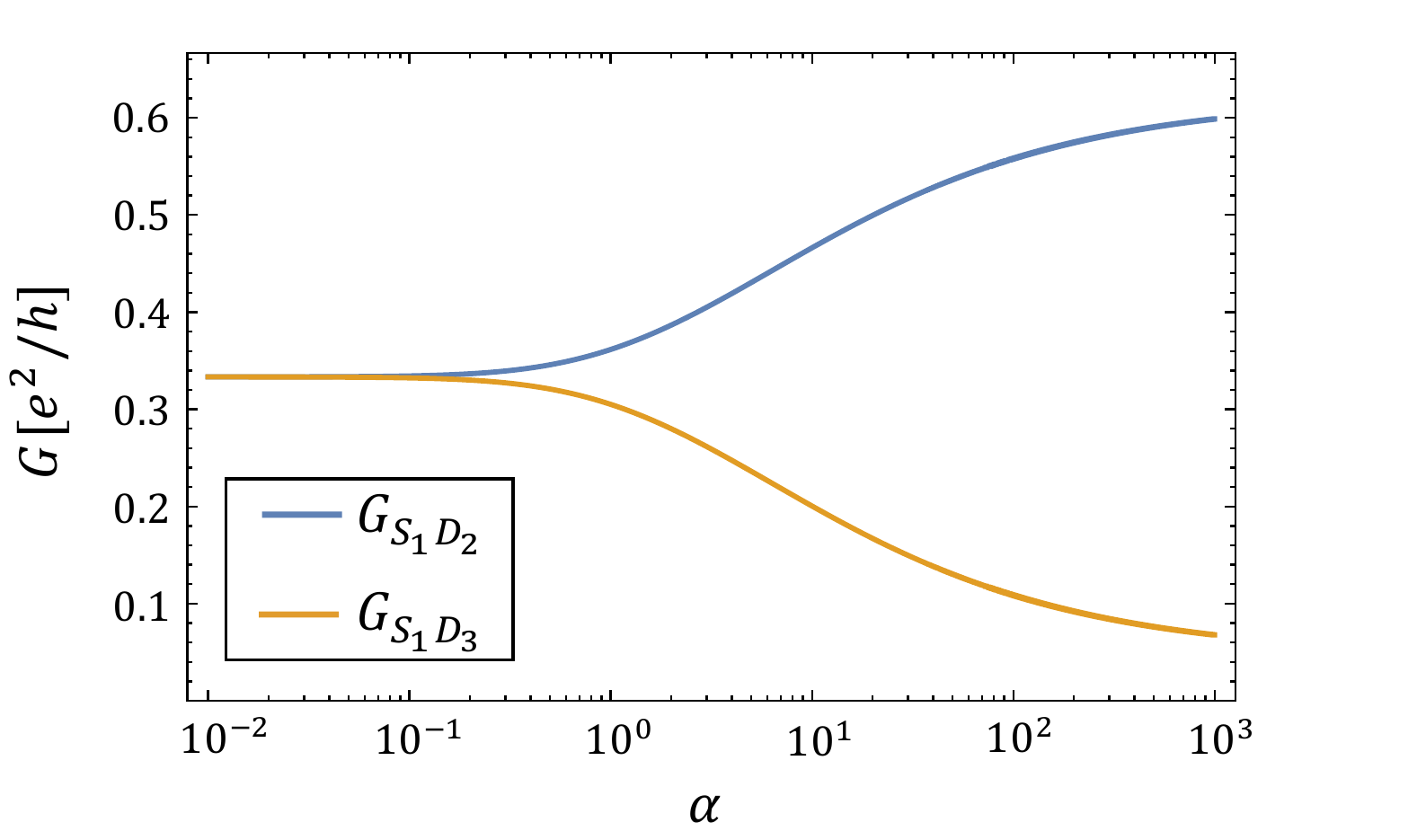}

}\qquad{}\subfloat[\label{fig: 6b WMG intermediate fixed pt single QPC}]{\includegraphics[width=8cm]{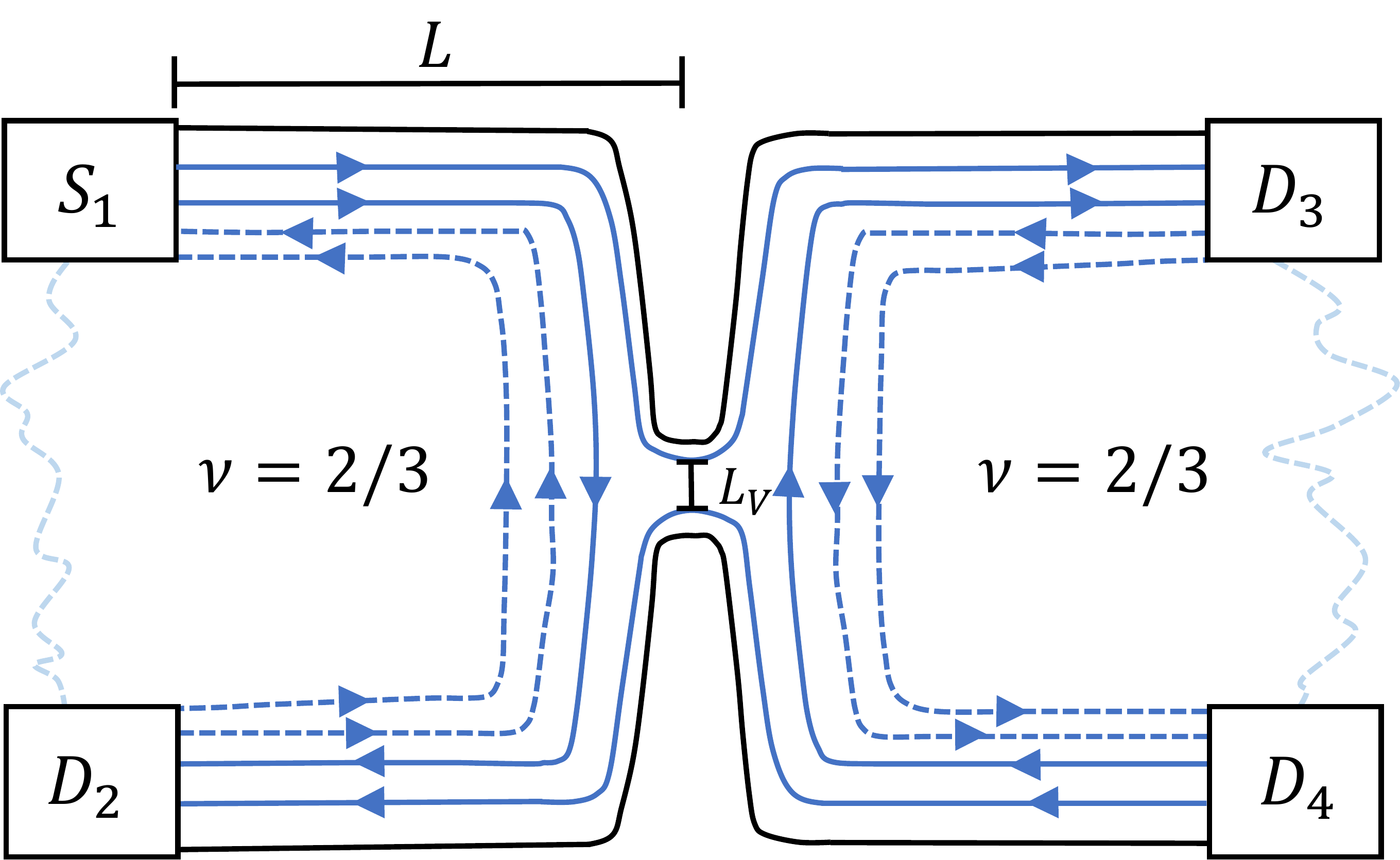}

}\qquad{}\subfloat[\label{fig: 6c KFP fixed pt single QPC}]{\includegraphics[width=8cm]{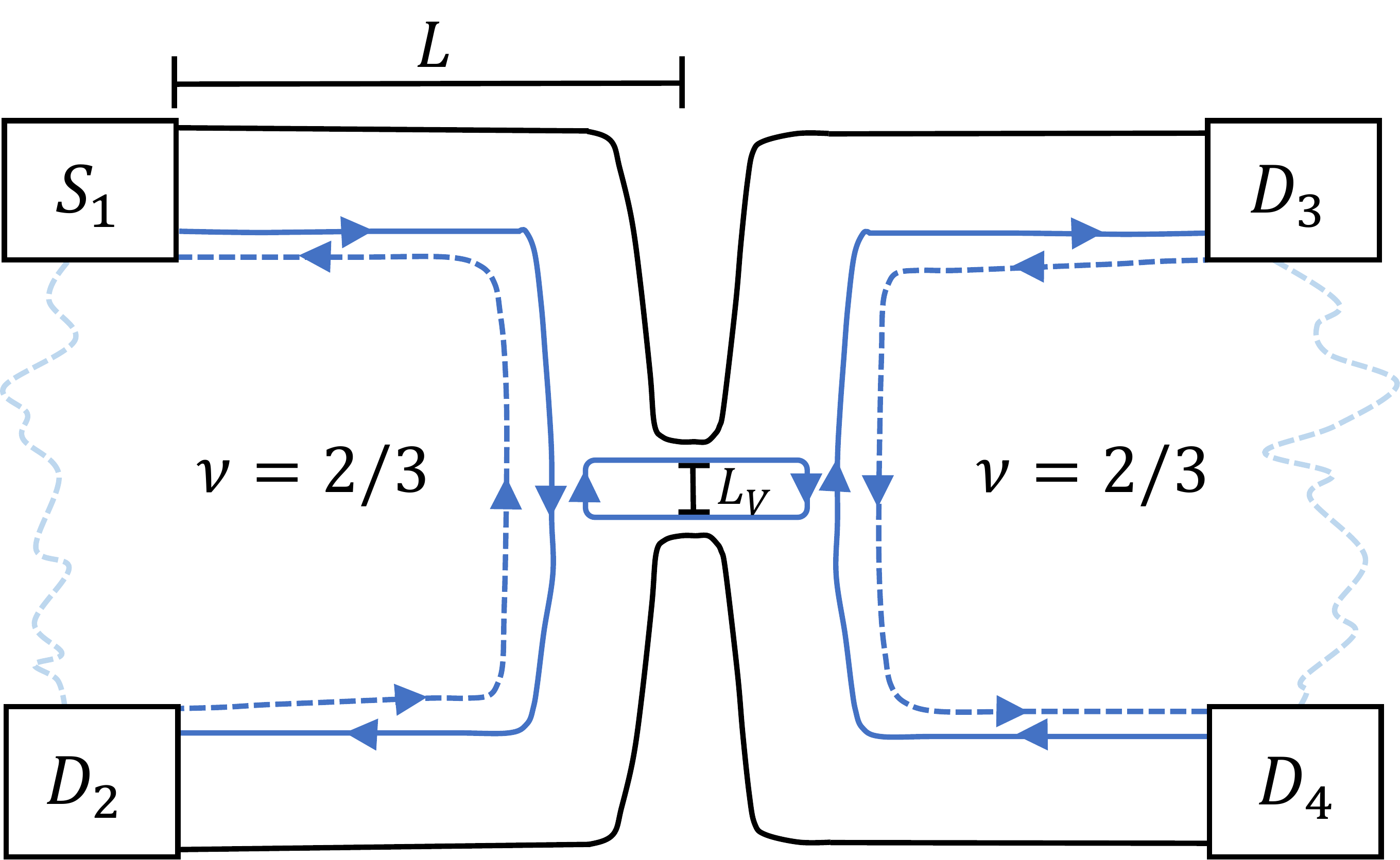}}\\
\caption{\textcolor{black}{$(a)$ Linear-log plot of the conductances $G_{S_{1}D_{2}},G_{S_{1}D_{3}}$
in the single QPC as a function of $\alpha=\ell_{1}/\ell_{2}$ for
$\ell_{1}=100a,L=1000a,L_{V}=5a,$ and $\alpha=.01-1000$. $(b)$
Renormalized edge modes of the single QPC at the intermediate fixed
point. This renormalized edge structure picture produces conductance
results qualitatively similar to our model in the regime that $\ell_{2}>L$
(the decay of the neutral modes is not shown). $(c)$ Renormalized
edge modes of the single QPC in the KFP (low-temperature) fixed point,
with a $\nu=1/3$ density QH droplet in the constriction. This renormalized
edge structure picture produces conductance results qualitatively
similar to our model in the limit that $\alpha\rightarrow\infty$
(the decay of the neutral mode is not shown).}}
\end{figure}

\subsubsection*{Results}

The results of the calculation can be characterized by several different
classes of behavior depending on the ratios of the lengths $L,L_{V},\ell_{1}$,
and $\ell_{2}$. We note that out of the six possible ratios of the
length scales $\left(\alpha,\ell_{1}/L,\ell_{1}/L_{V},\ell_{2}/L,\ell_{2}/L_{V},L/L_{V}\right)$,
the ratios which show up directly in the equations are $\alpha,\ell_{1}/L,\ell_{1}/L_{V},\ell_{2}/L$;
we choose these as our parameters to span parameter space. Then, the
parameters that we play with are $\alpha,\ell_{1}/L_{V},$ and $\ell_{2}/L,$
noting always that we are in the regime of $\ell_{1}\ll L$.\\
The main result of this calculation is that there are two distinct
regimes of conductance quantization: $(1)$ when $L_{V}\gg\ell_{1}$,
$G_{S_{1}D_{3}}$ is always quantized at $(1/3)e^{2}/h$ (up to exponentially
small corrections in $L/\ell_{1}$), regardless of the values of $\alpha$
or $\ell_{2}/L$, and $(2)$ when $L_{V}<\ell_{1}$ {[}see Fig. $6(a)${]}
\cite{footnote}. In the regime $(2)$, we identify three main points
of relevant behavior: $(a)$ the region in parameter space where $\alpha\ll1$,
$(b)$ a cross-over region where $\alpha\sim1,$ and $(c)$ where
$\alpha\gg1$. The conductance for these different regimes are shown
i\textcolor{black}{n Fig. $6(a)$.} To understand these conductance
results, we transform into the charge/neutral mode basis, in analogy
with the low temperature regime fixed points of Ref. \cite{=000023WMG2013}.
In the $\alpha\ll1$ regime, we note that this corresponds to the
incoherent analogue of the intermediate fixed point where the tunneling
between the outer-most mode and the inner three modes is ignored,
and there are two co-propagating $1/3$-charge modes which carry the
current, which give $G_{S_{1}D_{2}}=G_{S_{1}D_{3}}=(1/3)e^{2}/h$
{[}\textcolor{black}{see Fig.} $6(b)${]};\textcolor{red}{{} }\textcolor{black}{deviations
from the quantized value in the $\alpha\ll1$ regime characterize
the deviation from the intermediate fixed point. }In the $\alpha\rightarrow\infty$
limit, the corresponding edge structure is that of a $\nu=1/3$ low-density
constriction where the counter-propagating $1/3$-charge modes localize
each other, counter-propagating $2/3$ and neutral modes remain (KFP
fixed point)\textcolor{black}{{} {[}see Fig. $6(c)${]}.} In this case,
by using the results from a calculation of the counter-propagating
$2/3-1/3$ line junction (the calculation is straightforward and
we omit it here), we see that the conductance $G_{S_{1}D_{3}}=(1/3)e^{2}/h$
when $L_{V}\gg\ell_{1}$, (with vanishingly small corrections as a
function of $L_{V}/\ell_{1}$), and vanishes for $\ell_{1}\gg L_{V}$
(with vanishingly small corrections as a function of $L_{V}/\ell_{1}$).
In between these two limits, there is some continuous change between
the two fixed point pictures, with a change of inflection in the conductance
around the point $\alpha\sim1$ from concave to convex, which is the
transition point from $g_{1}$ processes being the strongest to $g_{2}$
processes being the strongest. We emphasize that the conductance behavior
as a function of $\alpha$ is conceptually simpler in the charge/neutral
mode picture than in the bare mode picture, and reproduces the results
of our numerical calculations in the $\alpha\ll1$ and $\alpha\rightarrow\infty$
limits.

\subsubsection*{Heat transport}

Applying boundary conditions for the heat transport equations in the
single QPC, there are again 22 equations in 22 unknowns, corresponding
to the different segments of the QPC \textcolor{black}{(see Appendix
D).} From this, we calculate numerically the conductances $K_{S_{1}D_{2}},K_{S_{1}D_{3}},K_{S_{1}D_{4}}$,
and the conductance associated with current backscattered to contact
$A$, namely $K_{{\rm back}}$. The total injected current contributes
a thermal conductance of $\left(4T_{0}\right)\pi^{2}k^{2}/6h$, so
energy conservation dictates the following condition:
\begin{equation}
K_{S_{1}D_{2}}+K_{S_{1}D_{3}}+K_{S_{1}D_{4}}+K_{{\rm back}}=\frac{\pi^{2}k^{2}}{6h}4T_{0}
\end{equation}
\\
\begin{figure}
\centering{}\subfloat[\label{7a heat conductance plot}]{\includegraphics[width=8cm]{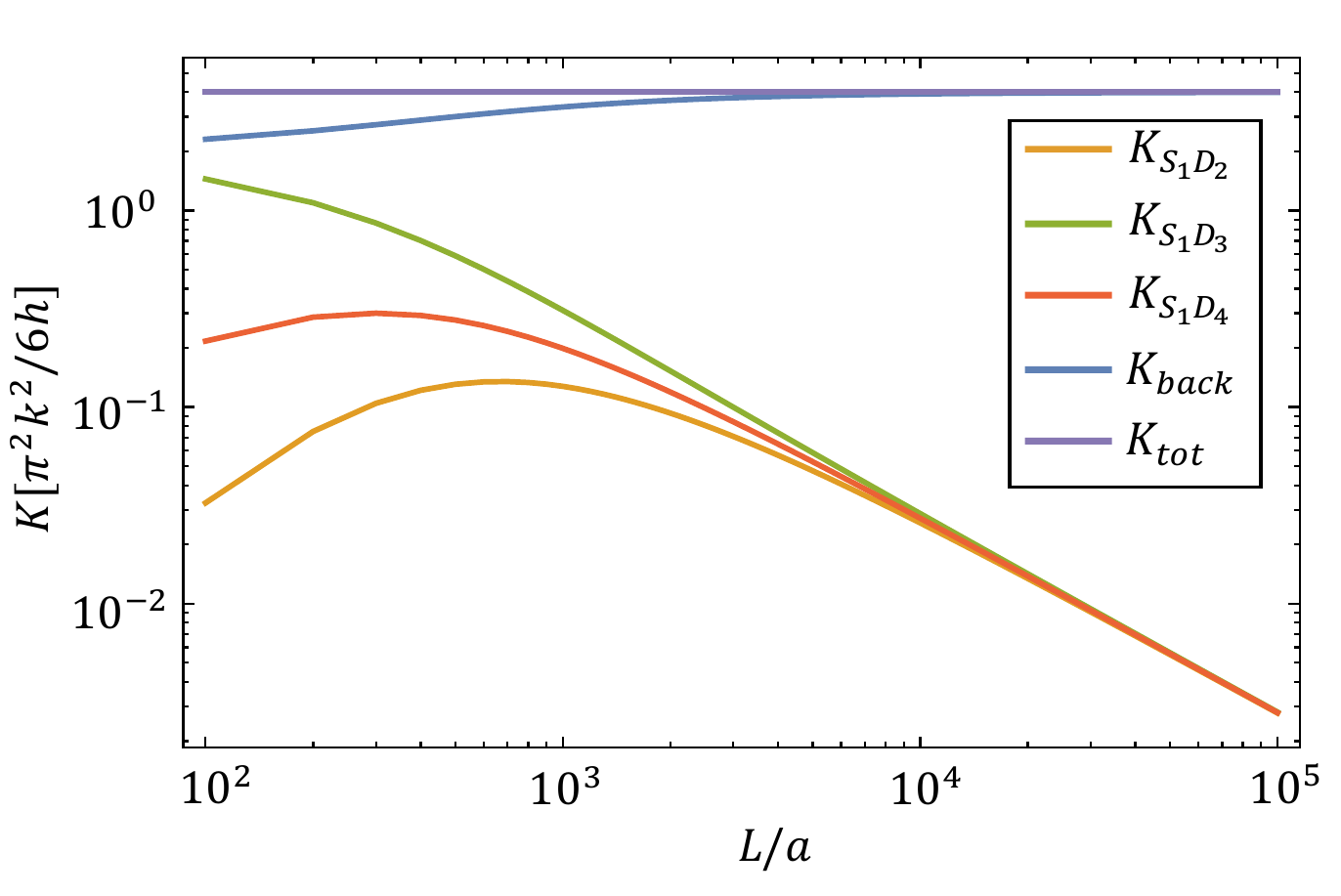}

}\qquad{}\subfloat[\label{7b heat conductance map}]{\includegraphics[width=7cm]{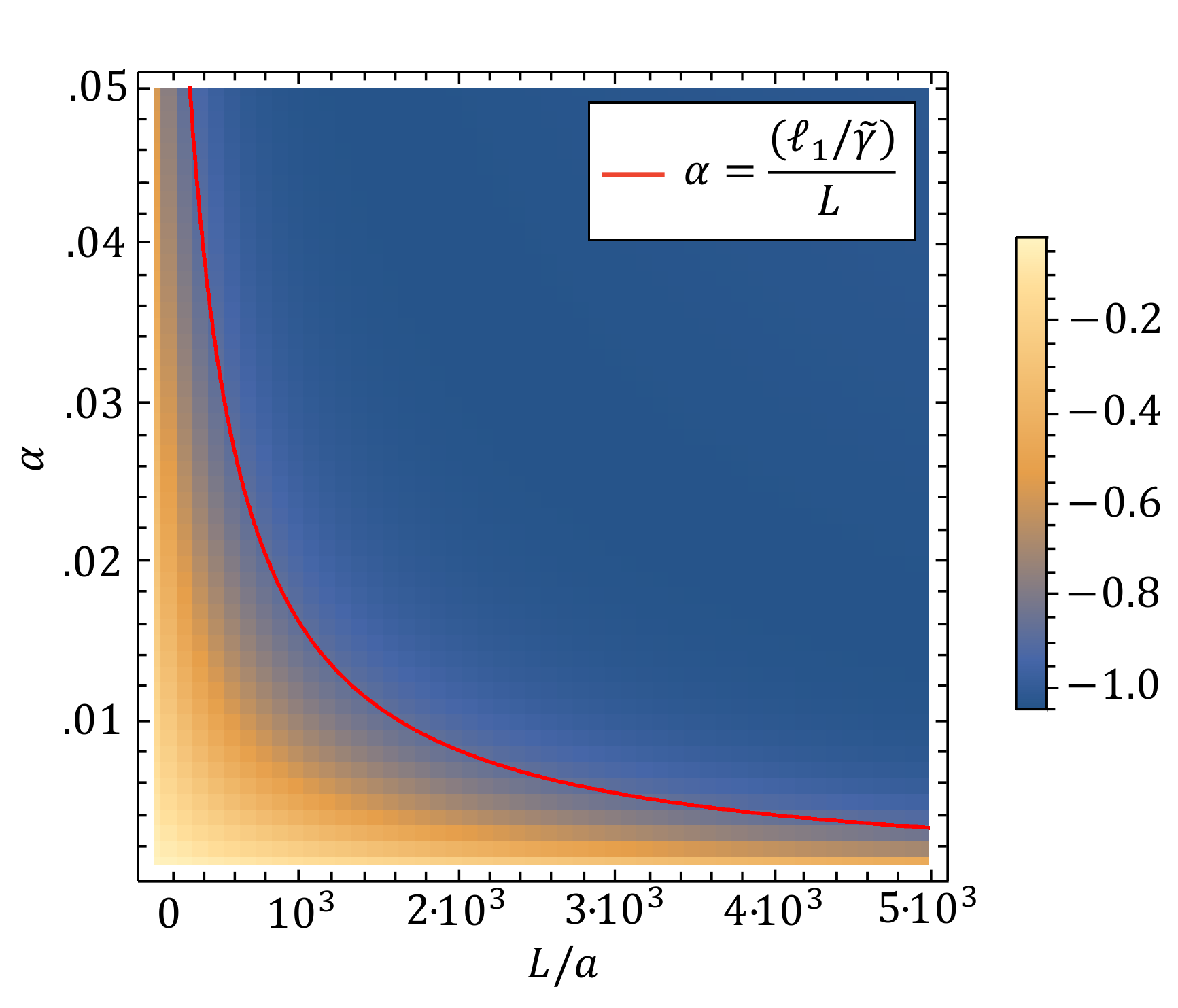}

}\caption{\textcolor{black}{$(a)$Log-log plot of the thermal conductances $K_{S_{1}D_{2}},K_{S_{1}D_{3}},K_{S_{1}D_{4}},K_{{\rm back}}$
in the single QPC as a function of $L$ for $\alpha=.03,\ell_{1}=10a,L_{V}=0a,$
and $L=10^{2}a-10^{5}a$. We note that there is a marked change in
the behavior of the conductance curves around $L\sim\bar{\ell}_{2}\sim560a$,
beyond which the power law in $L$ emerges; this is an indication
of equilibration between edge modes. $(b)$ Plot of the derivative
$d(\text{ln}(K_{S_{1}D_{3}}))/d(\text{ln}(L))$, which characterizes
the onset of the power law, as a function of $L$ and $\alpha$ for
$L=0-5000a$. We see that the onset of $1/L$ behavior for $K_{S_{1}D_{3}}$
occurs when $\bar{\ell}_{2}\lesssim L$, which happens when $\alpha\gtrsim\frac{\left(\ell_{1}/\tilde{\gamma}\right)}{L}$.
We overlay the curve $\alpha=\frac{\left(\ell_{1}/\tilde{\gamma}\right)}{L}$,
which matches qualitatively with the onset of $1/L$ behavior.}}
\end{figure}

\subsubsection*{Results}

For the QPC with the outermost channel transmitted, we expect now
that the relevant length scales will be $\bar{\ell}_{1},\bar{\ell}_{2},L$
and $L_{V}$, and we can look at how the conductances vary as function
of these parameters. We see several features: 
\begin{itemize}
\item We see a power law in $L$ in the conductances $K_{S_{1}D_{2}},K_{S_{1}D_{3}}$,
as we would naively expect from our previous analysis of heat transport
in different \textcolor{black}{geometries {[}see Fig. $7(a)${]}.
H}owever, this behavior only onsets once the the condition $\bar{\ell}_{1},\bar{\ell}_{2}\ll L$
is satisfied, which is the condition for full equilibration. In the
regime where $\bar{\ell}_{1}\ll L<\bar{\ell}_{2}$, there is some
non-trivial behavior where heat transport is non-vanishing, with a
peak in the backscattered heat current appearing around $L\sim\bar{\ell}_{2}$,
which indicates a change in transport from that with a non-equilibrated
outermost edge to that with an equilibrated outermost edge. Additionally,
we note that in the regime $\bar{\ell}_{2}>L$, the deviations which
$K_{S_{1}D_{2}},K_{S_{1}D_{3}}$ and $K_{S_{1}D_{4}}$ have from each
other drastically \textit{\textcolor{black}{decrease}} as $\alpha$
\textit{\textcolor{black}{increases}}.
\item Additionally, we note that the onset of the $1/L$ behavior is robust
with changes in $\alpha$, as can be see\textcolor{black}{n in Fig.
}$7(b)$.
\item The dependence of $K_{S_{1}D_{2}},K_{S_{1}D_{3}}$ and $K_{S_{1}D_{4}}$
on $L_{V}$ does not affect the onset of the power law in $L$. However,
it does affect the relative magnitudes of $K_{S_{1}D_{2}},K_{S_{1}D_{3}},K_{S_{1}D_{4}}$,
as we will describe in the following discussion below.
\end{itemize}
\textcolor{black}{It is informative to compare these results with
those of the MacDonald picture. In the MacDonald picture, the total
injected heat current contributes a thermal conductance of $\left(2T_{0}\right)\pi^{2}k^{2}/6h$,
so energy conservation dictates the following condition:
\begin{equation}
K_{S_{1}D_{2}}+K_{S_{1}D_{3}}+K_{S_{1}D_{4}}+K_{{\rm back}}=\frac{\pi^{2}k^{2}}{6h}2T_{0}
\end{equation}
In the MacDonald picture, we obtain the following thermal conductances
for the single QPC:}

\textcolor{black}{
\begin{align}
K_{S_{1}D_{2}} & =\frac{2L^{2}\bar{\ell}}{\left(2L+\bar{\ell}\right)\left(2L^{2}+2L\bar{\ell}+\bar{\ell}^{2}\right)}\frac{\pi^{2}k^{2}}{6h}T_{0}\nonumber \\
K_{S_{1}D_{3}} & =\frac{2\bar{\ell}\left(L+\bar{\ell}\right)^{2}}{\left(2L+\bar{\ell}\right)\left(2L^{2}+2L\bar{\ell}+\bar{\ell}^{2}\right)}\frac{\pi^{2}k^{2}}{6h}T_{0}\nonumber \\
K_{S_{1}D_{4}} & =\frac{2L\bar{\ell}\left(L+\bar{\ell}\right)}{\left(2L+\bar{\ell}\right)\left(2L^{2}+2L\bar{\ell}+\bar{\ell}^{2}\right)}\frac{\pi^{2}k^{2}}{6h}T_{0}\nonumber \\
K_{\text{back}} & =\frac{2L\left(4L^{2}+3L\bar{\ell}+\bar{\ell}^{2}\right)}{\left(2L+\bar{\ell}\right)\left(2L^{2}+2L\bar{\ell}+\bar{\ell}^{2}\right)}\frac{\pi^{2}k^{2}}{6h}T_{0}
\end{align}
We can now ask if there is a way to distinguish between the MacDonald
and WMG edge structures from these thermal conductance measurements.
From Eq. $(46)$, we see that the condition that $K_{S_{1}D_{3}}>K_{S_{1}D_{4}}>K_{S_{1}D_{2}}$
always holds for the MacDonald case. Comparing to the WMG case, if
we set $\bar{\ell}_{2}=\bar{\ell}$ and look at the regime where $\bar{\ell}_{1}\ll\bar{\ell}_{2}$,
we find that the thermal conductances of the WMG and MacDonald edges
are nearly identical; more generally, for the WMG case we find that
in the regime where $\bar{\ell}_{1}\ll\bar{\ell}_{2}$, the condition
$K_{S_{1}D_{3}}>K_{S_{1}D_{4}}>K_{S_{1}D_{2}}$ holds as we vary the
system size from a non-equilibrated regime $\left(L<\bar{\ell}_{1},\bar{\ell}_{2}\right)$
to a fully equilibrated one $\left(L\gg\bar{\ell}_{1},\bar{\ell}_{2}\right)$.
Thus in this regime, there is no robust signature in the thermal conductance
that would allow us to distinguish between the two edge structures.
However, in the regime where $\bar{\ell}_{1}\gg\bar{\ell}_{2},\ell_{1}\gg L_{V}$,
the following condition is met: $K_{S_{1}D_{2}}>K_{S_{1}D_{3}}>K_{S_{1}D_{4}}$,
differing from the MacDonald case. This condition holds as we vary
the system size from a non-equilibrated regime $\left(L<\bar{\ell}_{1},\bar{\ell}_{2}\right)$
to a fully equilibrated one $\left(L\gg\bar{\ell}_{1},\bar{\ell}_{2}\right)$,
and is therefore a robust signature which distinguishes the WMG picture
from the MacDonald picture. However, we note that the inequality $K_{S_{1}D_{2}}>K_{S_{1}D_{3}}>K_{S_{1}D_{4}}$
returns to $K_{S_{1}D_{3}}>K_{S_{1}D_{4}}>K_{S_{1}D_{2}}$ as we increase
$L_{V}\gtrsim\bar{\ell}_{1}$. Thus, the WMG edge is only distinguishable
in a robust manner when the system is in the limit of small constriction
width, i.e. $\ell_{1}\gg L_{V}$, with the additional condition that
$\bar{\ell}_{1}\gg\bar{\ell}_{2}$.}

\textcolor{red}{}

\begin{figure}[p]
\begin{centering}
\subfloat[\label{fig: 8a double QPC geometry 1}]{\includegraphics[width=8cm]{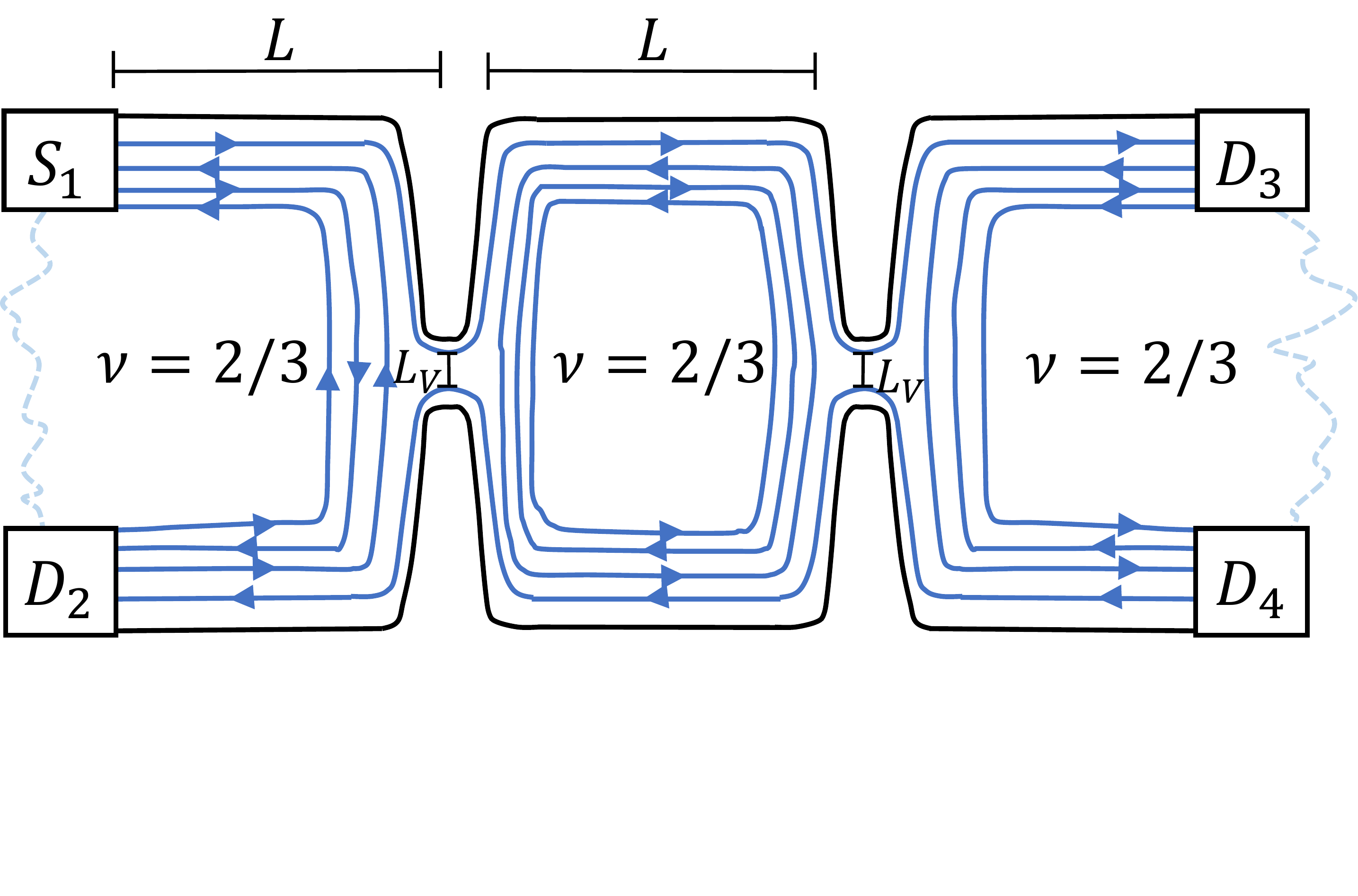}}\qquad{}\subfloat[\label{fig: 8b double QPC geometry 2}]{\includegraphics[width=8cm]{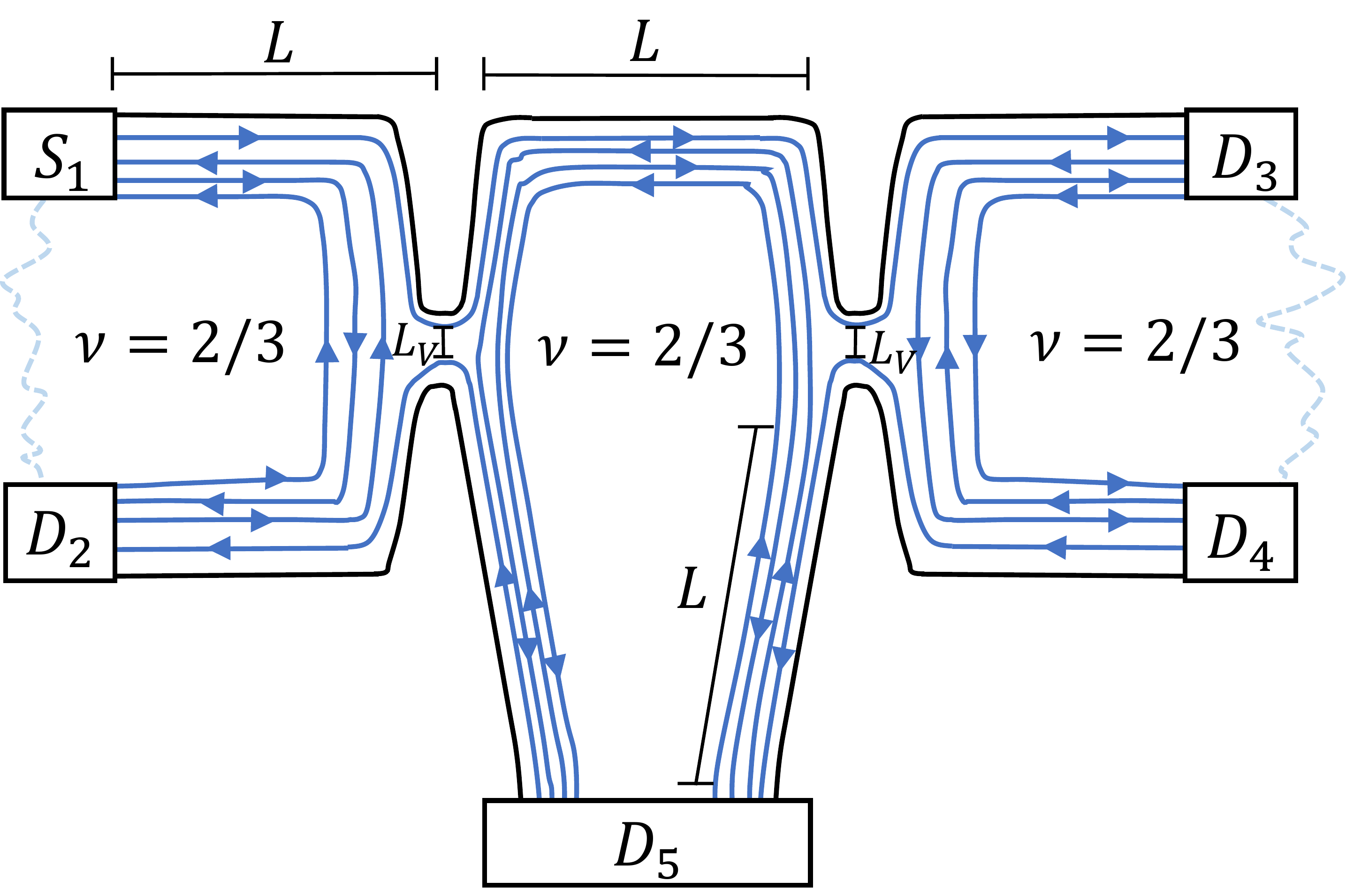}}
\par\end{centering}
\centering{}\subfloat[\label{fig:8c double QPC intermediate fixed pt}]{\includegraphics[width=8cm]{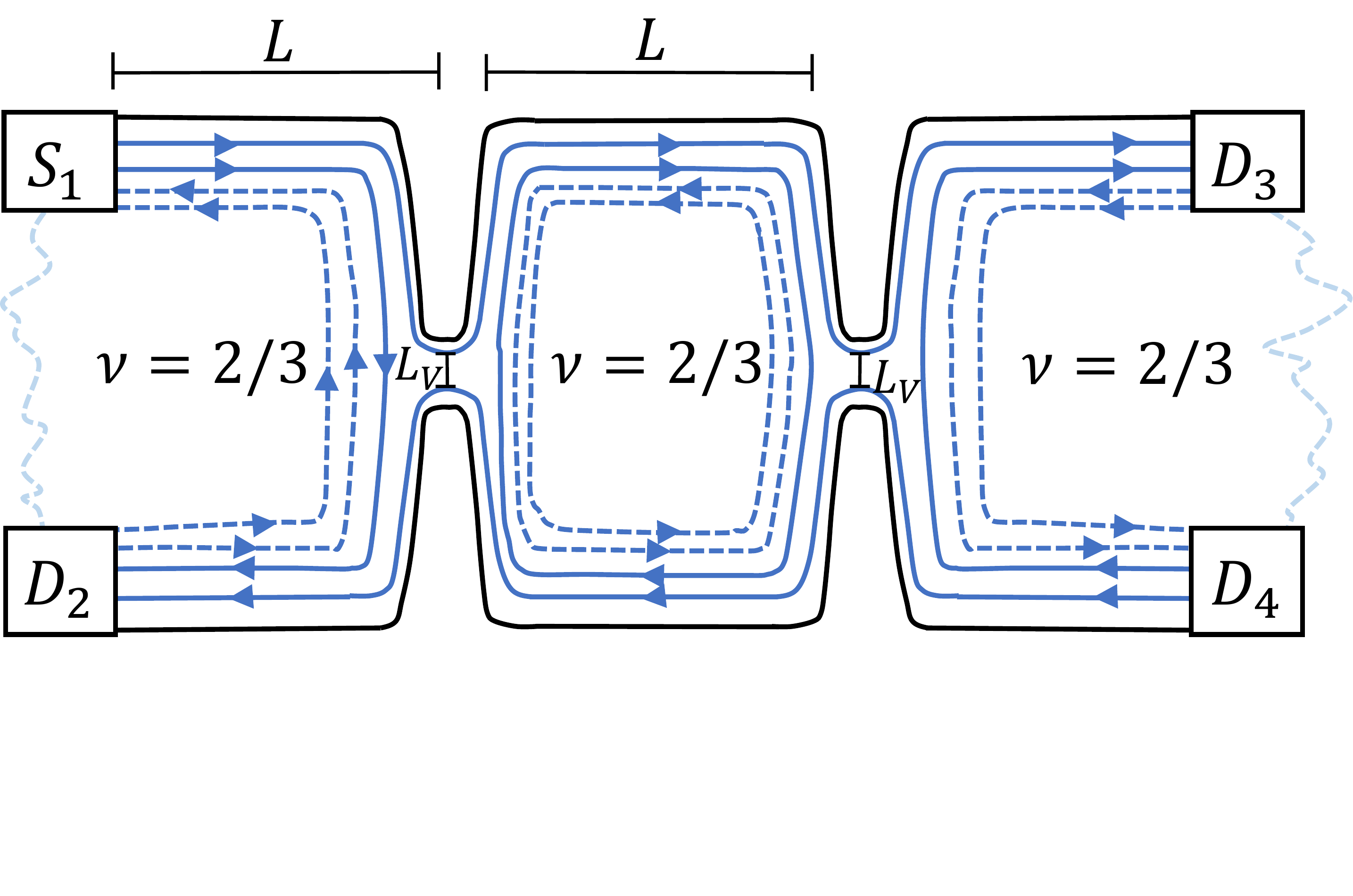}

}\qquad{}\subfloat[\label{fig:8d double QPC winding KFP fixed pt}]{\includegraphics[width=8cm]{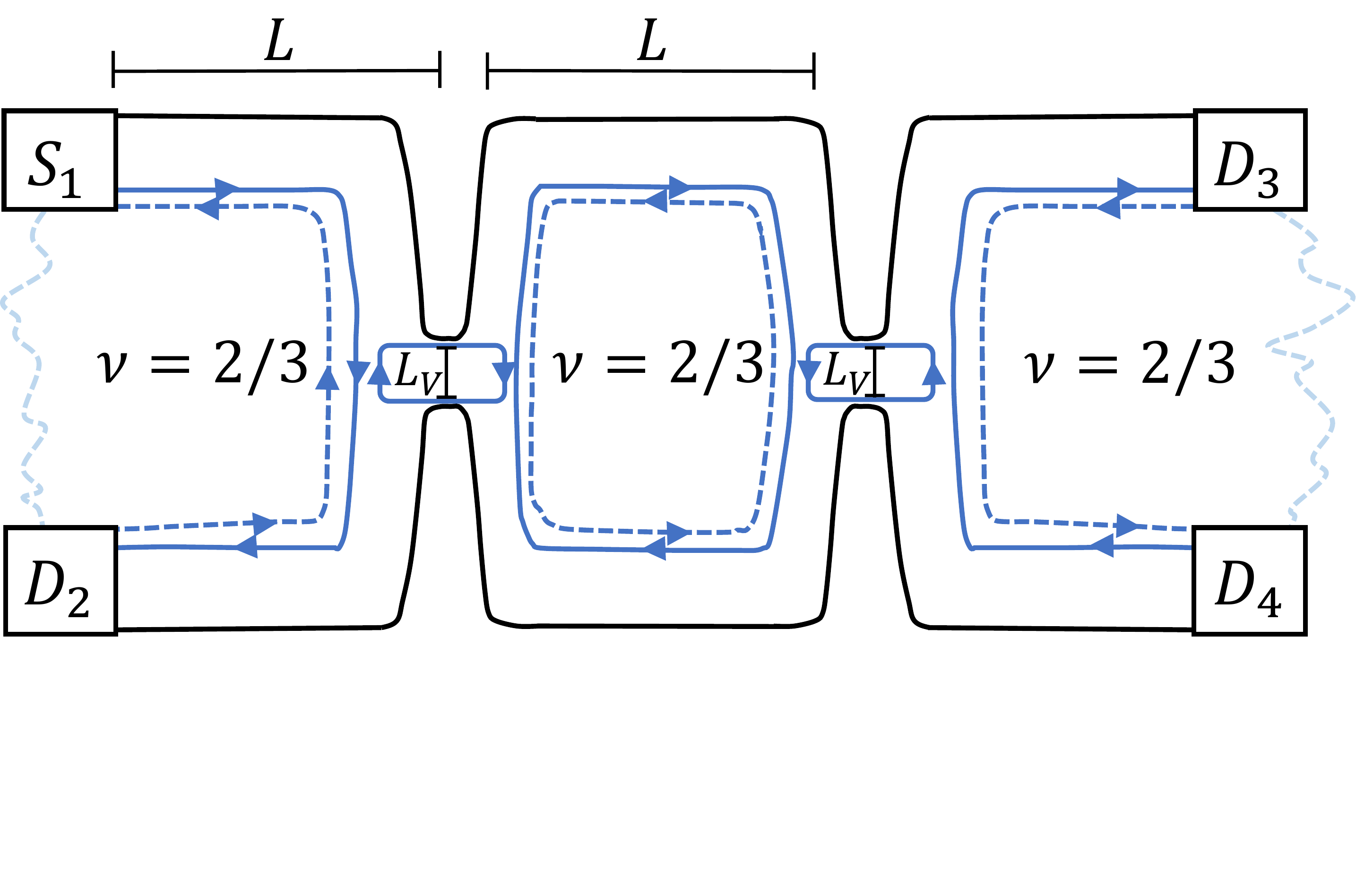}

}\caption{\textcolor{black}{$(a)$ Bare modes of the double QPC in the ``winding''
geometry. The width of the constriction is $L_{V}$, the length of
an arm of a QPC is $L$, and the circumference of the winding edge
states in between the two QPCs is $2L$. There are additional edge
structures for this geometry, in analogy with $(b)-(d)$ of this figure.
$(b)$ Bare modes of the double QPC in the ``no-winding'' geometry.
The length of the path that the edge states travel between the two
QPCs is $3L$ $(c)$ Incoherent analogue of the WMG intermediate fixed
point in the ``winding'' geometry (the decay of the neutral modes
is not shown). This picture reproduces the same conductance results
as our model in regime with $\ell_{2}>L$. $(d)$ Incoherent analogue
of the KFP fixed point in the ``winding'' geometry, with $\nu=1/3$
QH droplets in the constrictions (the decay of the neutral modes is
not shown). This picture reproduces the same conductance results as
our model in the limit that $\alpha\rightarrow\infty$.}}
\end{figure}

\subsection{Transport in a double QPC}

\subsubsection*{Electrical transport in the winding case}

In the case of electrical transport in a double QPC with circulating
edge states in between the two QPCs, we have the same sets of equations
as before, but now with 36 boundary conditions to match\textcolor{black}{{}
(see Appendix D). We refer to this geometry as the ``winding'' case
}{[}see Fig. $8(a)${]}\textcolor{black}{. }As in the case of electrical
transport in the single QPC, from the numerical calculations we see
two main regimes of behavior for the conductance $G_{S_{1}D_{3}}$:
$(1)$ for $\ell_{1}<L_{V}$, in the case where $\alpha\sim0$ (the
outermost edge is decoupled), the conductance is quantized at $(1/3)e^{2}/h$,
and drops down from this value as $\alpha$ is increased, dropping
to $(2/9)e^{2}/h$ as $\alpha\sim\ell_{1}/L\rightarrow\ell_{2}\sim L$.
Beyond this point, $\alpha$ is quantized at $(2/9)e^{2}/h$. $(2)$
In the case $\ell_{1}\gg L_{V}$, we see the same behavior of $G_{S_{1}D_{3}}$
up to $\ell_{2}\sim L$, but beyond this point, the conductance decreases
with $\alpha$ as in the single QPC case, \textcolor{black}{approaching
$0$ as $\alpha\rightarrow\infty$} {[}see Figs. $9(a,b)${]}. Again,
we can explain these behaviors in the charge/neutral mode basis in
the limits of $\alpha\ll1$ and $\alpha\rightarrow\infty$. In the
$\alpha\ll1$ case, where the edge structure is the incoherent analogue
of the intermediate fixed po\textcolor{black}{int {[}see Fig. $8(c)${]}},
we can see the transition from $G_{S_{1}D_{3}}=(1/3)e^{2}/h$ to $G_{S_{1}D_{3}}=(2/9)e^{2}/h$
as a function of $\ell_{2}/L$ {[}see Figs. $9(a,b)${]}; in the case
of $L\ll\ell_{2}$, nearly all of the injected current on the uppermost
mode i\textcolor{black}{s carried to drain $D_{3}$, and $G_{S_{1}D_{3}}\sim(1/3)e^{2}/h$
with deviations growing as $\ell_{2}$ decreases. Near $\ell_{2}\sim L$,
assuming full equilibration between the two charge modes in the island
between the QPCs, a short calculation yields $G_{S_{1}D_{3}}=(2/9)e^{2}/h$.
In the $\alpha\rightarrow\infty$ case, we again have the incoherent
analogue of the KFP edge structure with $1/3$ density constrictions
at the QPCs, as in Fig. $8(d)$. Using this effective edge structure,
and assuming equilibration between counter-propagating charge modes
(again using the results from the counter-propagating $2/3-1/3$ line
junction), one can easily calculate $G_{S_{1}D_{3}}=(2/9)e^{2}/h$.
Thus, as $\alpha$ is varied, we imagine that the picture of the edge
states is varied from that of the intermediate fixed point edge structure
to the KFP edge structure, just as in the case of the single QPC.
Additionally, we see that in the case $(2)$, we see the slow decay
in conductance as a function of $\alpha$ being controlled by the
competition of $g_{1}$ and $g_{2}$, as was seen in the single QPC
case. Thus around $\alpha\sim1$, there is a change of inflection
in the conductance curve as a function of $\alpha$ from concave to
convex, as in the single QPC case. In the limit of an extremely narrow
constriction with $L_{V}\rightarrow0$, the conductances $G_{S_{1}D_{3}}$
$\left(G_{S_{1}D_{2}}\right)$ do not saturate at the above described
bounds, but instead monotonically decrease (increase) as in Figs.
$9(a,b)$, eventually reaching the values $0$ $\left((2/3)e^{2}/h\right)$
in the $\alpha\rightarrow\infty$ limit.}

\subsubsection*{Electrical transport in the no-winding case}

In the presence of an Ohmic contact between the two QPCs, edge states
can no longer circulate in between the two QPCs. We refer to this
geometry as the ``no-winding'' case {[}see Fig. $8(b)${]}. The
results of the no-winding case are analogous to the winding case,
with the difference that $G_{S_{1}D_{3}}$ changes from $(1/3)e^{2}/h$
for $L\ll\ell_{2}$ to $(1/6)e^{2}/h$ for $L\geq\ell_{2}$ in both
the $\ell_{1}<L_{V}$ and the $\ell_{1}>L_{V}$ regimes (once again,
in the regime where $\ell_{1}>L_{V}$, as $\ell_{2}$ decreases beyond
$\ell_{2}\sim L$, $G_{S_{1}D_{3}}$ decreases, vanishing as $\alpha\rightarrow\infty$
{[}\textcolor{black}{see Fig. $8(c),8(d)$ f}or renormalized edges{]}).
We note that similar results were seen in a recent experiment \cite{=000023Heiblum2017}.
In this experiment, Sabo et al. \cite{=000023Heiblum2017} tested
two devices, one with a small separation between the QPCs ($\sim.4\mu m$)
and one with a large separation between the two QPCs ($\sim9\mu m$),
and the conductance $G_{S_{1}D_{3}}$ changed from $(1/3)e^{2}/h$
to $(1/6)e^{2}/h$, which is consistent with the transition from $L\ll\ell_{2}$
to $L\geq\ell_{2}$ in our model.\\
\begin{figure}[H]
\centering{}\subfloat[\label{fig:9a 2QPC no wind}]{\includegraphics[width=8cm]{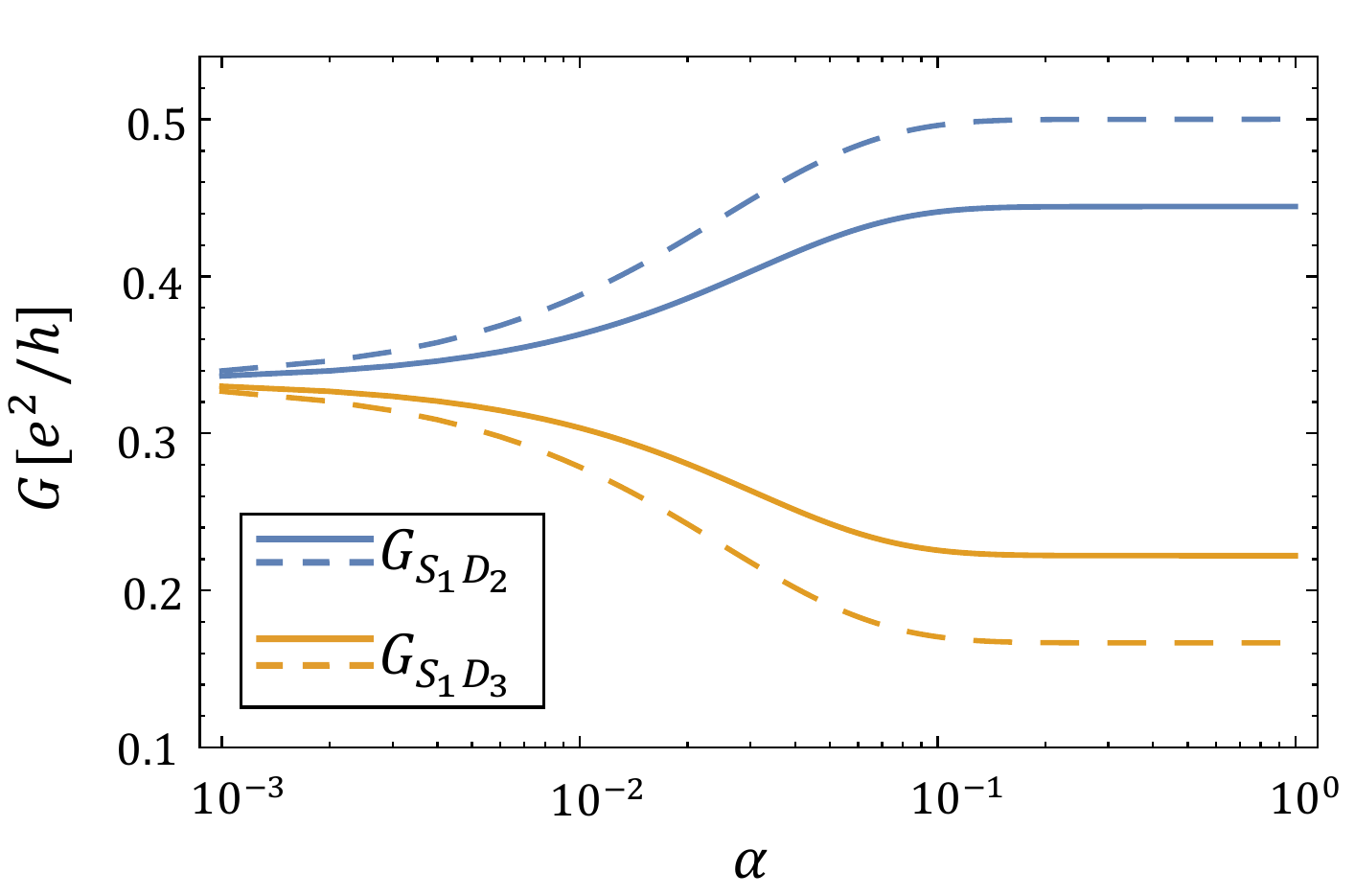}

}\qquad{}\subfloat[\label{fig:9b 2QPC wind}]{\includegraphics[width=8cm]{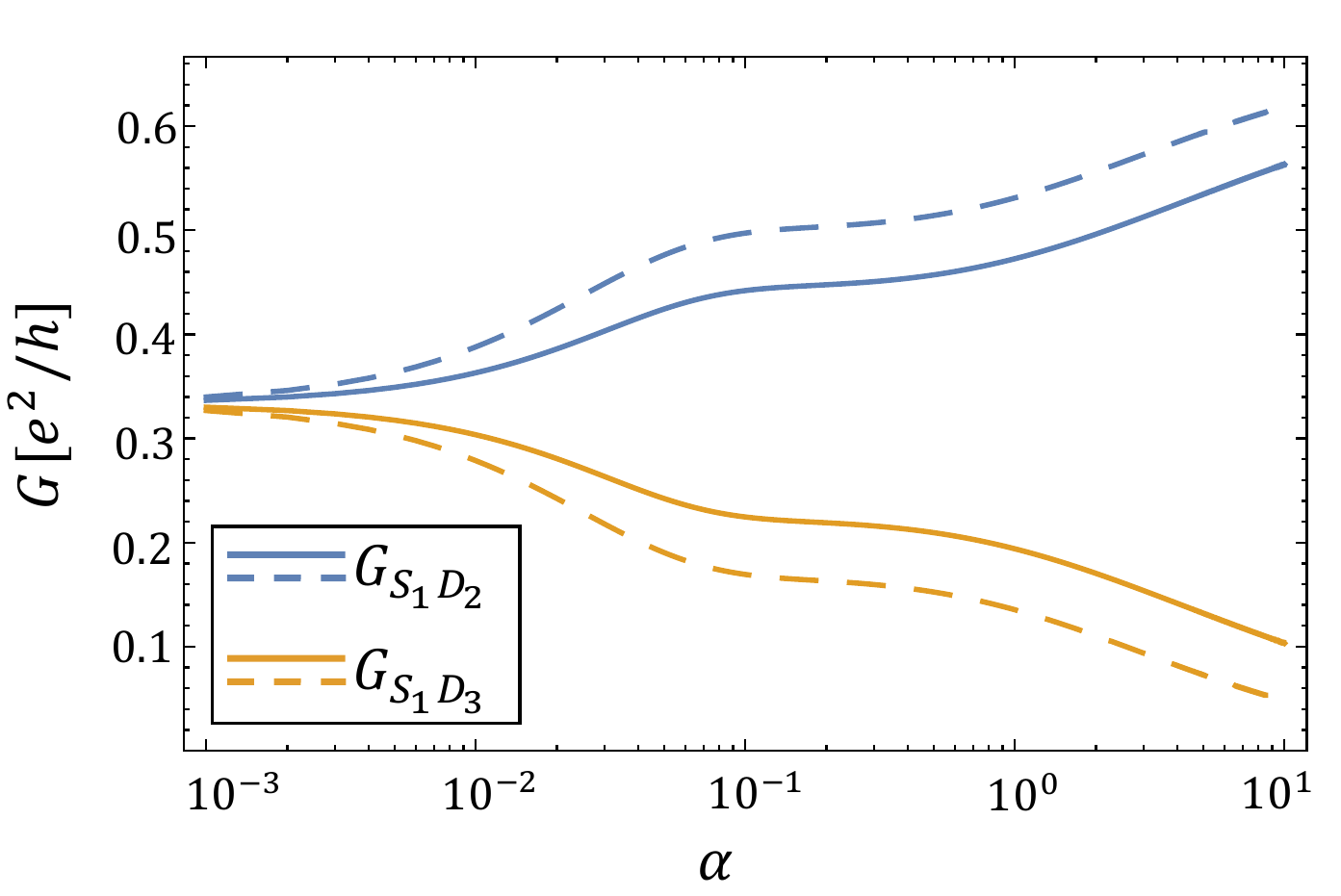}}\caption{\textcolor{black}{$(a)$ Log-linear conductance plot of the conductances
$G_{S_{1}D_{2}},G_{S_{1}D_{3}}$ as a function of $\alpha$ for $L=100a,L_{V}=10a,\ell_{1}=5a$;
the dashed lines correspond to the ``no-winding'' geometry, and
the solid lines correspond to the ``winding'' geometry. $(b)$ Log-linear
conductance plot of the conductances $G_{S_{1}D_{2}},G_{S_{1}D_{3}}$
as a function of $\alpha$ in the limit of a narrow constriction with,
$L=100a,L_{V}=0,\ell_{1}=5a$; the convention is the same as in $9(a)$.}}
\end{figure}

\section{Summary and Discussion}

\subsection{Summary}

In this work, we study transport along the QH edge for the $\nu_{\text{bulk}}=2/3$
FQH state, where impurity mediated tunneling between adjacent modes
induces equilibration. We assume that we are in the incoherent regime,
where subsequent tunneling events are incoherent with each other,
allowing us to ignore phase interference effects. Our model is a discrete
model consisting of counter-propagating modes linked by ``tunneling
bridges'' {[}\textcolor{black}{see Fig. }$2(a)${]}; the particular
form of the tunneling current can be calculated from the LL theory
using the Keldysh formalism, as discussed in Appendix B. Using this
form for the tunneling current and the local constitutive relations
for each mode (e.g. Eq. $(1)$), we then impose current and energy
conservation for each channel at each tunneling bridge, and go to
the continuum limit by taking the smallest scale of variation as $ia\rightarrow x$
and taking the number of tunneling bridges $n\rightarrow\infty$.
We finally apply boundary conditions from ideal contacts to obtain
temperature and voltage profiles, and thermal and electrical conductances. 

Within this scheme, we study two relevant models for the edge structure
at $\nu_{\text{bulk}}=2/3$ (MacDonald and WMG). The MacDonald model
(which should be relevant for a sharply-defined confining potential),
consists of two counter-propagating modes, a $\delta\nu=+1$ channel
and a counter-propagating $\delta\nu=-1/3$ channel. The WMG model
(which is relevant for a shallow confining potential), consists of
four channels, which are (counting from the bulk to the edge): a $\delta\nu=+1/3$
mode, a $\delta\nu=-1$ mode, a $\delta\nu=+1/3$ mode, and a final
$\delta\nu=-1/3$ mode. In the MacDonald picture, we consider only
electron tunneling between channels, and in the WMG picture, we consider
the most relevant (in the RG sense) electron and two-electron tunneling
operators between the inner three modes, and $1/3$-charge tunneling
between the outermost mode and the adjacent mode.

We now describe the main results of these calculations:
\begin{itemize}
\item MacDonald edge: For the MacDonald edge, we obtain quantized conductances
with corrections. In the line junction, the thermal conductance of
the transmitted channel $K\sim\left(\pi^{2}k^{2}T_{0}/3h\right)\left(\bar{\ell}/L\right)$
(vanishing thermal conductance with finite diffusive conductivity),
while the corresponding electrical conductance results reproduce the
quantization, but with \textit{exponentially} small corrections, i.e.
$G\sim\left(2e^{2}/3h\right)\left(1+e^{-\frac{2L}{\ell}}/3\right)$.
If we compare with the typical case of counter-propagating $\text{\ensuremath{\delta\nu}=+}1,-1$
modes, which corresponds to a quantum wire, we obtain vanishing electrical
conductance with diffusive $(\sim\ell/L)$ conductivity corrections.
Following KFP and switching to the charge/neutral basis, we see that
the neutral mode decays exponentially from the drain as $\sim e^{-\frac{2L}{\ell}}$,
and is proportional to the tunneling current; thus, the decay of the
neutral mode directly reflects the equilibration of the channels.
\textcolor{black}{This effect is also seen in the voltage induced
temperature profiles, where thermal equilibration occurs between the
$\delta\nu=+1$ and $\delta\nu=-1/3$ modes within a scattering length
$\sim\ell$ of the drain, and as a consequence generates heat near
the drain. Despite such heating effects, we find that the thermoelectric
response coefficients are identically zero}; this is a reflection
of the particle-hole symmetry of the linearized dispersion in the
LL model \cite{=000023Kane1996}. 
\item WMG: For the WMG edge structure, we numerically examine electrical
and heat transport in a single-QPC geometry, and electrical transport
in two different double-QPC geometries (the geometries are given in
\textcolor{black}{Figs. $8(a)$ and $8(b)$,} respectively). In this
model, we have two competing scattering lengths $\ell_{1}$ and $\ell_{2}$,
which are associated with tunneling between the inner-three modes
and the outer-mode tunneling, respectively; the competition between
these types of tunneling is controlled by the parameter $\alpha\equiv\ell_{1}/\ell_{2}$.
Additionally, the QPC is characterized by two length scales, $L_{V},L$,
which are the width of the QPC and the length of an arm of the QPC,
respectively. We see that the conductance results can be replicated
by a simplified analysis in the charge/neutral mode basis. For electrical
transport in the single-QPC geometry, we find that the conductance
from the source to drain $G_{S_{1}D_{3}}$ at $\alpha\ll1$ matches
well to the intermediate fixed point edge structure, while the $\alpha\rightarrow\infty$
behavior matches the KFP edge structure, with some crossover behavior;
in this setup, backscattering is strongly enhanced for $\ell_{1}>L_{V}$.
In the double QPC geometries, we obtain transitions from $(1/3)e^{2}/h$
to quantized conductances of $(2/9)e^{2}/h$ and $(1/6)e^{2}/h$ in
the ``winding'' and ``no-winding'' geometries, respectively \textcolor{black}{{[}see
Fig. $9${]}, with similarly enhanced backscattering for $\ell_{1}>L_{V}$;
again, all of the relevant features can be reproduced by the corresponding
fixed point edge structures in the $\alpha\ll1$ and $\alpha\rightarrow\infty$
regimes, with some continuous transition between them {[}see Figs.
$8(c,d)${]}. Finally, heat transport in the single QPC has $\sim1/L$
decay when the appropriate scattering lengths $\bar{\ell}_{1},\bar{\ell}_{2}\ll L$,
with some non-universal behavior occurring for $\bar{\ell}_{1}\ll L\lesssim\bar{\ell}_{2}$
{[}see Figs. $7(a,b)${]}. This behavior is nearly identical to the
results of the MacDonald model. However, it is possible to distinguish
between the two models in the regime where $\bar{\ell}_{1}\gg\bar{\ell}_{2},\ell_{1}\gg L_{V}$
by examining the thermal conductances at the three drains $D_{2},D_{3},D_{4}:$
in the MacDonald model we find that the thermal conductances always
obey the inequality $K_{S_{1}D_{3}}>K_{S_{1}D_{4}}>K_{S_{1}D_{2}}$,
whereas in this regime the WMG model follows $K_{S_{1}D_{2}}>K_{S_{1}D_{3}}>K_{S_{1}D_{4}}$.}
\end{itemize}

\subsection{Discussion}

We have seen from our incoherent, disordered transport model that
we can gain more insight into the process of equilibration between
counter-propagating modes in the $\nu_{\text{bulk}}=2/3$ FQH state.
In particular, we see that equilibration is essential for conductance
quantization, and that conductance quantization is achieved when the
system size is much larger than the appropriate scattering length,
i.e. $\ell\ll L$. In the case where there are multiple scattering
lengths and system size parameters (as in the case of the single and
double-QPC configurations with the outermost channel transmitted),
we have found that a variety of conductance results can be obtained,
and that the analysis can be simplified by switching to the charge/neutral
mode picture. Here we find that our model reproduces the fixed points
of the WMG model in the incoherent setting, with some crossover behavior.
For the thermal transport results, we can similarly see how equilibration
plays a role in the conductance quantization, as we obtain that the
thermal conductance vanishes as $\sim\bar{\ell}/L$ in the regime
that $\bar{\ell}\ll L$ in the case of two counter-propagating modes
(also as $\sim1/L$ with the condition that $\bar{\ell}_{1},\bar{\ell}_{2}\ll L$
in WMG case). Our model also admits local voltage and temperature
information on the edge, which shows how the equilibration process
takes place. 

Additionally, our calculation of the electrical conductance $G_{S_{1}D_{3}}$
in the two QPC geometry matches the experimental conductance crossover
from $(1/3)e^{2}/h$ to $(1/6)e^{2}/h$ as a function of the separation
between the QPCs \cite{=000023Heiblum2017}; in our model, this can
be understood simply as equilibration between an incoming fully-biased
mode and a grounded mode. This crossover behavior does not occur in
the simpler MacDonald model, and thus provides evidence for validity
of the WMG model. \textcolor{black}{Also, while recent experimental
progress with heat transport in the fractional regime has been achieved
in the experiments of Banerjee et al.}\textcolor{blue}{{} }\textcolor{black}{\cite{Banerjee2017,Banerjee 2018},
our predictions require more subtle experiments which can see how
thermal conductance scales with the system size $L$.}

However, there are some potential caveats concerning the results of
our calculations. The local information of the temperature and voltage
profiles is not easily accessible to experiment in the typical GaAs/AlGaAs
interface. There has been some progress in measuring temperature locally
using quantum dots \cite{Venkatachalam 2012,Altmiras 2009,Altmiras 2010,Gurman},
however this method is only applicable at single points along the
edge, and is not practical for measuring continually along the edge.
Additionally, on the model level, the temperature dependence of the
tunneling coefficients $g$ was ignored in our calculation, and the
electrical and energy tunneling currents were derived by assuming
certain constraints about the temperature and voltage biases; these
constraints are discussed in Appendix C.

\section{Acknowledgments}

\textcolor{black}{We thank Igor Gornyi, Alexander Mirlin, Dmitri Polyakov,
Ivan Protopopov, Yonathan Cohen, Wenmin Yang, and Moty Heiblum for
helpful discussions. We acknowledge financial support from DFG Grant
No. RO 2247/8-1, ISF Grant No. 1349/14, CRC 183 of the DFG, and the
Italia-Israel Collaborative Laboratory QUANTRA.}

\appendix
\renewcommand*{\thesection}{\appendixname~\Alph{section}}
\section{\textcolor{black}{Some bosonization details}}

The action for the bosonic edge fields is of the form:

\begin{equation}
S=\frac{\hbar}{4\pi}\int dxdt\sum_{ij}(K_{ij}\partial_{t}\phi_{i}\partial_{x}\phi_{j}-V_{ij}\partial_{x}\phi_{i}\partial_{x}\phi_{j}),
\end{equation}
where $K_{ij}$ an integer-valued symmetric matrix and $V_{ij}$ is
a positive definite matrix, with $K_{ij}=diag\left[3,-3,1,-3\right]$
in our case. The definition of the current is \cite{=000023Kane1995}
\begin{equation}
I_{i}=K_{il}^{-1}V_{lj}\rho_{j},
\end{equation}
with $\rho_{j}=-e\left(1/2\pi\right)\partial_{x}\phi_{j}$ the charge
density of the $j^{th}$ channel. The topological order of the state
is fully characterized by the matrix $K$ and the charge vector ${\bf t}$;
we work in the so-called ``symmetric basis'', where ${\bf t}=(1,1,1,1)^{T}$
and $K=diag\left[3,-3,1,-3\right]$. Additionally, the filling factor
is given by $\nu=\sum_{ij}t_{i}K_{ij}^{-1}t_{j}$ and the filling
factor discontinuities are given by $\delta\nu_{i}=\sum_{j}t_{i}K_{ij}^{-1}t_{j}$;
in the bare-mode basis, the filling factor discontinuities are $\delta\nu_{1}=+1/3,\delta\nu_{2}=-1/3,\delta\nu_{3}=+1,\delta\nu_{4}=-1/3$,
as noted in Fig. $1(b)$. The creation operator is of the form:
\begin{equation}
\hat{O}(x)=e^{i\sum_{j}n_{j}\phi_{j}(x)},
\end{equation}
which creates an excitation at point $x$ with charge $Q=\left(-e\right)\sum_{ij}n_{i}K_{ij}^{-1}t_{j}$.
From these definitions, we can see that the scattering vectors defined
by Eq. $(27)$ produce no net charge as required, and that neutral
modes should satisfy the conditions $Q=0$ and $\delta\nu=0$. Under
a change of basis, the above quantities transform as:
\begin{align}
K_{ij} & \rightarrow\tilde{K}_{ij}=U_{\alpha i}K_{\alpha\beta}U_{\beta j}\nonumber \\
V_{ij} & \rightarrow\tilde{V}_{ij}=U_{\alpha i}K_{\alpha\beta}U_{\beta j}\nonumber \\
t_{i} & \rightarrow\tilde{t}_{i}=U_{ji}t_{j}\nonumber \\
n_{i} & \rightarrow\tilde{n}_{i}=U_{ji}u_{j}\nonumber \\
\phi_{i} & \rightarrow\tilde{\phi}_{i}=U_{ij}^{-1}\phi_{j}
\end{align}

We can transform the fields $\phi_{i}$ into a new basis $\tilde{\phi}_{i}$.
The transformations are defined such that the action remains invariant:

\begin{equation}
\tilde{S}=\frac{\hbar}{4\pi}\int dxdt\sum_{ij}(\tilde{K}_{ij}\partial_{t}\tilde{\phi}_{i}\partial_{x}\tilde{\phi}_{j}-\tilde{V}_{ij}\partial_{x}\tilde{\phi}_{i}\partial_{x}\tilde{\phi}_{j}),
\end{equation}
with $\tilde{K}_{ij}\equiv U{}_{\alpha i}K_{\alpha\beta}U{}_{j\beta},\tilde{V}_{ij}\equiv U{}_{\delta i}V_{\delta\gamma}U{}_{j\gamma}$.
We see that the current in the new basis is of the form:
\begin{align}
\tilde{I}_{i} & =(\tilde{K}^{-1})_{il}\tilde{V}_{lj}\tilde{\rho}_{j}=\left[\left(U^{-1}\right)_{i\alpha}\left(K^{-1}\right)_{\alpha\beta}\left(U^{-1}\right)_{l\beta}\right]\left[U_{\gamma l}V_{\gamma\delta}U_{\delta j}\right]\left[\left(U^{-1}\right)_{j\omega}\rho_{\omega}\right]\nonumber \\
 & =\left(U^{-1}\right)_{i\alpha}\left(K^{-1}\right)_{\alpha\beta}V_{\beta\omega}\rho_{\omega}=(U^{-1})_{i\alpha}I_{\alpha}
\end{align}
We see that the current transforms in the same manner as the fields
$\phi_{i}$.

Additionally, we note that in the absence of tunneling, \textcolor{black}{we
assume a local current conservation condition even in the presence
of local electric fields $E_{x,i}=-\partial_{x}V_{i}(x)$. For the
discussion of the validity of this point, we refer the reader to the
section titled }\textbf{\textit{\textcolor{black}{Bulk currents}}}\textcolor{black}{{}
in Ref. \cite{=000023Kane1995}.}

\section{Calculation of tunneling currents}

In this Appendix, we derive electrical tunneling current (Eq.~(\ref{eq:3}))
and energy tunneling currents (Eq.~(\ref{eq:39})) between counter-propagating
$\delta\nu_{1}=+1$ and $\delta\nu_{\nu}=-\nu=-1/(2m+1)$ modes for
$m\in\mathbb{Z}^{+}$, or $\delta\nu_{1}=+\nu$ and $\delta\nu_{2}=-\nu$
modes at a single tunneling bridge $x=x_{\textrm{tun}}$ {[}see Fig.
$2(a)${]}. For the electric currents, we follow the similar scheme
t\textcolor{black}{o Ref. \cite{Martin 2005}. }We first consider
the case of the counter-propagating $\delta\nu_{1}=+1$ and $\delta\nu_{\nu}=-\nu$
modes. We assume that each mode is connected to each \textquotedbl{}reservoir\textquotedbl{}
with electro-chemical potential $\mu_{1(\nu)}=\mu_{0}+eV_{1(\nu)}$
and temperature $T_{1(\nu)}$. The total Hamiltonian is written as
$H=H_{0}+H_{V}+H_{\tau}$: $H_{0}$ is the Hamiltonian for the edge
modes as 
\begin{equation}
H_{0}=\frac{\hbar}{4\pi}\int\big[v_{1}(\partial_{x}\phi_{1})^{2}+\frac{v_{\nu}}{\nu}(\partial_{x}\phi_{\nu})^{2}\big]dx.\label{edgemodeHamiltonian}
\end{equation}
Here $\phi_{i=1,\nu}$ is bosonic field describing edge mode $i$,
and velocities $v_{1}$ and $v_{\nu}$ reflect the interaction within
each mode. We assume that the interaction between the modes is zero
for simplicity. The Hamiltonian $H_{\tau}$ describing electron tunneling
between the modes at $x=x_{\textrm{tun}}$ is written as 
\begin{equation}
H_{\tau}=\sum_{\epsilon=\pm}[\Gamma_{0}\Psi_{1}^{\dagger}(x_{\textrm{tun}})\Psi_{\nu}(x_{\textrm{tun}})]^{\epsilon}=\frac{1}{2\pi b}(\Gamma_{0}e^{-i(\phi_{1}(x_{\textrm{tun}})+\phi_{\nu}(x_{\textrm{tun}})/\nu)}+\textrm{H.c.}),\,\,\,\,\,\,\textrm{with}\,\Bigg\{\begin{matrix}[\Gamma_{0}\Psi_{1}^{\dagger}\Psi_{\nu}]^{+} & =\Gamma_{0}\Psi_{1}^{\dagger}\Psi_{\nu},\\{}
[\Gamma_{0}\Psi_{1}^{\dagger}\Psi_{\nu}]^{-} & =\Gamma_{0}^{*}\Psi_{\nu}^{\dagger}\Psi_{1},
\end{matrix}
\end{equation}
where $\Gamma_{0}$ is tunneling amplitude, a field operator $\Psi_{i}^{\dagger}(x)=\exp[-i\phi_{i}(x)/\delta\nu_{i}]/\sqrt{2\pi b}$
creates an electron at position $x$ on mode $i=1,\nu$ and $b$ is
the smallest length scale corresponding to the ultraviolet spatial
cutoff. Coupling each edge mode to each effective reservoir leads
to \textit{
\begin{equation}
H_{V}=-\frac{1}{2\pi}\sum_{i=1,\nu}\int\mu_{i}\partial_{x}\phi_{i}dx.
\end{equation}
}

The tunneling current operator $I_{\tau}$ is derived from the Heisenberg
equation of motion as 
\begin{equation}
I_{\tau}(t)=\frac{d}{dt}\Big(\frac{e}{2\pi}\int dx\partial_{x}\phi_{1}\Big)=-\frac{ie}{\hbar}\sum_{\epsilon=\pm}\epsilon[\Gamma_{0}\Psi_{1}^{\dagger}(x_{\textrm{tun}})\Psi_{\nu}(x_{\textrm{tun}})]^{\epsilon},\label{currentoperator}
\end{equation}
where we employed the commutation relation of $[\phi_{i}(x),\phi_{i}(x')]=i\pi\delta\nu_{i}\textrm{sgn}(x-x')$.
The tunneling Hamiltonian $H_{\tau,H_{0}+H_{V}}$ and the tunneling
current operator $I_{\tau,H_{0}+H_{V}}$ in the interaction picture
with respect to $H_{0}+H_{V}$ are written in terms of operators in
the interaction picture with respect to $H_{0}$ (denoted as subscript
$H_{0}$) as 
\begin{align}
H_{\tau,H_{0}+H_{V}}(t) & =\sum_{\epsilon=\pm}e^{-i\epsilon e(V_{1}-V_{\nu})t/\hbar}[\Gamma_{0}\Psi_{1,H_{0}}^{\dagger}(x_{\textrm{tun}},t)\Psi_{\nu,H_{0}}(x_{\textrm{tun}},t)]^{\epsilon},\nonumber \\
I_{\tau,H_{0}+H_{V}}(t) & =-\frac{ie}{\hbar}\sum_{\epsilon=\pm}\epsilon e^{-i\epsilon e(V_{1}-V_{\nu})t/\hbar}[\Gamma_{0}\Psi_{1,H_{0}}^{\dagger}(x_{\textrm{tun}},t)\Psi_{\nu,H_{0}}(x_{\textrm{tun}},t)]^{\epsilon}.\label{interactingpicture}
\end{align}
In the Keldysh formalism, the tunneling current is written as 
\begin{equation}
\langle I_{\tau}(t)\rangle=\frac{1}{2}\sum_{\eta=\pm1}\langle T_{C}{I_{\tau,H_{0}+H_{V}}(t^{\eta}})e^{-\frac{i}{\hbar}\int_{C}dt_{1}H_{\tau,H_{0}+H_{V}}(t_{1}^{\eta_{1}})}\rangle,\label{backscatteringcurrent}
\end{equation}
where $t^{\eta}$ denotes time $t$ on a upper (lower) branch $\eta=\pm1$
of Keldysh contour $C$ and is ordered in Keldysh ordering: Keldysh
ordering is defined as $t_{1}^{-}>t_{2}^{+}$ for all $t_{1}$ and
$t_{2}$, $t_{1}^{+}>t_{2}^{+}$ for $t_{1}>t_{2}$, and $t_{1}^{-}>t_{2}^{-}$
for $t_{1}<t_{2}$. Keldysh ordering operator $T_{C}$ arranges operators
in a sequence of their time argument's Keldysh ordering. Making use
of Eq.~(\ref{interactingpicture}), Eq.~(\ref{backscatteringcurrent})
in the second order in $\Gamma_{0}$ becomes 
\begin{align}
\langle I_{\tau}(t)\rangle & \simeq\frac{e|\Gamma_{0}|^{2}}{2\hbar^{2}}\sum_{\eta,\eta_{1}=\pm1}\sum_{\epsilon=\pm}\epsilon\eta_{1}\int_{-\infty}^{\infty}dt_{1}e^{i\epsilon e(V_{1}-V_{\nu})(t-t_{1})/\hbar}\nonumber \\
 & \times\langle T_{C}[\Psi_{1,H_{0}}^{\dagger}(x_{\textrm{tun}},t^{\eta})\Psi_{\nu,H_{0}}(x_{\textrm{tun}},t^{\eta})]^{\epsilon}[\Psi_{1,H_{0}}^{\dagger}(x_{\textrm{tun}},t_{1}^{\eta_{1}})\Psi_{\nu,H_{0}}(x_{\textrm{tun}},t_{1}^{\eta_{1}})]^{-\epsilon}\rangle,\nonumber \\
 & =\frac{ie|\Gamma_{0}|^{2}}{h^{2}b^{2}}\sum_{\eta,\eta_{1}=\pm1}\eta_{1}\int_{-\infty}^{\infty}dt_{1}\big[\sin\{\frac{e}{\hbar}(V_{1}-V_{\nu})(t-t_{1})\}e^{G_{1}^{\eta\eta_{1}}(t-t_{1})}e^{G_{\nu}^{\eta\eta_{1}}(t-t_{1})/\nu^{2}}\big].\label{tunnelingcurrentcal}
\end{align}
Here we introduce the Keldysh Green's function of mode $\phi_{i=1,\nu}$,
$G_{i}^{\eta_{1}\eta_{2}}(x,t_{1}-t_{2})=\langle T_{C}\phi_{i}(x,t_{1}^{\eta_{1}})\phi_{i}(0,t_{2}^{\eta_{2}})\rangle-\langle T_{C}\phi_{i}(0,t_{1}^{\eta_{1}})\phi_{i}(0,t_{1}^{\eta_{1}})\rangle$,
which is calculated as 
\begin{equation}
G_{i=1,\nu}^{\eta_{1}\eta_{2}}(x,t)=-|\delta\nu_{i}|\ln\bigg[\frac{\sin[\pi(b+i(v_{i}t\mp x)\chi_{\eta_{1}\eta_{2}}(t))/{\hbar v_{i}\beta_{i}}]}{\pi b/\hbar v_{i}\beta_{i}}\bigg],
\end{equation}
where $\beta_{i}\equiv1/(k_{B}T_{i})$ is the inverse temperature
of mode $i=1,\nu$, $-(+)$ represents chirality of mode $i=1(\nu)$,
and $\chi_{\eta_{1}\eta_{2}}(t)\equiv[(\eta_{1}+\eta_{2})/2]\textrm{sgn}(t)-(\eta_{1}-\eta_{2})/2$.
Because $G_{i}^{\eta\eta}(t)$ is symmetric with respect to $t=0$,
$G_{i}^{\eta\eta}(t)$ contribution in Eq.~(\ref{tunnelingcurrentcal})
is zero, and then Eq.~(\ref{tunnelingcurrentcal}) becomes 
\begin{align}
\langle I_{\tau}(t)\rangle & =-\frac{ie|\Gamma_{0}|^{2}}{h^{2}b^{2}}\sum_{\eta=\pm1}\eta\int_{-\infty}^{\infty}dt'\sin\Big(\frac{et'}{\hbar}(V_{1}-V_{\nu})\Big)e^{G_{1}^{\eta-\eta}(t')}e^{G_{\nu}^{\eta-\eta}(t')/\nu^{2}},\nonumber \\
 & =-\frac{ie|\Gamma_{0}|^{2}}{h^{2}b^{2}}\sum_{\eta=\pm1}\eta\int_{-\infty}^{\infty}dt'\sin\Big(\frac{et'}{\hbar}(V_{1}-V_{\nu})\Big)\bigg(\frac{\pi b/\hbar v_{1}\beta_{1}}{\sin[\pi(b-i\eta v_{1}t')/{\hbar v_{1}\beta_{1}}]}\bigg)\bigg(\frac{\pi b/\hbar v_{\nu}\beta_{\nu}}{\sin[\pi(b-i\eta v_{\nu}t')/{\hbar v_{\nu}\beta_{\nu}}]}\bigg)^{1/\nu}.
\end{align}
When the change of variable $t'\rightarrow t'-ib\eta/v_{\nu}+i\eta\hbar\beta_{\nu}/2$
is performed, the integral becomes simplified as 
\begin{align}
\langle I_{\tau}(t)\rangle & =-\frac{ie|\Gamma_{0}|^{2}}{h^{2}b^{2}}\sum_{\eta=\pm1}\eta\int_{-\infty+ib\eta/v_{\nu}-i\eta\hbar\beta_{\nu}/2}^{\infty+ib\eta/v_{\nu}-i\eta\hbar\beta_{\nu}/2}dt'\sin\Big[\frac{e(t'+i\eta\hbar\beta_{\nu}/2)}{\hbar}(V_{1}-V_{\nu})\Big]\nonumber \\
 & \times\bigg(\frac{\pi b/\hbar v_{1}\beta_{1}}{\sin\big[\frac{\pi}{\hbar v_{1}\beta_{1}}\big\{ b\big(1-\frac{v_{1}}{v_{\nu}}\big)-i\eta v_{1}t'+\frac{\hbar\beta_{\nu}v_{1}}{2}\big\}\big]}\bigg)\bigg(\frac{\pi b/\hbar v_{\nu}\beta_{\nu}}{\sin\big(\frac{\pi}{2}-\frac{i\eta\pi t'}{\hbar\beta_{\nu}}\big)}\bigg)^{1/\nu}\nonumber \\
 & =-\frac{ie|\Gamma_{0}|^{2}}{h^{2}b^{2}}\sum_{\eta=\pm1}\eta\int_{-\infty}^{\infty}dt'\sin\Big[\frac{e(V_{1}-V_{\nu})(t'+i\eta\hbar\beta_{\nu}/2)}{\hbar}\Big]\bigg(\frac{\pi b/\hbar v_{\nu}\beta_{\nu}}{\cosh(\pi t'/\hbar\beta_{\nu})}\bigg)^{1/\nu}\bigg(\frac{\pi b/\hbar v_{1}\beta_{1}}{\sin\big(\frac{\pi\beta_{\nu}}{2\beta_{1}}-\frac{i\pi\eta t'}{\hbar\beta_{1}}\big)}\bigg).\label{tunnelingcurrentcal2}
\end{align}
In the second equality of Eq.~(\ref{tunnelingcurrentcal2}), we changed
the integral range back to $-\infty$ to $\infty$ because in the
integrand of the first right hand side term of Eq.~(\ref{tunnelingcurrentcal2}),
there is no pole inside the rectangular contour with vertices $t'=\infty$,
$-\infty$, $-\infty+ib\eta/v_{\nu}-i\eta\hbar\beta_{\nu}/2$, and
$\infty+ib\eta/v_{\nu}-i\eta\hbar\beta_{\nu}/2$. under the condition
of $\hbar v_{\nu}\beta_{\nu}>b$. In the limit of $|T_{1}-T_{\nu}|\ll\overline{T}\equiv(T_{1}+T_{\nu})/2$,
the leading term of Eq.~(\ref{tunnelingcurrentcal2}) in $|T_{1}-T_{\nu}|$
is obtained, replacing $T_{1}$ and $T_{\nu}$ by $\overline{T}$,
as 
\begin{align}
\langle I_{\tau}(t)\rangle & \simeq\frac{2e|\Gamma_{0}|^{2}}{h^{2}b^{2}}\sinh\Big(\frac{e(V_{1}-V_{\nu})}{2k_{B}\overline{T}}\Big)\int_{-\infty}^{\infty}dt'\cos\Big(\frac{e(V_{1}-V_{\nu})t'}{\hbar}\Big)\bigg(\frac{\pi bk_{B}\overline{T}/\hbar v_{\nu}}{\cosh(k_{B}\overline{T}\pi t'/\hbar)}\bigg)^{1/\nu}\bigg(\frac{\pi bk_{B}\overline{T}/\hbar v_{1}}{\cosh(k_{B}\overline{T}\pi t'/\hbar)}\bigg)\nonumber \\
 & =\frac{2e|\Gamma_{0}|^{2}}{h^{2}b^{2}}\Big(\frac{\pi bk_{B}\overline{T}}{\hbar v_{\nu}}\Big)^{1/\nu}\Big(\frac{\pi bk_{B}\overline{T}}{\hbar v_{1}}\Big)\sinh\Big(\frac{e(V_{1}-V_{\nu})}{2k_{B}\overline{T}}\Big)\int_{-\infty}^{\infty}dt'\frac{\cos[e(V_{1}-V_{\nu})t'/\hbar]}{[\cosh(k_{B}\overline{T}\pi t'/\hbar)]^{1/\nu+1}}\nonumber \\
 & =\frac{2^{1+1/\nu}e|\Gamma_{0}|^{2}}{h^{2}bv_{1}}\Big(\frac{\pi bk_{B}\overline{T}}{\hbar v_{\nu}}\Big)^{1/\nu}\sinh\Big(\frac{e(V_{1}-V_{\nu})}{2k_{B}\overline{T}}\Big)\frac{|\Gamma(\frac{1}{2}+\frac{1}{2\nu}+i\frac{e(V_{1}-V_{\nu})}{2\pi k_{B}\overline{T}})|^{2}}{\Gamma(1+1/\nu)},%%\Big(\frac{\piak_{B}\overline{T}}{\hbarv_{1}}\Big)
\end{align}
where $\Gamma(x)$ is a gamma function and we used the integral equality
$\int_{-\infty}^{\infty}dt\cosh(2yt)/\cosh^{2x}t=2^{2x-1}\Gamma(x+y)\Gamma(x-y)/\Gamma(2x)$
in the case of $\textrm{Re}\,x>|\textrm{Re}\,y|$ and $\textrm{Re}\,x>0$.
%Notice that the transport coefficent, $d\langle I_{\tau} \rangle/ d( T_l - T_u ) |_{V_0 \rightarrow 0}$ is zero due to the linearlization of the electronic band structure near the Fermi %level. 
In the case of $e(V_{1}-V_{\nu})\ll k_{B}\overline{T}$, the tunneling
current is proportional to the voltage difference $(V_{1}-V_{\nu})$
as 
\begin{equation}
\langle I_{\tau}(t)\rangle=g_{1}\frac{e^{2}}{h}(V_{1}-V_{\nu}),\,\,\,\,\,\textrm{where}\,\,g_{1}=\frac{\left(2\pi\right)^{\frac{1}{\nu}-1}\left[\Gamma\left(\frac{1}{2}+\frac{1}{2\nu}\right)\right]^{2}}{\Gamma\left(1+\frac{1}{\nu}\right)}\left(\frac{\left|\Gamma_{0}\right|^{2}/b^{2}}{\left(\hbar v_{1}/b\right)k_{B}\overline{T}}\right)\left(\frac{k_{B}\overline{T}}{\left(\hbar v_{\nu}/b\right)}\right)^{\frac{1}{\nu}}.\label{Tunnelingcurrent}
\end{equation}
Note that a dimensionless effective conductance $g$ scales as $\overline{T}^{(1/\nu-1)}$.
It proves Eq.~(\ref{eq:3}) in the main text.

Let us a similar calculation for the energy current $J_{\tau}$ in
the same geometry as the above. The tunneling energy current is written
as 
\begin{align}
J_{\tau} & =-\frac{1}{2}\frac{d}{dt}\big(H_{L}-H_{R}\big)=\frac{i}{4}\sum_{\epsilon=\pm}\epsilon\big\{(v_{1}\partial_{x}\phi_{1}|_{x=x_{\textrm{tun}}}+\frac{v_{\nu}}{\nu}\partial_{x}\phi_{\nu}|_{x=x_{\textrm{tun}}}-\frac{\mu_{1}+\mu_{\nu}}{\hbar}),[\Gamma_{0}\Psi_{1}^{\dagger}(x_{\textrm{tun}})\Psi_{\nu}(x_{\textrm{tun}})]^{\epsilon}\big\},
\end{align}
where $\{A,B\}=AB+BA$ and we defined $H_{L}$ and $H_{R}$ as 
\begin{align}
H_{L}=\frac{\hbar v_{1}}{4\pi}\int(\partial_{x}\phi_{1})^{2}dx-\frac{1}{2\pi}\int\mu_{1}\partial_{x}\phi_{1}dx,\,\,\,\,\,\,\,\,H_{R}=\frac{\hbar v_{\nu}}{4\pi\nu}\int(\partial_{x}\phi_{\nu})^{2}dx-\frac{1}{2\pi}\int\mu_{\nu}\partial_{x}\phi_{\nu}dx.
\end{align}
The tunneling energy current operator $J_{\tau,H_{0}+H_{V}}$ in the
interaction picture with respect to $H_{0}+H_{V}$ is related to operators
in the interaction picture with respect to $H_{0}$ (denoted as subscript
$H_{0}$) as 
\begin{equation}
J_{\tau,H_{0}+H_{V}}=\frac{i}{4}\sum_{\epsilon=\pm}\epsilon e^{-i\epsilon e(V_{1}-V_{\nu})t/\hbar}\big\{(v_{1}\partial_{x}\phi_{1,H_{0}}|_{x=x_{\textrm{tun}}}+\frac{v_{\nu}}{\nu}\partial_{x}\phi_{\nu,H_{0}}|_{x=x_{\textrm{tun}}}-\frac{\mu_{1}+\mu_{\nu}}{\hbar}),[\Gamma_{0}\Psi_{1,H_{0}}^{\dagger}(x_{\textrm{tun}},t)\Psi_{\nu,H_{0}}(x_{\textrm{tun}},t)]^{\epsilon}\big\},
\end{equation}
In the Keldysh formalism, the tunneling energy current is written
as 
\begin{equation}
\langle J_{\tau}^{E}(t)\rangle=\frac{1}{2}\sum_{\eta=\pm1}\langle T_{C}{J_{\tau,H_{0}+H_{V}}(t^{\eta}})e^{-\frac{i}{\hbar}\int_{C}dt_{1}H_{\tau,H_{0}+H_{V}}(t_{1}^{\eta_{1}})}\rangle.\label{backscatteringenergycurrent}
\end{equation}
In the second order in $\Gamma_{0}$, the tunneling energy current
becomes 
\begin{align}
\langle J_{\tau}(t)\rangle & \simeq\frac{|\Gamma_{0}|^{2}}{8\hbar}\sum_{\eta,\eta_{1}=\pm1}\sum_{\epsilon=\pm}\epsilon\eta_{1}\int_{-\infty}^{\infty}dt_{1}e^{-i\epsilon e(V_{1}-V_{\nu})(t-t_{1})/\hbar}\langle T_{C}\big\{(v_{1}\partial_{x}\phi_{1,H_{0}}|_{x=x_{\textrm{tun}}}+\frac{v_{\nu}}{\nu}\partial_{x}\phi_{\nu,H_{0}}|_{x=x_{\textrm{tun}}}-\frac{\mu_{1}+\mu_{\nu}}{\hbar})\nonumber \\
 & ,[\Psi_{1,H_{0}}^{\dagger}(x_{\textrm{tun}},t^{\eta})\Psi_{\nu,H_{0}}(x_{\textrm{tun}},t^{\eta})]^{\epsilon}\big\}[\Psi_{1,H_{0}}^{\dagger}(x_{\textrm{tun}},t_{1}^{\eta_{1}})\Psi_{\nu,H_{0}}(x_{\textrm{tun}},t_{1}^{\eta_{1}})]^{-\epsilon}\rangle,
\end{align}
and is decomposed into $\langle J_{\tau}^{Q}\rangle$ and $\langle J_{\tau}^{N}\rangle$
as 
\begin{align}
\langle J_{\tau}(t)\rangle & =\langle J_{\tau}^{Q}(t)\rangle+\langle J_{\tau}^{N}(t)\rangle,
\end{align}
where 
\begin{align}
\langle J_{\tau}^{Q}(t)\rangle & =-\frac{|\Gamma_{0}|^{2}}{8\hbar}\sum_{\eta,\eta_{1}=\pm1}\sum_{\epsilon=\pm}\epsilon\eta_{1}\int_{-\infty}^{\infty}dt_{1}e^{i\epsilon e(V_{1}-V_{\nu})(t-t_{1})/\hbar}\nonumber \\
 & \times\langle T_{C}\big\{ v_{1}\partial_{x}\phi_{1,H_{0}}|_{x=x_{\textrm{tun}}}+\frac{v_{\nu}}{\nu}\partial_{x}\phi_{\nu,H_{0}}|_{x=x_{\textrm{tun}}},[\Psi_{1,H_{0}}^{\dagger}(x_{\textrm{tun}},t^{\eta})\Psi_{\nu,H_{0}}(x_{\textrm{tun}},t^{\eta})]^{\epsilon}\big\}[\Psi_{1,H_{0}}^{\dagger}(x_{\textrm{tun}},t_{1}^{\eta_{1}})\Psi_{\nu,H_{0}}(x_{\textrm{tun}},t_{1}^{\eta_{1}})]^{-\epsilon}\rangle,\nonumber \\
\langle J_{\tau}^{N}(t)\rangle & =\frac{|\Gamma_{0}|^{2}(\mu_{1}+\mu_{\nu})}{4\hbar^{2}}\sum_{\eta,\eta_{1}=\pm1}\sum_{\epsilon=\pm}\epsilon\eta_{1}\int_{-\infty}^{\infty}dt_{1}e^{i\epsilon e(V_{1}-V_{\nu})(t-t_{1})/\hbar}\nonumber \\
 & \times\langle T_{C}[\Psi_{1,H_{0}}^{\dagger}(x_{\textrm{tun}},t^{\eta})\Psi_{\nu,H_{0}}(x_{\textrm{tun}},t^{\eta})]^{\epsilon}[\Psi_{1,H_{0}}^{\dagger}(x_{\textrm{tun}},t_{1}^{\eta_{1}})\Psi_{\nu,H_{0}}(x_{\textrm{tun}},t_{1}^{\eta_{1}})]^{-\epsilon}\rangle.
\end{align}
As we will show now, $\langle J_{\tau}^{Q}\rangle$ corresponds to
tunneling heat current in the case of $e(V_{1}-V_{\nu})=0$ and $\langle J_{\tau}^{N}\rangle$
corresponds to tunneling energy current associated with the tunneling
charge current. Through a similar calculation to that of $\langle I_{\tau}(t)\rangle$
(from Eq.~(\ref{tunnelingcurrentcal}) to Eq.~(\ref{Tunnelingcurrent})),
$\langle J_{\tau}^{N}(t)\rangle$ is easily computed as 
\begin{align}
\langle J_{\tau}^{N}(t)\rangle & =\frac{i(\mu_{1}+\mu_{\nu})|\Gamma_{0}|^{2}}{2h^{2}b^{2}}\sum_{\eta,\eta_{1}=\pm1}\eta_{1}\int_{-\infty}^{\infty}dt_{1}\sin\Big(\frac{e}{\hbar}(V_{1}-V_{\nu})(t-t_{1})\Big)e^{G_{1}^{\eta\eta_{1}}(t-t_{1})}e^{G_{\nu}^{\eta\eta_{1}}(t-t_{1})/\nu^{2}}\nonumber \\
 & \simeq\frac{e}{2h}g_{1}(\mu_{1}+\mu_{\nu})(V_{1}-V_{\nu})=g_{1}\frac{\mu_{1}^{2}-\mu_{\nu}^{2}}{2h}.
\end{align}
Using the relation $\langle B_{1}e^{B_{2}}\rangle=\langle B_{1}B_{2}\rangle e^{\langle B_{2}^{2}\rangle/2}$
where $B_{1}$ and $B_{2}$ are linear operators of free bosons, $\langle J_{\tau}^{Q}(t)\rangle$
is written in terms of Green's functions of the modes as 
\begin{align}
\langle J_{\tau}^{Q}\rangle & =-\frac{i|\Gamma_{0}|^{2}}{4\pi hb^{2}}\sum_{\eta,\eta_{1}=\pm1}\eta_{1}\int_{-\infty}^{\infty}dt_{1}\cos\Big(\frac{e}{\hbar}(V_{1}-V_{\nu})(t-t_{1})\Big)e^{G_{\nu}^{\eta\eta_{1}}(t-t_{1})/\nu^{2}}e^{G_{1}^{\eta\eta_{1}}(t-t_{1})}\nonumber \\
 & \times\big(v_{1}\partial_{x}G_{1}^{\eta\eta_{1}}(x-x_{\textrm{tun}},t-t_{1})|_{x=x_{\textrm{tun}}}+\frac{v_{\nu}}{\nu^{2}}\partial_{x}G_{\nu}^{\eta\eta_{1}}(x-x_{\textrm{tun}},t-t_{1})|_{x=x_{\textrm{tun}}}\big)\nonumber \\
 & =-\frac{\pi|\Gamma_{0}|^{2}}{2h^{2}b^{2}}\sum_{\eta,\eta_{1}=\pm1}\eta_{1}\int_{-\infty}^{\infty}dt'\cos\Big(\frac{e}{\hbar}(V_{1}-V_{\nu})t'\Big)\chi_{\eta\eta_{1}}(t')\bigg(\frac{\pi b/\hbar v_{\nu}\beta_{\nu}}{\sin[\frac{\pi}{\hbar v_{\nu}\beta_{\nu}}(b+iv_{\nu}t'\chi_{\eta\eta_{1}}(t'))]}\bigg)^{1/\nu}\nonumber \\
 & \times\bigg(\frac{\pi b/\hbar v_{1}}{\sin[\frac{\pi}{\hbar v_{1}\beta_{1}}(b+iv_{1}t'\chi_{\eta\eta_{1}}(t'))]}\bigg)\bigg(\frac{1}{\beta_{1}}\frac{\cos[\frac{\pi}{\hbar v_{1}\beta_{1}}(b+iv_{1}t'\chi_{\eta\eta_{1}}(t'))]}{\sin[\frac{\pi}{\hbar v_{1}\beta_{1}}(b+iv_{1}t'\chi_{\eta\eta_{1}}(t'))]}-\frac{1}{\beta_{\nu}\nu}\frac{\cos[\frac{\pi}{\hbar v_{\nu}\beta_{\nu}}(b+iv_{\nu}t'\chi_{\eta\eta_{1}}(t'))]}{\sin[\frac{\pi}{\hbar v_{\nu}\beta_{\nu}}(b+iv_{\nu}t'\chi_{\eta\eta_{1}}(t'))\}}\bigg)\nonumber \\
 & =-\frac{\pi|\Gamma_{0}|^{2}}{2h^{2}b^{2}}\sum_{\eta=\pm1}\int_{-\infty}^{\infty}dt'\cos\Big(\frac{e}{\hbar}(V_{1}-V_{\nu})t'\Big)\bigg(\frac{\pi b/\hbar v_{\nu}\beta_{\nu}}{\sin[\frac{\pi}{\hbar v_{\nu}\beta_{\nu}}(b-i\eta v_{\nu}t')]}\bigg)^{1/\nu}\bigg(\frac{\pi b/\hbar v_{1}\beta_{1}}{\sin[\frac{\pi}{\hbar v_{1}\beta_{1}}(b-i\eta v_{1}t')]}\bigg)\nonumber \\
 & \times\bigg(\frac{1}{\beta_{1}}\frac{\cos[\frac{\pi}{\hbar v_{1}\beta_{1}}(b-i\eta v_{1}t')]}{\sin[\frac{\pi}{\hbar v_{1}\beta_{1}}(b-i\eta v_{1}t')]}-\frac{1}{\beta_{\nu}\nu}\frac{\cos[\frac{\pi}{\hbar v_{\nu}\beta_{\nu}}(b-i\eta v_{\nu}t')]}{\sin[\frac{\pi}{\hbar v_{\nu}\beta_{\nu}}(b-i\eta v_{\nu}t')]}\bigg).
\end{align}
The last equality comes from the fact that only $\chi_{\eta-\eta}(t)$
contribution to the integral survives because the $\chi_{\eta\eta}$
contribution makes the integrand odd with respect to $t=0$. When
the change of variable $t'\rightarrow t'-ib\eta/v_{\nu}+i\eta\hbar\beta_{\nu}/2$
is performed, the integral becomes simplified as 
\begin{align}
\langle J_{\tau}^{Q}\rangle & =-\frac{\pi|\Gamma_{0}|^{2}}{2h^{2}b^{2}}\sum_{\eta=\pm1}\int_{-\infty+ib\eta/v_{\nu}-i\eta\hbar\beta_{\nu}/2}^{\infty+ib\eta/v_{\nu}-i\eta\hbar\beta_{\nu}/2}dt'\cos\Big[\frac{e(V_{1}-V_{\nu})(t'+i\eta\hbar\beta_{\nu}/2)}{\hbar}\Big]\bigg(\frac{\pi b/\hbar v_{\nu}\beta_{\nu}}{\sin\big(\frac{\pi}{2}-\frac{i\eta\pi t'}{\hbar\beta_{\nu}}\big)}\bigg)^{1/\nu}\nonumber \\
 & \times\bigg(\frac{\pi b/\hbar v_{1}\beta_{1}}{\sin\big[\frac{\pi}{\hbar v_{1}\beta_{1}}\big\{ b\big(1-\frac{v_{1}}{v_{\nu}}\big)-i\eta v_{1}t'+\frac{\hbar\beta_{\nu}v_{1}}{2}\big\}\big]}\bigg)\bigg(\frac{1}{\beta_{1}}\frac{\cos\big[\frac{\pi}{\hbar v_{1}\beta_{1}}\big\{ b\big(1-\frac{v_{1}}{v_{\nu}}\big)-i\eta v_{1}t'+\frac{\hbar\beta_{\nu}v_{1}}{2}\big\}\big]}{\sin\big[\frac{\pi}{\hbar v_{1}\beta_{1}}\big\{ b\big(1-\frac{v_{1}}{v_{\nu}}\big)-i\eta v_{1}t'+\frac{\hbar\beta_{\nu}v_{1}}{2}\big\}\big]}-\frac{1}{\beta_{\nu}\nu}\frac{\cos\big(\frac{\pi}{2}-\frac{i\eta\pi t'}{\hbar\beta_{\nu}}\big)}{\sin\big(\frac{\pi}{2}-\frac{i\eta\pi t'}{\hbar\beta_{\nu}}\big)}\bigg)\nonumber \\
 & =-\frac{\pi|\Gamma_{0}|^{2}}{2h^{2}b^{2}}\sum_{\eta=\pm1}\int_{-\infty}^{\infty}dt'\cos\Big[\frac{e(V_{1}-V_{\nu})(t'+i\eta\hbar\beta_{\nu}/2)}{\hbar}\Big]\bigg(\frac{\pi b/\hbar v_{\nu}\beta_{\nu}}{\cosh(\pi t'/\hbar\beta_{\nu})}\bigg)^{1/\nu}\nonumber \\
 & \times\bigg(\frac{\pi b/\hbar v_{1}\beta_{1}}{\sin\big(\frac{\pi\beta_{\nu}}{2\beta_{1}}-\frac{i\pi\eta t'}{\hbar\beta_{1}}\big)}\bigg)\bigg(\frac{1}{\beta_{1}\tan\big(\frac{\pi\beta_{\nu}}{2\beta_{1}}-\frac{i\pi\eta t'}{\hbar\beta_{1}}\big)}-\frac{i\eta}{\beta_{\nu}\nu}\tanh(\frac{\pi t'}{\hbar\beta_{\nu}})\bigg).%\frac{\pi}{\hbarv_{1}\beta_{1}}\big\{ i\etav_{1}t'+\frac{\hbar\beta_{\nu}v_{1}}{2}\big\}\big]\bigg).
\end{align}
Let us compute $\langle J_{\tau}^{Q}\rangle$ in the first order in
$\Delta T=T_{1}-T_{\nu}$ in the limit of $|T_{1}-T_{\nu}|\ll\overline{T}$.
In addition, we ignore the term in the second order of $(V_{1}-V_{\nu})$
($O((V_{1}-V_{\nu})^{2})$). %Next, we will consider the first order in $\Delta T =T_1 - T_\nu$ while keeping $e(V_1-V_{\nu})$ zero. Notice that
%the zeroth order is zero because there is no temperature difference ($\Delta T = 0$) and voltage difference ($V_1- V_{\nu} = 0$).
The first order in $\Delta T$ contributes to $\langle J_{\tau}^{Q}\rangle$
as 
\begin{align}
\langle J_{\tau}^{Q}\rangle & \simeq\frac{\pi^{2}|\Gamma_{0}|^{2}k_{B}\Delta Tv_{\nu}}{2h^{2}b^{2}v_{1}}\Big(\frac{\pi bk_{B}\overline{T}}{\hbar v_{\nu}}\Big)^{1+1/\nu}\int_{-\infty}^{\infty}dt'\bigg(\frac{(1/\nu-1)[\cosh(\pi k_{B}\overline{T}t'/\hbar)]^{2}+(2-1/\nu)}{[\cosh(\pi k_{B}\overline{T}t'/\hbar)]^{1/\nu+3}}\bigg)%%\frac{\pi}{\hbarv_{1}\beta_{1}}\big\{-i\etav_{1}t'+\frac{\hbar\beta_{\nu}v_{1}}{}
\nonumber \\
 & =\frac{\pi^{2}(k_{B}\overline{T})(k_{B}\Delta T)}{h}\frac{1/\nu}{1/\nu+2}\frac{\left(2\pi\right)^{\frac{1}{\nu}-1}\left[\Gamma\left(\frac{1}{2}+\frac{1}{2\nu}\right)\right]^{2}}{\Gamma\left(1+\frac{1}{\nu}\right)}\left(\frac{\left|\Gamma_{0}\right|^{2}/b^{2}}{\left(\hbar v_{1}/b\right)k_{B}\overline{T}}\right)\left(\frac{k_{B}\overline{T}}{\left(\hbar v_{\nu}/b\right)}\right)^{\frac{1}{\nu}}\nonumber \\
 & =g_{1}\frac{\pi^{2}k_{B}^{2}(T_{1}^{2}-T_{\nu}^{2})}{6h}\frac{3/\nu}{1/\nu+2},
\end{align}
and it corresponds to the heat current in the linear response regime.
Note that in the non-interacting case ($\nu=1$), we reproduce Wiedemann-Franz
law $\langle J_{\tau}^{Q}\rangle=g\pi^{2}k_{B}^{2}(T_{1}^{2}-T_{\nu}^{2})/6h$.
The factor $(3/\nu)/(1/\nu+2)\neq1$ for $\nu\neq1$ is a correction
of Wiedemann-Franz law in the quantum Hall line junction system consisting
of $\delta\nu_{1}=+1$ and $\delta\nu_{2}=-\nu$ edge modes. %The total energy current is the summation of Eq.~\eqref{energycurrent1} and Eq.~\eqref{energycurrent2},
%\begin{equation}
%\langle I^{E}_{\tau} \rangle = \langle I^{E}_{\tau, 0} \rangle + \langle I^{E}_{\tau, p} \rangle = g \bigg [ \frac{\pi^2 k_B^2  (T_1^2 - T_\nu^2)}{6h} \frac{3/\nu}{1/\nu + 2} + \frac{ (\mu_l^2 - \mu_u^2)}{2h} \bigg ].
%\end{equation}
%The total energy current measured at $D_l$ is given as 
%\begin{equation}
%\langle I_{D_l}^{E} \rangle  = \frac{\mu_l^2}{2h} + \frac{\pi^2 k_B^2 T_l^2}{6h}  - \langle I_{\tau}^{E}\rangle. 
%\end{equation}

We do the similar calculation of the electric tunneling current and
the energy tunneling current between downstream $\nu=1/(2m+1)$ and
upstream $-\nu$ edge modes for $m\in Z^{+}$ (these correspond to
modes $1$ and $2$ of the WMG edge structure in the main text). The
only difference we have is that only quasiparticle tunneling between
the edge modes is considered. % with the following Hamiltonian
%\begin{align}
%H_{\tau}=\frac{1}{2\pi b}(\Gamma_{0}e^{-i(\phi_{\textrm{u}}(x_{\textrm{tun}})+\phi_{\textrm{l}}(x_{\textrm{tun}}))}+\textrm{H.c.}).
%\end{align}
%Here $\phi_{\textrm{u} (\textrm{l})}$ is the bosonic field for the upper (lower) edge mode.
The electric tunneling current $\langle I_{\tau}\rangle$ mode $1$
to mode $2$ is written as 
\begin{align}
\langle I_{\tau}\rangle=g_{2}\frac{\nu e^{2}}{h}(V_{\textrm{1}}-V_{2}),\label{Supple:electriccurrent}
\end{align}
where 
\begin{align}
g_{2}=\nu\left(2\pi\right)^{(2\nu-2)}\frac{\left(\Gamma(\nu)\right)^{2}}{\Gamma(2\nu)}\left(\frac{\left(\left|\Gamma_{0}\right|/b\right)}{k_{B}\overline{T}}\right)^{2}\left(\frac{k_{B}\overline{T}}{\left(\hbar v_{1}/b\right)}\right)^{\nu}\left(\frac{k_{B}\overline{T}}{\left(\hbar v_{2}/b\right)}\right)^{\nu},
\end{align}
$v_{1(\textrm{2})}$ is the velocity of the downstream (upstream)
edge mode, and $V_{1(\textrm{2})}$ is applied voltage to the downstream
(upstream) edge mode. The energy tunneling current $\langle J_{\tau}\rangle$
is written as 
\begin{align}
\langle J_{\tau}\rangle=\frac{\nu e^{2}}{2h}(V_{\textrm{1}}^{2}-V_{2}^{2})+g_{2}\frac{\pi^{2}k_{B}^{2}(T_{1}^{2}-T_{2}^{2})}{6h}\frac{3\nu}{2\nu+1}.\label{Supple:energycurrent}
\end{align}
Eqs.~(\ref{Supple:electriccurrent}) and (\ref{Supple:energycurrent})
are the first equation in Eq.~(\ref{eq:32}) and (\ref{eq:39}) in
the main text. 

\section{Range of validity of approximations }

The approximations made in the calculation of the tunneling currents
set restrictions on the range of validity of the temperature and voltage
profiles. In particular, the approximations are somewhat uncontrolled
(since they must be valid at every point $x$), so we can only minimize
their inaccuracies by putting certain restrictions on the energy scales
involved. From Eqs. $(B.12)$ and $(B.29)$, in order for $g_{1},g_{2}$
to be small, we require that
\begin{equation}
(\left|\Gamma_{0}\right|/b)\ll k_{B}\overline{T}(x)\ll(\hbar v/b)
\end{equation}
In other words, we require that the average temperature $\overline{T}(x)$
at each point be much smaller than the bandwidth $\hbar v/b$ of the
gapless modes on the edge, which are typical experimental conditions,
and we also require that both of these scales be much larger than
the tunneling energy; this corresponds to the weak tunneling limit.
We also note that we neglect the dependence of the $g_{1},g_{2}$
on $\overline{T}(x)$, noting that from the microscopic calculations
of Appendix B we have $g_{1}\propto\overline{T}^{2}$, and $g_{2}\propto\overline{T}^{-\frac{4}{3}}$.
We can minimize the errors due to this approximation by requiring
that:
\begin{equation}
\Delta T_{i}(x)\ll T_{0},
\end{equation}
where $\Delta T_{i}(x)\equiv T_{i}(x)-T_{0}$ is the deviation from
the ambient temperature $T_{0}$ on channel $i$.

\section{\textcolor{black}{Boundary conditions}}

\begin{figure}[H]
\begin{centering}
\subfloat[\label{fig:E.10.a single QPC}]{\includegraphics[width=7.5cm]{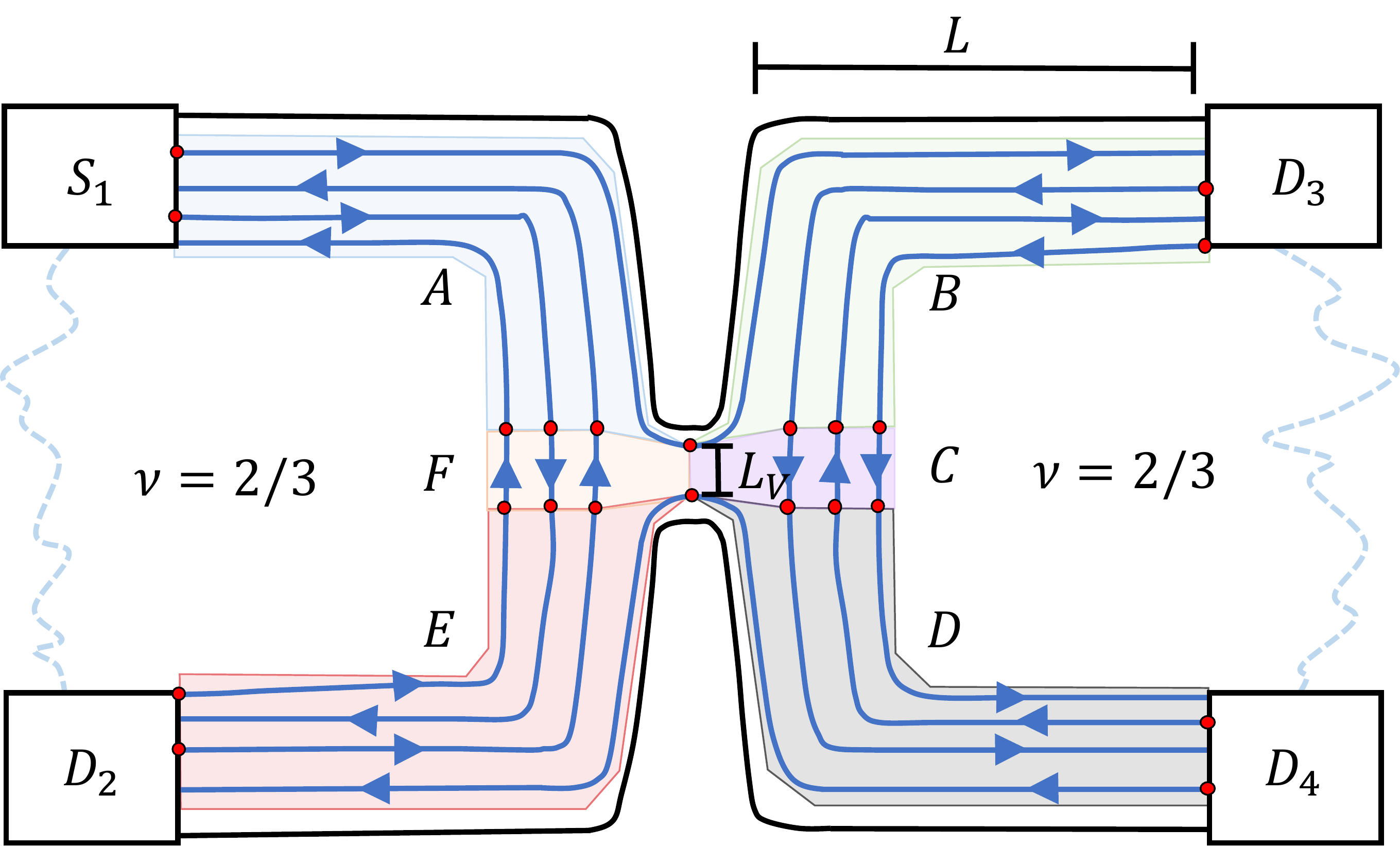}}
\par\end{centering}
\begin{centering}
\subfloat[\label{fig:E.10.b wind}]{\includegraphics[width=8cm]{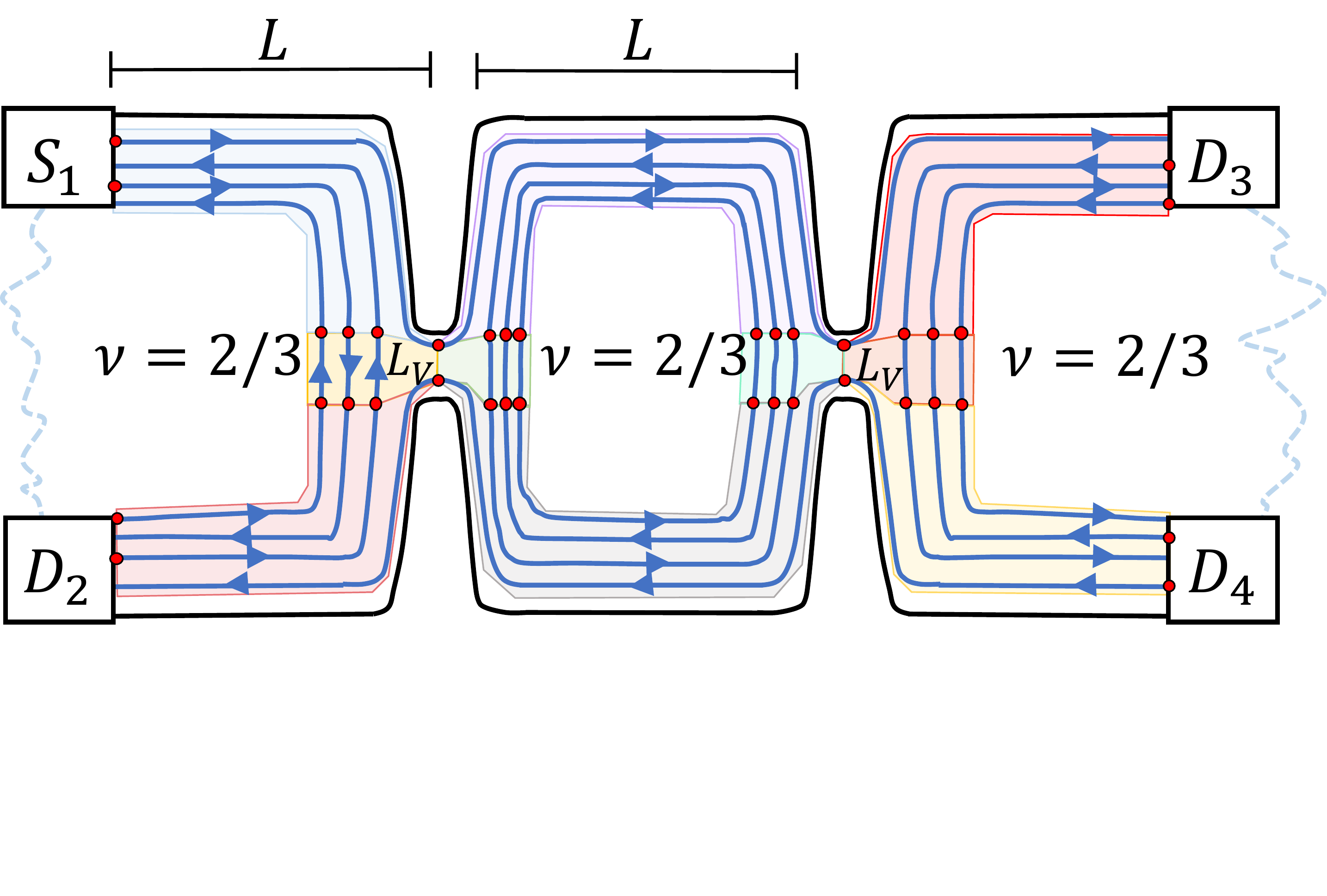}

}\qquad{}\subfloat[\label{fig:E.10.c no wind}]{\includegraphics[width=8cm]{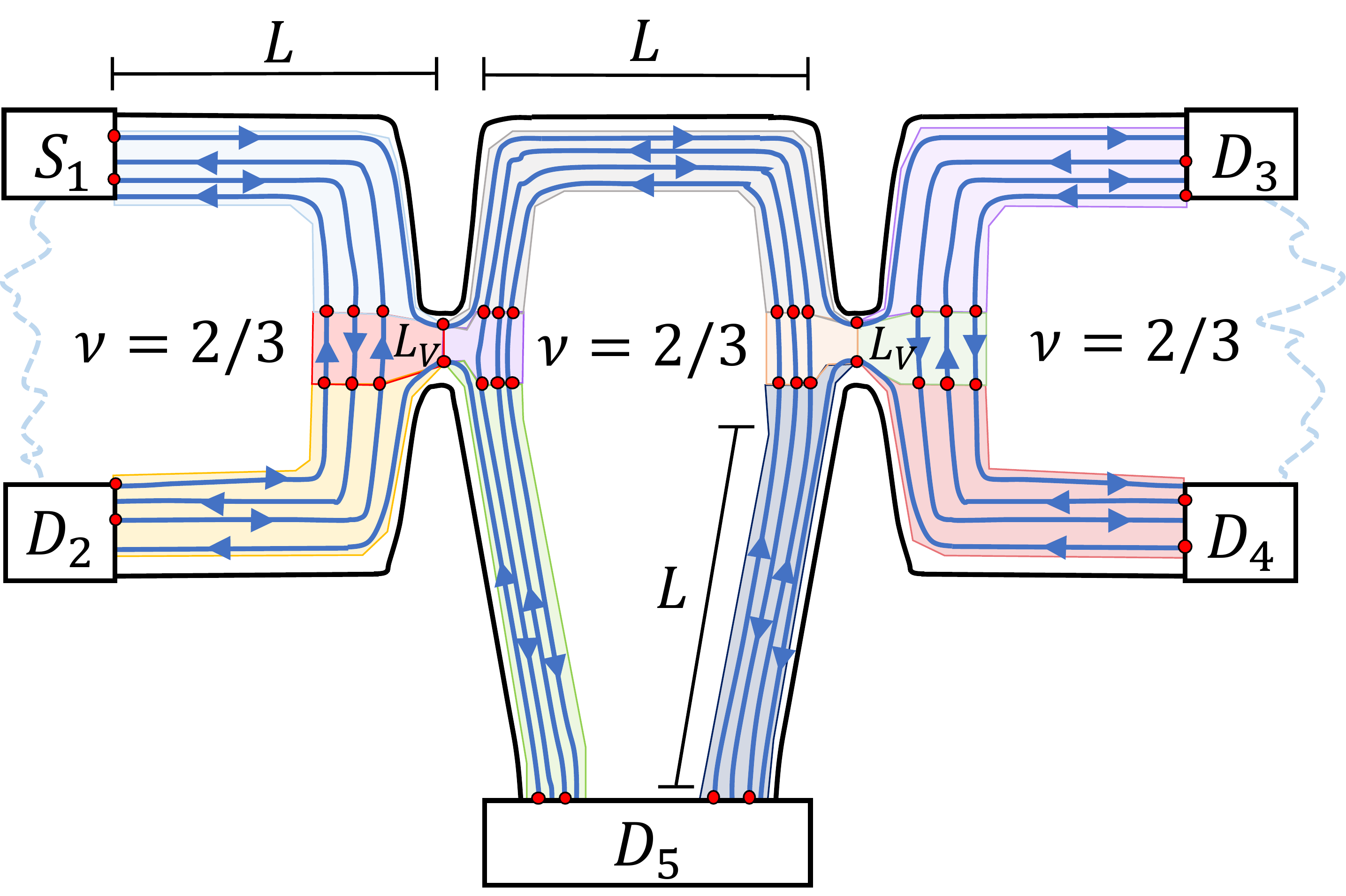}

}
\par\end{centering}
\centering{}\caption{(a) Single QPC geometry with points of boundary condition matching
shown (red dots); there are $22$ such points. The single QPC geometry
consists of six different regions ($A,B,C,D,E,F$ as shown in the
figure) where the solutions are matched; these regions are indicated
with translucent colorings. (b) Winding geometry with points of boundary
condition matching shown (red dots); there are $36$ such points.
The different regions where the solutions are matched are indicated
with translucent colorings. (c) No-winding geometry with points of
boundary condition matching shown (red dots); there are $40$ such
points. The different regions where the solutions are matched are
indicated with translucent colorings.}
\end{figure}
The boundary conditions are matched from the given solutions Eqs.
$(34),(35)$ in the electrical case, and Eqs. $(41),(42)$ in the
heat transport case. For example, for electrical transport in the
single QPC geometry, we have 6 regions (from upper left going clockwise,
we have regions $A,B,C,D,E,F$) to match the boundary conditions from
the solutions Eqs. $(34),(35)$. The direction of the coordinate $x$
always follows the chirality of the $\delta\nu_{3}=+1$ channel, so
that in regions $A,B$ the $x$ coordinate goes from left to right,
in region $C$ the $x$ coordinate goes from down to up, in regions
$D,E$ the $x$ coordinate goes from right to left and in region $F$
the $x$ coordinate goes from up to down. The boundary conditions
are given by the sources of the channels for each region, as shown
by the red dots in Figure $E.10(a)$; in this case, there are $22$.
Eventually, we get a set of coupled linear equations of the form:
\begin{equation}
M\cdot\vec{a}=\vec{I}_{ext},
\end{equation}
 with $\vec{I}_{ext}$ as the vector of the currents induced by the
external voltage $V_{0}$, $M$ a $22\times22$ matrix encoding the
current matching conditions, and $\bar{a}$ the vector of the coefficients
in each region. The boundary conditions $\vec{a}$ are plugged back
into the solutions, and the conductances $G_{S_{1}D_{i}}$ for $i=2,3,4$
are given by the sum of the incoming currents to the given drains
$D_{2,3,4}$, divided by the external voltage $V_{0}$. Solving for
the boundary conditions can be done analytically; however, the analytical
expressions for the final conductances are intractably large and impractical
to simplify, so therefore we numerically solved the system as a function
of the parameters $L,L_{V},\ell_{1},\alpha$.

\section{\textcolor{black}{Charge/neutral mode description of the edge}}

We have the following equation for the WMG edge:
\begin{equation}
\frac{1}{\ell_{1}}\begin{pmatrix}-\alpha & -\alpha & 0 & 0\\
\alpha & (6+\alpha) & 3 & 3\\
0 & -9 & -6 & -9\\
0 & 3 & 3 & 6
\end{pmatrix}\begin{pmatrix}I_{1}(x)\\
I_{2}(x)\\
I_{3}(x)\\
I_{4}(x)
\end{pmatrix}=\partial_{x}\begin{pmatrix}I_{1}(x)\\
I_{2}(x)\\
I_{3}(x)\\
I_{4}(x)
\end{pmatrix},
\end{equation}
which can be written in the simple form of $A\bar{I}(x)=\partial_{x}\bar{I}(x)$.
We wish to examine the problem in the neutral mode basis for the intermediate
fixed point $(\alpha\ll1)$ and the KFP fixed point $(\alpha\rightarrow\infty)$. 

\subsection*{Intermediate fixed point $(\alpha\ll1)$}

In this picture, we want to describe the problem in terms of the neutral
mode basis for $\alpha=0$. Contributions from the fact that $\alpha\neq0$
are interpreted as effective tunneling terms.

\subsubsection*{Neutral mode basis of inner three modes}

For $\alpha=0$, we can solve the problem through diagonalization
of $A$:
\begin{align}
A=UDU^{-1}, & \vec{\tilde{I}}(x)=U^{-1}\vec{\tilde{I}}(x)\nonumber \\
U^{-1}=\begin{pmatrix}0 & 1 & 1 & 2\\
0 & 3 & 2 & 3\\
0 & 1 & 1 & 1\\
1 & 0 & 0 & 0
\end{pmatrix}, & U=\begin{pmatrix}0 & 0 & 0 & 1\\
-1 & 1 & -1 & 0\\
0 & -1 & 3 & 0\\
1 & 0 & -1 & 0
\end{pmatrix},D=\begin{pmatrix}3 & 0 & 0 & 0\\
0 & 3 & 0 & 0\\
0 & 0 & 0 & 0\\
0 & 0 & 0 & 0
\end{pmatrix}
\end{align}
The rows of $U^{-1}$ can directly be read off as transformations
of the $\phi_{i}$ into the charge/neutral mode basis: the first and
second rows are neutral modes with $\delta\tilde{\nu}_{1,2}=0$, and
the third and fourth rows are downstream charge modes with $\delta\tilde{\nu}_{3,4}=+1/3$.
\textcolor{black}{}\\
\textcolor{black}{We can apply the same transformation to the matrix
$A(\alpha\neq0)$, i.e. transform the problem to the charge/neutral
mode basis of the incoherent fixed point; the picture that emerges
is that there are effective tunneling terms between all of the modes
on top of the incoherent fixed point at $\alpha=0$:}

\begin{align}
U^{-1}A(\alpha)U\vec{\tilde{I}}(x)\equiv\tilde{A}\vec{\tilde{I}}(x) & =\partial_{x}\vec{\tilde{I}}(x),\nonumber \\
\frac{1}{\ell_{1}}\begin{pmatrix}(3-\alpha) & \alpha & -\alpha & \alpha\\
-3\alpha & 3(1+\alpha) & -3\alpha & 3\alpha\\
-\alpha & \alpha & -\alpha & \alpha\\
\alpha & -\alpha & \alpha & -\alpha
\end{pmatrix}\begin{pmatrix}\tilde{I}_{1}(x)\\
\tilde{I}_{2}(x)\\
\tilde{I}_{3}(x)\\
\tilde{I}_{4}(x)
\end{pmatrix} & =\partial_{x}\begin{pmatrix}\tilde{I}_{1}(x)\\
\tilde{I}_{2}(x)\\
\tilde{I}_{3}(x)\\
\tilde{I}_{4}(x)
\end{pmatrix}
\end{align}
We note that the sum $\tilde{I}_{3}+\tilde{I}_{4}$ is conserved.
Solving the equation $\tilde{A}\vec{\tilde{I}}(x)=\partial_{x}\vec{\tilde{I}}(x)$,
we get the following solution:\\
\textcolor{black}{
\begin{align}
\vec{\tilde{I}}(x) & =a_{1}\begin{pmatrix}1\\
1\\
0\\
0
\end{pmatrix}e^{\frac{3x}{\ell_{1}}}+a_{2}\begin{pmatrix}0\\
0\\
-1\\
-1
\end{pmatrix}+a_{3}\begin{pmatrix}\frac{1}{6}\left(3-\sqrt{3\left(3+8\alpha\right)}\right)\\
\frac{1}{2}\left(3-\sqrt{3\left(3+8\alpha\right)}\right)\\
-\frac{1}{6}\left(3+\sqrt{3\left(3+8\alpha\right)}\right)\\
\frac{1}{6}\left(3+\sqrt{3\left(3+8\alpha\right)}\right)
\end{pmatrix}e^{\frac{\left(3-\sqrt{3\left(3+8\alpha\right)}\right)x}{2\ell_{1}}}\nonumber \\
 & +a_{4}\begin{pmatrix}-\frac{1}{6}\left(3+\sqrt{3\left(3+8\alpha\right)}\right)\\
\frac{1}{2}\left(3+\sqrt{3\left(3+8\alpha\right)}\right)\\
-\frac{1}{6}\left(3-\sqrt{3\left(3+8\alpha\right)}\right)\\
\frac{1}{6}\left(3-\sqrt{3\left(3+8\alpha\right)}\right)
\end{pmatrix}e^{\frac{\left(3+\sqrt{3\left(3+8\alpha\right)}\right)x}{2\ell_{1}}}
\end{align}
}From Eqs. $(34)$ and $(D.4)$, we have that $\tilde{I}_{3}+\tilde{I}_{4}=I_{1}+I_{2}+I_{3}+I_{4}$,
as required. Thus $\tilde{I}_{3},\tilde{I}_{4}$ carry all of the
physical current, and are the charge modes. \textcolor{black}{}\\
\textcolor{black}{The form further simplifies in the $\alpha\ll1$
limit:}\textcolor{red}{}

\begin{equation}
\vec{\tilde{I}}(x)\approx a_{1}\begin{pmatrix}1\\
1\\
0\\
0
\end{pmatrix}e^{\frac{3x}{\ell_{1}}}+a_{2}\begin{pmatrix}0\\
0\\
-1\\
-1
\end{pmatrix}+a_{3}\begin{pmatrix}-\frac{2\alpha}{3}\\
-2\alpha\\
-1\\
1
\end{pmatrix}e^{-\frac{2\alpha x}{\ell_{1}}}+a_{4}\begin{pmatrix}-1\\
3\\
\frac{2\alpha}{3}\\
-\frac{2\alpha}{3}
\end{pmatrix}e^{\frac{3x}{\ell_{1}}}
\end{equation}
To further understand the implications of this form for the transport
results, we look at the boundary conditions in the limit that $\ell_{1}/L\ll1,\alpha\ll1$.
In general, the boundary conditions can be written in the following
manner:
\begin{align}
\begin{pmatrix}0 & -1 & \frac{1}{6}\left(3+\sqrt{3\left(3+8\alpha\right)}\right) & \frac{1}{6}\left(3-\sqrt{3\left(3+8\alpha\right)}\right)\\
0 & 1 & \frac{1}{6}\left(9-\sqrt{3\left(3+8\alpha\right)}\right)e^{\frac{\left(3-\sqrt{3\left(3+8\alpha\right)}\right)L}{2\ell_{1}}} & \frac{1}{6}\left(9+\sqrt{3\left(3+8\alpha\right)}\right)e^{\frac{\left(3+\sqrt{3\left(3+8\alpha\right)}\right)L}{2\ell_{1}}}\\
-1 & -3 & -3 & -3\\
e^{\frac{3L}{\ell_{1}}} & 1 & e^{\frac{\left(3-\sqrt{3\left(3+8\alpha\right)}\right)L}{2\ell_{1}}} & e^{\frac{\left(3+\sqrt{3\left(3+8\alpha\right)}\right)L}{2\ell_{1}}}
\end{pmatrix}\begin{pmatrix}a\\
b\\
c\\
d
\end{pmatrix} & =\begin{pmatrix}I_{1}(0)\\
I_{2}(L)\\
I_{3}(0)\\
I_{4}(L)
\end{pmatrix}\nonumber \\
\xrightarrow{\alpha\ll1}\begin{pmatrix}0 & -1 & 1 & -\frac{2\alpha}{3}\\
0 & 1 & e^{-\frac{2\alpha L}{\ell_{1}}} & 2e^{\frac{3L}{\ell_{1}}}\\
-1 & -3 & -3 & -3\\
e^{\frac{3L}{\ell_{1}}} & 1 & e^{-\frac{2\alpha L}{\ell_{1}}} & e^{\frac{3L}{\ell_{1}}}
\end{pmatrix}\begin{pmatrix}a\\
b\\
c\\
d
\end{pmatrix} & =\begin{pmatrix}I_{1}(0)\\
I_{2}(L)\\
I_{3}(0)\\
I_{4}(L)
\end{pmatrix},
\end{align}
where $I_{1}(0),I_{2}(L),I_{3}(0),I_{4}(L)$ are the currents at the
boundaries in the original basis; the currents in each channel are
bounded by $\delta\nu_{i}\left(e^{2}/h\right)V_{0}$. \textcolor{black}{We
see from the form of the solutions that in order to suppress exponential
blow up, the boundary conditions for $a,d\sim e^{-\frac{3L}{\ell_{1}}}$
(up to signs), and we expect then also that $b,c\sim1$ (up to signs),
without the exponential suppression. Plugging these in to the charge/neutral
mode basis, the modes $\tilde{I}_{3},\tilde{I}_{4}$ are charge modes
with a decay term $\sim e^{-\frac{2\alpha x}{\ell_{1}}}$, with the
long decay characterizing the tunneling between them mediated by the
$g_{2}$ process; the $\sim e^{\frac{3x}{\ell_{1}}}$ part characteristic
of the neutral mode is largely suppressed (and is additionally suppressed
by $\alpha$). The neutral modes are characterized by their $\sim e^{\frac{3x}{\ell_{1}}}$
decay, with an additional slower $\sim e^{-\frac{2\alpha x}{\ell_{1}}}$
decay. Thus, in the ``middle'' of a segment of the edge (within
$>\ell_{1}$ of the edge of a boundary), we can characterize transport
purely by the charge modes tunneling between each other with the long
scale of $\sim e^{-\frac{2\alpha x}{\ell_{1}}}$, i.e. scattering
length of $\ell_{2}$, and in the $\alpha\ll1$ limit we can ignore
the $\sim e^{\frac{3x}{\ell_{1}}}$ decay term in the charge modes
altogether. In the case that $\ell_{2}>L$, the mixing between the
two downstream charge modes causes the conductance $G_{S_{1}D_{3}}$
to deviate from $\left(1/3\right)e^{2}/h$; as $\ell_{2}\lesssim L$,
the two modes can fully equilibrate with each other, and current mixes
equally between them; this is the regime where $G_{S_{1}D_{3}}\rightarrow\left(2/9\right)e^{2}/h$
for the winding case, and $G_{S_{1}D_{3}}\rightarrow\left(1/6\right)e^{2}/h$
for the no-winding case. }

\subsection*{KFP fixed point $(\alpha\rightarrow\infty)$}

Diagonalization of the problem leads to the following solution:
\begin{align}
\vec{\tilde{I}}(x) & =a_{1}\begin{pmatrix}1\\
0\\
0\\
0
\end{pmatrix}e^{\frac{3x}{\ell_{1}}}+a_{2}\begin{pmatrix}0\\
1\\
0\\
0
\end{pmatrix}+a_{3}\begin{pmatrix}0\\
0\\
1\\
0
\end{pmatrix}e^{\frac{\left(3-\sqrt{3(3+8\alpha)}\right)x}{2\ell_{1}}}+a_{4}\begin{pmatrix}0\\
0\\
0\\
1
\end{pmatrix}e^{\frac{\left(3+\sqrt{3(3+8\alpha)}\right)x}{2\ell_{1}}}\nonumber \\
 & \stackrel[\alpha\gg1]{}{\approx}a_{1}\begin{pmatrix}1\\
0\\
0\\
0
\end{pmatrix}e^{\frac{3x}{\ell_{1}}}+a_{2}\begin{pmatrix}0\\
1\\
0\\
0
\end{pmatrix}+a_{3}\begin{pmatrix}0\\
0\\
1\\
0
\end{pmatrix}e^{-\frac{\sqrt{6\alpha}x}{2\ell_{1}}}+a_{4}\begin{pmatrix}0\\
0\\
0\\
1
\end{pmatrix}e^{\frac{\sqrt{6\alpha}x}{2\ell_{1}}}
\end{align}
From this equation, we can see that in the $\alpha\rightarrow\infty$
limit, the two lower modes become localized near the boundaries, and
the electrical transport will be determined by the single $\delta\nu=+2/3$
charge mode, realizing an incoherent analogue of the KFP fixed point.
The problem in the diagonal basis is of the following form:
\begin{equation}
A=UDU^{-1},D=\left(\frac{1}{\ell_{1}}\right)diag\left[3,0,\left(3-\sqrt{3(3+8\alpha)}\right)/2,\left(3+\sqrt{3(3+8\alpha)}\right)/2\right],
\end{equation}
with the matrix $U^{-1}$ of the following form:
\begin{align}
U^{-1} & =\begin{pmatrix}0 & 0 & \frac{1}{2} & \frac{3}{2}\\
-\frac{1}{2} & -\frac{1}{2} & -\frac{1}{2} & -\frac{1}{2}\\
\frac{1}{4}\left(1+\frac{3}{\sqrt{3\left(3+8\alpha\right)}}\right) & \frac{1}{4}\left(1-\frac{9}{\sqrt{3\left(3+8\alpha\right)}}\right) & -\frac{3}{2\sqrt{3\left(3+8\alpha\right)}} & -\frac{3}{2\sqrt{3\left(3+8\alpha\right)}}\\
\frac{1}{4}\left(1-\frac{3}{\sqrt{3\left(3+8\alpha\right)}}\right) & \frac{1}{4}\left(1+\frac{9}{\sqrt{3\left(3+8\alpha\right)}}\right) & \frac{3}{2\sqrt{3\left(3+8\alpha\right)}} & \frac{3}{2\sqrt{3\left(3+8\alpha\right)}}
\end{pmatrix},
\end{align}
At the $\alpha\rightarrow\infty$ fixed point, the essential physics
happens around the constrictions, where there are counter-propagating
$\delta\nu=+2/3$ and $\delta\nu=-1/3$ modes which tunnel between
each other, with an additional neutral mode. To see the physics near
the constriction, we wish to transform the $3\times3$ problem into
a suitable basis. We first require that $\tilde{I}_{2}$ is a $1/3$
mode, $\tilde{I}_{3}$ is a $2/3$ mode and $\tilde{I}_{4}$ is a
neutral mode. The discontinuity in the filling factor is given by
$\delta\nu_{i}=\sum_{j}t_{i}K_{ij}^{-1}t_{j}$, so that under the
change of basis we require that:
\begin{align}
\delta\tilde{\nu}_{2} & =\sum_{j}\tilde{t}_{2}\tilde{K}_{ij}^{-1}\tilde{t}_{j}=-\frac{1}{3}\nonumber \\
\delta\tilde{\nu}_{3} & =\sum_{j}\tilde{t}_{3}\tilde{K}_{ij}^{-1}\tilde{t}_{j}=+\frac{2}{3}\nonumber \\
\delta\tilde{\nu}_{4} & =\sum_{j}\tilde{t}_{4}\tilde{K}_{ij}^{-1}\tilde{t}_{j}=0
\end{align}
Additionally, we restrict the charges of the neutral mode and the
$-1/3$ mode excitations:
\begin{align}
\tilde{Q}_{3} & =\left(-e\right)\sum_{ij}\left(U_{3\times3}\right)_{2i}^{-1}K_{ij}^{-1}t_{j}=\frac{e}{3}\nonumber \\
\tilde{Q}_{4} & =\left(-e\right)\sum_{ij}\left(U_{3\times3}\right)_{3i}^{-1}K_{ij}^{-1}t_{j}=0
\end{align}
Finally, we require local current conservation between the two charge
modes: this forces the matrix $A_{3\times3}$ in its most general
form to be: 
\begin{equation}
\tilde{A}_{3\times3}=U^{-1}A_{3\times3}U=\frac{1}{\ell_{1}}\begin{pmatrix}a & b & c\\
-a & -b & -c\\
d & e & f
\end{pmatrix}
\end{equation}
for some undetermined constants $a,b,c,d,e,f$. Putting these constraints
together, we find that $\tilde{A}_{3\times3}=\frac{1}{\ell_{1}}\begin{pmatrix}6 & 3 & 0\\
-6 & -3 & 0\\
0 & 0 & 3
\end{pmatrix}$. Additionally, we require that the transformation obeys global current
conservation:
\begin{equation}
\tilde{I}_{2}+\tilde{I}_{3}=I_{2}+I_{3}+I_{4}
\end{equation}
The solution of the untransformed equation $A_{3\times3}\vec{I}(x)=\partial_{x}\vec{I}(x)$
is 
\begin{equation}
\vec{I}(x)=b_{1}\begin{pmatrix}-1\\
0\\
1
\end{pmatrix}e^{\frac{3x}{\ell_{1}}}+b_{2}\begin{pmatrix}-1\\
1\\
0
\end{pmatrix}e^{\frac{3x}{\ell_{1}}}+b_{3}\begin{pmatrix}1\\
-3\\
1
\end{pmatrix}.
\end{equation}
Following the aforementioned constraints, the transformed solution
becomes: 
\begin{equation}
\vec{\tilde{I}}(x)=\frac{b_{1}}{X}\begin{pmatrix}(u_{3}+u_{9})\\
-(u_{3}+u_{9})\\
-(2+u_{2}+u_{8})
\end{pmatrix}e^{\frac{3x}{\ell_{1}}}+\frac{b_{2}}{X}\begin{pmatrix}u_{9}\\
-u_{9}\\
-\left(1+u_{8}\right)
\end{pmatrix}e^{\frac{3x}{\ell_{1}}}+b_{3}\begin{pmatrix}1\\
-2\\
0
\end{pmatrix},
\end{equation}
with
\begin{equation}
U_{3\times3}=\begin{pmatrix}1+2u_{2} & u_{2} & u_{3}\\
-\left(1+2u_{2}+2u_{8}\right) & \left(1-u_{2}-u_{8}\right) & -\left(u_{3}+u_{9}\right)\\
1+2u_{8} & u_{8} & u_{9}
\end{pmatrix},
\end{equation}
and $X\equiv u_{3}(1+u_{8})-u_{9}(1+u_{2})$. Solving for $\vec{\tilde{I}}(x)$
and $U_{3\times3}$ fully requires matching boundary conditions in
the new basis, which is prohibitively difficult (as it will involve
expressing the solution in terms of the original coefficients, which
are complicated). The main observations about Eq. $(D.15)$ are that:
$(1)$ the neutral mode consists of a purely decaying part with decay
length $\ell_{1}/3$, which is a reflection of the decoupling of the
neutral mode from the charge modes in $\tilde{A}_{3\times3}$, $(2)$
the tunneling between the charge modes $\delta\tilde{\nu}_{2},\delta\tilde{\nu}_{3}$
induces equilibration while the sum $\tilde{I}_{2}+\tilde{I}_{3}=I_{2}+I_{3}+I_{4}$
as required, and $(3)$ in the absence of tunneling between charge
modes, the charge modes obey the condition that $\tilde{I}_{2}=-2\tilde{I}_{3}$,
which is a reflection of the restriction on the filling factors $\delta\tilde{\nu}_{2}=-1/3,\delta\tilde{\nu}_{3}=+2/3$.


\begin{thebibliography}{10}
\bibitem[35]{footnote} We note that while our numerical parameters
clearly violate the condition that $g_{1},g_{2}\ll1$, this is not
a problem, since in principle all of the lengths in the problem can
be rescaled by some factor $c$, such that $g_{1}\rightarrow g_{1}/c,g_{2}\rightarrow g_{2}/c$,
and $g_{1}/c,g_{2}/c\ll1$, and this rescaling will not change any
of the calculable results.
\end{thebibliography}
\end{document}